%% file: ms.tex
\documentclass[emulateapj]{emulateapj}

\begin{document}
\shorttitle{h \& $\chi$ Persei}
\shortauthors{Currie, T. et al.}
\title{Spitzer/IRAC and JH$K_{s}$ Observations of h \& $\chi$ Persei: Constraints on 
Protoplanetary Disk and Massive Cluster
 Evolution at $\sim 10^{7}$ yr}

\author{Thayne Currie\altaffilmark{1}, Zoltan Balog\altaffilmark{2,3}, S. J. Kenyon\altaffilmark{1},
,G. Rieke\altaffilmark{2}, L. Prato\altaffilmark{4},
E. T. Young \altaffilmark{2}, J. Muzerolle\altaffilmark{2}, D.P. Clemens\altaffilmark{5}, M. Buie\altaffilmark{4}, 
D. Sarcia \altaffilmark{5}, A. Grabau\altaffilmark{5}, E. V. Tollestrup\altaffilmark{5}, B. Taylor\altaffilmark{4}, 
E. Dunham\altaffilmark{4}, \& G. Mace\altaffilmark{4}} 
\altaffiltext{1}{Harvard-Smithsonian Center for Astrophysics, 60 Garden St. Cambridge, MA 02140}
\altaffiltext{2}{Steward Observatory, University of Arizona,  933 N. Cherry Av. Tucson, AZ 85721}
\altaffiltext{3}{on leave from Dept. of Optics and Quantum Electronics, University of Szeged, H-6720, Szeged, Hungary}
\altaffiltext{4}{Lowell Observatory}
\altaffiltext{5}{Institute for Astrophysical Research, Boston University, 725 Commonwealth Avenue, Boston, MA 02215}
\email{tcurrie@cfa.harvard.edu}
\begin{abstract}
We describe IRAC 3.6-8 $\mu m$ observations and ground-based near-IR JH$K_{s}$ photometry from Mimir and 2MASS 
of the massive double cluster 
h \& $\chi$ Persei complete to J=15.5 ($M\sim 1.3 M_{\odot}$). 
Within 25' of the cluster centers we detect 
$\sim 11,000$ sources with J$\le$ 15.5, $\sim 7000$ sources with [4.5] $\le$ 15, and $\sim 5000$ sources
with [8]$\le$ 14.5.  
 In both clusters, the surface density profiles derived from the
2MASS data decline with distance from the cluster centers as expected
for a bound cluster.
 Within 15' of the cluster centers,
$\sim$ 50\% of the stars lie on a reddened $\sim$ 13 Myr isochrone;
at 15'-25' from the cluster centers, $\sim$ 40\% lie on this isochrone.
Thus, the optical/2MASS color-magnitude diagrams indicate that h \& $\chi$
Per are accompanied by a halo population with roughly the same age and
distance as the two dense clusters.
The double cluster lacks any clear IR excess sources for J$\le 13.5$ 
($\sim 2.7 M_{\odot}$).  Therefore, disks around high-mass stars 
disperse prior to $\sim 10^{7}$ yr.  At least $2-3\%$ of the fainter 
cluster stars have strong IR excess at both [5.8] and [8].
About  $4-8\%$ of sources slightly more massive than the Sun ($\sim 1.4 
M_{\odot}$) have IR excesses at [8].  Combined with the lack of detectable 
excesses for brighter stars, this result suggests that disks around lower-mass stars 
have longer lifetimes.
The IR excess population also appears to be larger at longer IRAC bands ([5.8], [8]) than at shorter 
IRAC/2MASS bands ($K_{s}$, [4.5]), a result consistent with
an inside-out clearing of disks.  
\end{abstract}
\keywords{Galaxy: Open Clusters and Associations: Individual: NGC Number: NGC 869, Galaxy: Open Clusters and Associations: Individual: NGC Number: NGC 884, Stars: Circumstellar Matter, Infrared: Stars}
\section{Introduction}
The evolution of circumstellar disks sets strong constraints on the initial conditions for terrestrial 
and gas giant planet formation.
In young stars, the presence of circumstellar disks is inferred by near-to-mid infrared (IR) emission by 
dust in excess of the output of 
stellar photospheres.  
While the vast majority of stars in $\sim$ $1$ Myr clusters have disks, 
the frequency 
of disks declines on $\sim$ 1-10 Myr timescales (e.g. Hillenbrand 2005; Young et al. 2004; Mamajek et al. 2004; Haisch et al. 2001). 
Because circumstellar dust is the evolutionary precursor to $\gtrsim$ 1 km-sized planetesimals, the 
timescale for the disappearance of dust emission, the 'disk evolution timescale', is an important
constraint for planet formation models (e.g. Strom et al. 1989; Hillenbrand 2005; Alibert et al. 2005; Currie 2005).

Recent observations of young ($\lesssim$ 10 Myr) stars suggest that
the evolution timescale may depend on stellar properties and locations within the disk. 
K band ($\sim 2$$\mu m$) excess from the inner ($r\lesssim 0.1 AU$) disk 
is relatively uncommon around both massive and very low mass stars (Hillenbrand et al. 1998). 
The $\sim$ 5 Myr Upper Sco OB association also exhibits a mass-dependent frequency of disks: 
sources with IR excess at 8 $\mu m$ and at 16 $\mu m$ appear to be more 
prevalent around K/M and B/A stars than for F and G stars (Carpenter et al. 2006).
Spatially resolved, mid-IR observations of $\sim 3-5$ Myr T Tauri stars (McCabe et al. 2006) 
show that 
 many disks with strong mid-IR emission lack near-IR emission, providing evidence for a location-dependent 
disk evolution. 
The drop in near-IR circumstellar dust emission may be explained by grain growth, which is probably faster 
in the inner disk regions due to higher midplane densities and orbital frequencies (e.g. Dullemond \& Dominik 2005). 
Circumstellar gas dispersal by UV photoevaporation (Clarke et al. 2001; Alexander et al. 2006) 
and planetesimal formation by gravitational instability (Youdin \& Shu 2002; Youdin \& Chiang 2004; Currie 2005) 
 may also result in a location-dependent evolution and, thus, a wavelength-dependent timescale for the disappearance of near-IR emission.  

Multiwavelength near-to-mid IR observations of evolved ($\gtrsim 10$ Myr), massive ($\gtrsim$ 1000 members) 
clusters are required to measure the disk evolution timescale as a function of stellar mass and 
disk properties. 
The double cluster h \& $\chi$ Persei, the most massive, evolved open cluster within 3 kpc, 
provides a rich laboratory to
study disk evolution in more detail. Starting with the
initial study of Oosterhoff (1937), there has been much debate
concerning the age, distance, and stellar content of the clusters
(e.g. Borgman \& Blaauw 1964; Wildey 1964; Schild 1967; Crawford et al. 1970; Vogt 1971; 
Tapia et al. 1984; Marco \& Benabeau 2001; Capilla \& Fabregat 2002). 
 Recent work (Keller et al. 2001; Slesnick et al. 2002, hereafter S02; 
Bragg \& Kenyon 2005, hereafter BK05) has converged on a nearly
identical age for both clusters, $\sim$ 12-13 Myr,
a common distance of $\sim$ 2.34 kpc (a distance
modulus $\sim$ 11.85), a low extinction of E(B-V)$\sim$0.5 uniform across the clusters (BK05), 
and an initial mass function
(IMF) of massive stars consistent with results for other
young, massive clusters (Massey 2003; Bragg 2004). Although the clusters have
similar masses of $\gtrsim$ 4000 $M_{\odot}$ (h Per) and $\gtrsim$
3000 $M_{\odot}$ ($\chi$ Per), the internal dynamical structure of the
clusters may be very different (BK05), with some evidence
for mass segregation in both clusters (S02, BK05).

While numerous optical photometric and spectroscopic studies of h \& $\chi$ Persei 
exist, the cluster has yet to be explored in detail in the near-to-mid IR.
Previous IR observations have been shallow, J$\sim$K
$\lesssim$ 11-13 (Tapia et al. 1984; Bragg \& Kenyon 2002, respectively), and
concentrated on understanding the population of Be stars.
While some constraints have been placed on the stellar population 
of h \& $\chi$ Per, the population of \textit{circumstellar disks} remains 
unprobed. Thus, its potential to inform disk evolution 
theories remains untapped.

In this paper, we describe the analysis of the first deep IR survey
of h \& $\chi$ Persei. The survey combines 2MASS All-Sky Survey (2MASS) and 
Infrared Array Camera (IRAC) data covering
the entire double cluster with deeper near-IR data of selected
cluster regions from the Mimir camera at Lowell Observatory. 
With approximate completeness limits in JH$K_{s}$ about 3-5 magnitudes deeper 
than BK02 and Tapia et al. (1984), respectively,
the 2MASS survey reaches stars with masses
smaller than the limits of all optical surveys except for Keller et al. (2001)
with less interference from the large line-of-sight extinction.
When combined with longer wavelength (3.6, 4.5, 5.8, \& 8 $\mu m$) Spitzer/IRAC
data, 2MASS and deeper near-IR surveys allow probes of disk evolution
for stars with ages $\sim$ 10-15 Myr and masses $\gtrsim$ 1.3 $M_{\odot}$.
From our analysis we hope to provide valuable input for 
circumstellar disk and massive cluster evolution models.

Our results provide good evidence for two populations of $\sim$
13 Myr old stars in the direction of h \& $\chi$ Per. In both clusters, the
sky surface density declines approximately inversely with distance at $\le$ 20'
from the cluster centers, which is consistent with results derived from optical
data (Bragg \& Kenyon 2005).  At distances beyond $\sim$ 20' from the cluster
centers, the surface density is consistent with the background density.
CMDs indicate that nearly half of the stars within 15'
of the cluster centers lie on a reddened 13 Myr isochrone. In regions of
lower surface density 15'--25' from the cluster centers, however, roughly
40\% of the stars lie on the same isochrone. Thus, the double cluster may
be accompanied by an extensive halo population, much like the Orion
star-forming region, that has roughly the same age and distance as
{\it bone fide} cluster members.

While the 
majority of sources have photospheric IRAC colors through 8$\mu m$, the cluster and halo population associated
with h \& $\chi$ Persei have a small IR excess population.
Disks around stars with $J\le 13.5$ ($\sim 2.7 M_{\odot}$) 
are extremely rare: disks around massive stars disperse by $\sim 10^{7}$ years.  The 
IR excess population is larger for fainter sources down to J=15.5; disks around lower mass stars 
 may have longer lifetimes.  The IR excess population also grows progressively 
larger with longer IRAC wavelengths, a result consistent with an inside-out clearing of protoplanetary disks.

We begin with a description of the data and reduction in Section 2 and discuss
the analysis of the near-IR data in Sections 3 and 4. In Section 5, we develop our technique
to identify IR excess souces and use this approach to make several
estimates of the fraction of IR excess sources in the cluster.
We conclude with a brief summary in Section 6.
\section{Observations and Data Reduction}
\subsection{Near-Infrared Ground-Based J,H, \& $K_{s}$ data from 2MASS}
2MASS is an all-sky survey with uniform, complete photometry
at J, H, and K$_{s}$ (Skrutskie et al. 2006). The survey
has 10$\sigma$ sensitivity for point sources with J $\gtrsim$
15.8, H $\gtrsim$ 15.1, and K$_s$ $\gtrsim$ 14.3. For sources with
10$\sigma$ detections, the survey is more than 99\% complete
and more than 99.95\% reliable, with 1$\sigma$ astrometric
accuracies of at least 0.1 arcsec relative to the {\it Hipparcos}
reference frame for sources with K$_s$ $\lesssim$ 14. Although the
2MASS survey used cameras with 2 arcsec pixels, multiple observations
of each sky position yields images with a nominal resolution of
1 arcsec.

We downloaded data for h \& $\chi$ Persei from the 2MASS point
source catalog (PSC) using the IRSA web interface at
IPAC\footnote{http:$//$irsa.ipac.caltech.edu$/$}. The catalog
includes $\sim$ 31,000 sources with 5$\sigma$ detections within
1 deg of $\alpha_{J2000} = 2^{h} 20^{m} 2.9^{s}$
$\delta_{J2000}= 57^{o} 5' 41.7"$ (l=135$^{o}$, b=-3.7$^{o}$),
which is a point midway between the two clusters.
A skymap from the 2MASS PSC clearly shows the two clusters as
dense concentrations of stars (Figure 1). There are $\sim$ 11,000
sources within 25' of either cluster center.

Figure 2 shows the magnitude and error distributions for J and K$_s$
using stars within 25' of the cluster centers. At J ($K_{s}$), the number
counts monotonically increase to J = 15.5 (K$_s$ = 14.8) and then
turn over.  Both of these peaks are somewhat brighter than the nominal
10$\sigma$ sensitivity limits, suggesting that confusion of sources
in the galactic plane or the cluster centers might limit the source
counts. However, both clusters show secondary peaks at J $\approx$ 14.5
and at K$_s \approx$ 14. We show below that stars within the two
clusters produce these peaks.
\subsection{Near-Infrared Ground-Based J, H, \& $K_{s}$ data from Mimir}
The relatively low spatial resolution and the shallow peaks in the JHK$_s$ magnitude distributions
of the 2MASS data prompted us to acquire additional near-IR photometry.  
We used Mimir, a multifunction IR instrument (http://people.bu.edu/clemens/mimir/) at the f/17
focus of the 1.8m Perkins telescope at Lowell Observatory (Table 1).
Mimir uses a Mauna Kea JHK filter set and covers a 3'$\times$3' field with 0.18" pixels in this
configuration.  
The total on-cluster coverage was
$\sim 144$ square arc-minutes (Figure 1).
A series of flats, darks, and
biases was taken at the beginning and end of each night.  The telescope was dithered by 
30" for each pointing where a series of three 10-second exposures were taken.
In most cases the pointings were too crowded to construct sky frames from median filtering.
  Therefore, we took a series of three 10-second dithered pairs of exposures for sky frames
7' in right ascension and declination away from each dithered pair of on-cluster pointings.

We followed a standard image processing procedure.  First, we subtracted a dark frame from each image and divided by
a dark-subtracted and normalized flat field. 
Next, we subtracted each object frame by
the appropriate median-filtered sky frame.  We used a custom IDL procedure to interpolate over
bad pixels as well as image/detector
artifacts unique to each frame.
We used another custom IDL procedure (similar to 'imcombine' in IRAF) to stack 10-second frames together.  
Up to 18 individual frames were stacked together for each pointing
yielding a total integration time of $\sim 3$ minutes each in a given field.  
We obtained longer integrations of $\sim$ 7 minutes for two of the fields, one each adjacent to the h \& $\chi$ Per 
centers,
 typically improving the completeness limits by $\sim 0.3$ magnitudes.
  Each stacked image was visually inspected for errors in offset computation and residual
image artifacts.

We identified sources and extracted aperture photometry using
SourceExtractor (SE; Bertin \& Arnouts 1996).  Each potential source was convolved with a 9x9 gaussian filter (5-pixel 
FWHM) and sources on image edges were removed to eliminate spurious detections.  
We typically used a 12 pixel diameter for
aperture photometry in all bands with a tendency to use smaller apertures for K data (due to smaller point-spread 
functions) and subtracted the background from a filtered, global background map.  
Sources from each dithered pair were matched to generate a final list of detections in each
band.   

To derive an absolute calibration, we matched Mimir sources with
30-40 2MASS zero-color sources on each frame. Once we established
the zero-point, these sources had $\lesssim$ 1\% offset as a function
of JHK$_s$. For redder sources, we measured a reliable offset between
K(Mimir) and K$_{s}$(2MASS) of $\sim$ 3\% on sources with J-K $\sim$ 1.
Applying the conversions from Carpenter et al. (2001) to correct
Mimir colors to 2MASS colors eliminated the offset. For the ensemble
of Mimir sources, we estimate absolute uncertainties of $\lesssim$
0.025 mag relative to 2MASS in all bands.

To derive coordinates for the Mimir sources, we relied on accurate
2MASS astrometry. Although tests with standard packages, such as
WCStools, showed that the Mimir fields were well-oriented north-south
and showed little distortion and rotation, the small Mimir fields
made robust astrometric solutions difficult. Because our main goals
for astrometry were matching 2MASS sources and estimating incompleteness,
we computed J2000 coordinates for Mimir sources by matching several 2MASS
sources per field and deriving relative coordinates using the known
pixel scale. Comparisons of all Mimir sources with 2MASS counterparts
yields an average positional offset of 0.8" $\pm$ 0.4", with little
evidence for systematic offsets as a function of position on the Mimir
detector. Although better coordinate accuracy might be possible with
complete astrometric solutions for all Mimir fields, this positional
uncertainty is sufficient for robust source matching even in the centers of
each cluster.

Figure 3 shows the J and K$_s$ magnitude distributions for the Mimir data.
These data reach $\sim$ 0.5 mag deeper than the 2MASS data, with clear
peaks in the counts at J = 16 and at K$_s$ = 15.5. Of the $\sim$ 1000
Mimir detections, $\sim$ 650 sources with J $\le$ 16 have 5$\sigma$
detections in all three bands. Multiple observations of $\sim$ 50 sources
verified these uncertainties. At the 2MASS 10$\sigma$ sensitivity limit,
we recover all 2MASS sources.
At this limit, the typical magnitude
difference was $\delta$J $\approx$ 0.1 mag and $\delta$K$_s \approx$ 0.15 mag, 
which is consistent with the expected error distribution.

Figure 3 shows the same secondary peaks as the 2MASS data. For a cluster
age of 13 Myr, a distance modulus of 11.85, the Siess et al. (2000)
isochrones, and the color conversion table from Kenyon \& Hartmann (1995),
the peaks at J = 14.5 and K$_s$ = 14.25 correspond to stars with masses
$M = 1.6 M_{\odot}$. With these assumptions, stars at the J = 16 limit
have masses of 1.1--1.2 $M_{\odot}$.
In Section 3, we restrict our analysis of the Mimir data to those 
650 sources detected at all bands.
\subsection{IRAC 3.6-8 $\mu m$ data}
We obtained observations of h \& $\chi$ Persei on January 18, 2004 with IRAC
(Fazio et al. 2004) on the Spitzer Space Telescope.
The IRAC survey covers about $\sim 0.75$ $\deg^{2}$ centered on
$\alpha_{2000}$ = 02:20:29.166, $\delta_{2000}$ = +57:12:27.54 (Figure 1). There is a $\sim$ 7
arcmin offset between the channel 1/3 and channel 2/4 mosaics.
We used the 12s high
dynamic range mode to obtain two frames in
each position, with 0.4s and 10.4s integration times. The
observation of each field was repeated with a small offset, which
allowed 20.8s integration time for each position. We identified image artifacts 
and cosmic ray hits by comparing the two observations.  

   The data were taken during a period of above-normal solar activity.  To 
remove cosmic rays we evaluated two approaches.
First, we took the conventional steps of mosaicing and extracting sources,
relying on the cosmic ray circumvention features of the software. 
We used PhotVis (v. 1.09)
 for source finding and aperture photometry (see Gutermuth et al. 2004).
The radii of the source
aperture, and the inner and outer sky annuli were 2.4, 2.4, and 7.2
arc-seconds respectively. 
In the second approach, we used a custom IDL routine developed by T. C. to extract sources individually from the BCD
data prior to mosaicing, applied array-location-dependent photometric corrections 
(see Quijada et al. 2004; http://ssc.spitzer.caltech.edu/irac/locationcolor), removed sources near image edges, and 
interpolated over pixels flagged as cosmic ray hits/image artifacts in the 
first method.
This approach degraded the detection limit but allowed better
control of the cosmic ray artifacts than by mosaicing alone. 
Source Extractor aperture photometry was performed using a 4 pixel diameter aperture;
the background count level and rms was computed from the filtered, global background pixel map.

We compared the results of the two methods and examined the raw data to determine the best method 
for each band. This comparison included examining the number of sources with colors bluer
than Rayleigh Jeans and the scatter in color-color plots, both taken as indicators of
cosmic ray effects.
The source extraction prior to mosaicing provided the most reliable results
in bands 1 and 2 ([3.6] and [4.5]), while mosaicing first was better in bands 3 and 4 ([5.8] and [8]). 
The slightly larger
PSFs in the latter two bands provided enough pixels for the circumvention software to distinguish 
cosmic rays from real point sources.  
Since the intrinsic ratio of signal to noise is higher at [3.6] and [4.5]
than [5.8] and [8], the degradation from extracting sources prior to mosaicing was 
not serious for the shorter wavelength bands.

We calibrated the photometry using large aperture
measurements of standard stars obtained during h \& $\chi$ Per observations, applying an aperture
correction for each channel to account for the difference
between the aperture sizes used for standard and h \& $\chi$ Per photometry (see Reach et al. 2005 
for calibration details).  
The
brightest sources ($\le 10$th magnitude) are saturated even in the short exposure frames.
Figure 4 shows the mosaic image of the [3.6] channel at low contrast.

We bandmerged the data using a sub-arcsecond matching radius to minimize the
contamination of any residual cosmic ray hits mistakenly identified as 'sources'
by the source extraction algorithms of SE and PhotVis.  

To remove sources that are likely AGN or galaxies with aromatic emission, we rely on published optical data, our JH$K_{s}$ 
data, and IRAC colors.  Requiring 2MASS, Mimir, or optical counterparts eliminated most highly reddened extragalactic sources.  
In the IRAC bands, typical field AGN have [4.5]$>$ 14 and [4.5]-[8] $\gtrsim$ 1.25 
(Gutermuth, unpublished); galaxies with aromatic emission lie to the 
right of a line from ([3.6]-[5.8],[4.5]-[8]) $\sim$ (0,1) to $\sim$ (1.5,3) 
 with [4.5]-[8] $\gtrsim$ 1 as well as 
([4.5]-[5.8],[5.8]-[8]) $\sim$ (0,1) to $\sim$ (1,2.25) with [5.8]-[8] $\gtrsim 1$.  
We identify very few sources with optical/near-IR counterparts and [4.5]-[5.8] $\ge$ 1 or [5.8]-[8] $\ge 1$.  Thus, extragalactic 
sources have negligible impact on our analysis.

Figure 5 shows magnitude distributions for the four IRAC bands and error distributions for [4.5] and [8].
At [3.6] and [4.5], the distributions have a monotonic rise to
[3.6] $\approx$ [4.5] $\sim$ 14, a broad plateau at [3.6] $\approx$
[4.5] $\sim$ 14-15, and a sharp drop at [3.6] $\approx$ [4.5] $>$ 15.
The longer wavelength IRAC bands have steeper magnitude distributions
and brighter magnitudes with peak count levels ([5.8] $\sim$ 14.25 and [8] $\sim$
13.75).  The errors shown in the IRAC bands demonstrate that the average photometric 
uncertainty is well below 0.2 ($\sim$ 5$\sigma$) at least through [4.5]=15 and 
[8]=14.5. 

As with the 2MASS and Mimir data, the IRAC data show evidence for a
secondary peak in the magnitude distributions. This peak lies at
[3.6] $\approx$ [4.5] $\sim $ 14 and [5.8] $\sim$ 13.75. Adopting
the Siess et al (2000) isochrone and the Kenyon \& Hartmann (1995)
colors for main sequence stars, the peaks at JH$K_{s}$ and [3.6]-[5.8]
are consistent with cluster stars having masses $\sim$ 1.3 $M_{\odot}$ 
at a distance modulus $\approx$ 11.85.

Figure 6 shows our estimate of sample completeness, 
where we plot the ratio of sources detected in both 
IRAC and 2MASS to those detected in 2MASS for each IRAC band as a function of 
2MASS J magnitude (for J$\le$ 15.0).  
The sample oscillates between 90 and 100\% complete in each IRAC band at magnitudes 
brighter than expected based on where the source counts peak in each band.  
This discrepancy, most noticeable in the [4.5] band, is due to the very small 
matching radius employed in combining the 2MASS and IRAC data sets, not photometric 
errors.  If the matching radius is increased, the completeness level in the IRAC bands 
increases at the expense of larger cosmic ray contamination.  
Because our goal is to estimate the fraction of sources with small IR excesses, we prefer 
to analyze a smaller sample with more robust colors. 
When the $K_{s}$-[IRAC] colors are analyzed in Section 5, we 
consider the effect that uneven sample completeness has on our estimates of the IR 
excess population.
Table 2 lists our photometry 
from 2MASS and IRAC.
\section{Ground-based JH$K_{s}$ Data Analysis}

To make reliable estimates for the fraction of cluster
members with IR excesses, we must (i) derive the fraction
of stars in either cluster, (ii) derive robust criteria
for defining an IR excess, and (iii) combine the two
criteria into clear estimates for the excess fraction.
In this section, we consider two methods for estimating
the fraction of stars within each cluster, number counts
and model isochrones. The first method measures the number
of stars in the clusters relative to the stars in the
background population. Model isochrones allow us to estimate
the fraction of stars with magnitudes and colors that are
consistent with the magnitudes and colors of stars on a 13 Myr
isochrone (S02). Both approaches yield similar results.

With probabilities for cluster membership established, we then
consider whether any cluster stars have near-IR excesses.These
data demonstrate that the vast majority of sources brighter than
the completeness limits have colors consistent with photospheric
colors at 1-2 $\mu$m. Thus, we find no evidence for near-IR excess
sources in h \& $\chi$ Per for $J\le 15.5$, which correspond
to cluster stars with masses $\gtrsim$ 1.3 $M_{\odot}$.
\subsection{2MASS archival JH$K_{s}$ data: density distribution and cluster membership}
\subsubsection{Density Distribution of Sources on the Sky}
To investigate spatial inhomogeneities and other structure, we computed the projected sky surface density of stars, a standard 
approach for deriving the properties of star clusters (e.g.
Binney \& Tremaine 1987).  We restrict our analysis here and in Section 3.1.2 to sources brighter than J=15.5.
To derive the surface density of stars (top panel of
Figure 7), we counted stars in 1.5 $\times$ 1.5 arcmin bins, displaying the density in 
10$\%$ increments from 0-90$\%$ of the 
peak surface density of 15 arcmin$^{-2}$
in the center of h Per.   This surface density is higher than the $\sim$ 7-8 arcmin$^{-2}$ peak found by BK05, who 
were restricted to sources with spectral types earlier than A5. Based on J magnitude to spectral type conversions using the Siess et al. (2000) 
isochrones and preliminary MMT/Hectospec spectroscopy  (Currie et al. in prep), our population includes sources with 
spectral types earlier than 
$\sim$ K0.  

Aside from the strong peaks, each cluster has considerable small-scale
structure. BK05 report a kidney-shaped isodensity contour in $\chi$
Per, a fairly symmetric inner core surrounded by a rectangular isodensity
contour in h Per, and other structures $\sim$ 5-10 arcmin from the
cluster centers. With the deeper 2MASS data, these structures are clearly visible
at $\sim 40-60\%$ of the peak density and well above the median background level of $\sim$ 2.7 arcmin$^{-2}$.
The mean level is $\sim 2.8$ arcmin$^{-2}$ and the background noise fluctuates by $\sigma$$\sim$0.65 arcmin$^{-2}$.

Other asymmetric structures are also apparent $\sim$ 10-15 arcmin
from the cluster centers, which may suggest that both clusters are
asymmetric on large scales, $\sim$ 10 pc at a distance of 2.34 kpc.  
However, features $\gtrsim$ 15'-25' from the cluster centers have amplitudes 
comparable to the noise level 
and thus might not be real.  Deeper near-IR 
data could verify the existence of lower-amplitude structure.

The bottom panel of Figure 7 shows the radial surface density plots
derived from the 2D map, using
 1'-wide half-annuli (facing away from midway point of h \& $\chi$ Per).
 In both clusters, the
surface density drops rapidly from $\sim$ 2.5 to 10 arcmin and then slowly
merges into the apparent background level $\sim 30$ arcmin from the cluster centers.
The surface density reaches the median background
density level of 2.7 arcmin$^{-2}$ ($\sigma$=0.278 arcmin$^{-2}$; J $\le$ 15.5) $\sim$ 20-25 arcmin
away from the h\& $\chi$ Per centers at 
$2^{h} 18^{m} 56.4^{s}$, $57^{o}8'25"$ and $2^{h} 22^{m} 4.3^{s}$, $57^{o} 8'35"$
respectively (BK05).  
Although the clusters clearly
are not symmetric, we can derive a reasonable estimate of the fraction
of stars within the clusters by integrating the surface density above
this background level. This approach yields a cluster population of $\sim$
2000 stars, $\sim$ 35\% of the total population within 15 arcmin of
the cluster centers.  

Although the median background level is reached $\sim 15-20$' there is a small gradient in the 
radial surface density profile from $\sim 10-35$ arcmin (e.g. to the limit of our sample), though 
no clear general gradient across the field along constant right ascension or declination.  There are also 
is at least one region beyond 10' from 
either cluster center (from $\sim$ 12 to 20' away from $\chi$ Per) that may extend above its surroundings
background level on scales larger than size scale of the background fluctuations. 
 This behavior motivates us to investigate the stellar content within the 'background' (beyond
$\sim 15$' from the cluster centers).
Thus, we now compare the color-magnitude (CMD) diagrams
of the cluster-dominated and low-density regions.
\subsubsection{Possible Evidence for a Halo Population of $\sim$ $10^{7}$ year old sources in the vicinity of h \& $\chi$ Persei}
With no evidence for a significant age spread (S02) or patchy extinction (BK05) across the clusters,
we can use the cluster's isochrone to constrain cluster membership.  
We construct V/V-J and J/J-H color-magnitude diagrams, using data from S02 and 2MASS.
For both color-magnitude diagrams, we adopt an age of 13 Myr for the cluster 
(S02) using theoretical isochrones from Siess et al. (2000).  For the J/J-H diagram we also used pre-main sequence tracks from 
Bernasconi et al. (1996) for comparison, which show excellent agreement.  
To compute the cluster reddening, S02 restricted their analysis to the cluster nuclei and found a median 
reddening of E(B-V) $\sim$ 0.56.  Using a larger spatial sample 
of sources, BK05 computed a slightly smaller E(B-V) of $\sim$ 0.52.  Since the larger spatial sample of BK05 is more 
similar to ours, we consider this estimate more appropriate for our sample and adopt it. This choice has negligible bearing 
on our results since the reddening is low and yields an age within 1$\sigma$ of S02's value 
(Bragg 2004).  
Converting from optical to infrared extinction
 via Bessel \& Brett (1988) yields E(J-H) $\sim$ 0.185 for E(B-V) $\sim$ 0.52.   

If the clusters have a small spread in age, the fraction of stars identified as being on the 13 Myr isochrone 
should be similar to the $\sim$ $35\%$ estimate derived from the number counts in Section 3.1.1.   
 For objects with J $\le$ 14.5, the V/V-J diagram provides an efficient method to test for
cluster membership. The S02 V data have corresponding 2MASS J limits of $\sim$ 14.5; the diagonal slope (see Figure 8, top panels) 
of the isochrone on V/V-J allows good detection of blue, early spectral type (typically) background, non-member sources and very 
red, foreground non members.  
For objects with J $>$ 14.5, the V data become
incomplete. However, as shown in Figure 8 (bottom panels), the J/J-H CMD isochrone is sensitive to membership in
this magnitude range and thus provides a good substitute for the optical CMD.
The two CMDs are used together to identify sources of the same age and distance of h \& $\chi$ Per 
brighter than J = 15.5.
We divide our sample into two main populations: an 'on cluster' population corresponding to sources 
within 15' of the cluster centers and a low-density population for sources between 15' and 25' away.  
We chose 15' for the first population because it corresponds to $\sim$ 3-4 cluster core radii (BK05). Based on
our density distribution analysis, this distance corresponds to the point where the typical stellar density begins to approach the median 
background level.  
The 25' outer radius for low-density regions is chosen primarily 
because it fully samples our entire 
IRAC coverage.
According to the modified Hubble law distribution ($\rho$ $\sim$ $r^{-2}$) of cluster members, the cluster density 
$\sim$ 4 core radii away should be $\sim$ 6\% of the peak density.  If the number of stars in a circle of 
diameter 1 arcmin at the center is N, an annulus of 1 arcmin width
at 15 arcmin should have roughly 4-5 N stars in the cluster
and roughly 25-30 times that many in the `background'.  The fraction of stars in the cluster beyond 
15' to 25' should be over twice as small ($\rho$ $\sim$ 0.01-0.03$\rho_{peak}$).
Therefore, if h \& $\chi$ Persei sit in a background of non-related field stars, the low-density regions beyond 15' will be
dominated by sources of different ages and distances than sources in h \& $\chi$ Per.  The density of sources 
tracking the isochrone should be 
\textit{far} larger within 15' than outside 15'.  

The V/V-J diagram for sources within 15' of the h \& $\chi$ Persei cluster centers (Figure 8, top left panel) clearly 
shows a distribution of sources tracking the 13 Myr isochrone for V=12-16.  The isochrone slightly bends  
at V $\sim$ 15-15.5.  Foreground sources are clearly visible above the isochrone.
Figure 8 (bottom left panel) shows the J/J-H CMD diagram for all sources within 15'.
Sources with J $\le$ 14.5 track the reddened isochrone well except for some 
sources with J-H $\gtrsim$ 0.5 that are probably foreground M dwarfs or background
M supergiants.  The isochrone 'bends' horizontally
at J $\sim 14.5$.  
Almost all sources to J = 15-15.5 fall along the 13 Myr isochrone or are redder.
Although many sources with J $\ge$ 15.5 may follow the isochrone, the larger errors ($\sigma$ $>$ 0.1) 
and increasing population of blue sources make it hard to measure the cluster population.  Thus, we 
restrict our analysis to stars with J $\le$ 15.5.

The top-right panel of Figure 8 shows the V/V-J diagram for sources both
15'-25' from each cluster center and greater than 15' from both centers.
While sources appear quite scattered around the isochrone for V $\le$ 14,
there is a significant population of sources between the two isochrones
(for single and binary stars) for V = 14 - 16.5.  There are also many
sources between the corresponding isochrones in the J, J-H CMD (Figure 8,
bottom right panel). From analysis of the projected radial density
distribution, regions $\ge$ 15' from either center should contain mostly
background or foreground, non-member sources. The fact that the fainter of
these sources loosely follow the 13Myr isochrone for the cluster suggests
that there may be a larger region of enhanced formation of low mass stars
in the same general direction. This behavior would be similar to the
situation with the Orion cluster, for example.
 
We now quantify the fraction of sources in the h \& $\chi$ Persei 2MASS coverage that are consistent with cluster membership
by counting the number of stars that are, within photometric errors, lying on the V/V-J and J/J-H isochrones and the number lying 
outside them.  We require that sources be within 0.3 magnitudes ($\sim$ 5$\sigma$ added in quadrature) of the 
isochrone to count as a member.  This procedure neglects colors expected for unresolved binaries, which would 
increase the source brightness for a given color.  We compare the number of h \& $\chi$ Per sources less than 15' away from either 
cluster core and the number between 15' and 25' away.  Sources $\ge$ 15' from one cluster and $\le$ 15'
from the other cluster were not counted.  

From the V/V-J diagram, $\sim$ $33\%$ of sources within 15' of the cluster 
centers have a CMD position consistent with cluster membership
($\sim$ 767 with J $\le$ 14.5).  In the sample of stars at 15'-25' from the cluster centers,  $\sim$ $23\%$ lie in a similar zone
on the CMD.  About $\sim$ $47\%$ of sources within 15' have a CMD position consistent with a 13 Myr isochrone ($\sim$ 2600
sources with J $\le$ 15.5) on the J/J-H diagram.  Stars near the 13 Myr isochrone between 15' and 25' away 
from either center make up a comparable fraction $\sim$ $41\%$.   If we group the two populations together  
$\sim$ $44\%$ of the $\sim$ 11,000 sources are of the same age, distance and reddening as h \& $\chi$ Persei.  
Therefore, within 25' of the cluster centers the
 population of sources with J $\le$ 15.5 lying near a 13 Myr isochrone is $\sim$ 4700: $\sim$ 2500 closer to
 h Persei and $\sim$ 2200 closer to
$\chi$ Persei.  Estimates for the fraction of 13 Myr sources from the J/J-H diagram are comparable 
($\sim$ 27$\%$ and 20$\%$ for $\le$ 15' and 15-25' respectively) to those from the V/V-J diagram 
in the appropriate limit (J $\le$ 14.5): the lower fractional memberships for J $\le$ 14.5
 are likely caused by a high number of foreground 
K and M stars and background M supergiants.
Thus, to within $\sim$ 10-15$\%$, both the V/V-J (J $\le$ 14.5) and J/J-H (J $\le$ 14.5, 15.5) CMDs predict comparable estimates of 
the percentage 13 Myr sources at the distance of h \& $\chi$ Persei.
Furthermore, the percentage of sources lying on the 13 Myr isochrone in the low-density regions (15'-25') and high-density regions ($\le$ 15') are consistent within $\sim$ 10$\%$ of one another. 
If the 'background' were dominated by foreground and background sources then the percentage of 13 Myr sources in the high-density regions 
should be much larger than in the low-density regions. 

It is difficult for a random distribution of foreground/background stars with 
a wide range of ages and spectral types to mimic a $\sim$ $10^{7}$ yr isochrone, so we conclude that at least 
\textit{some} of the 'background' 
population includes young stars with nearly identical ages and distances as those in the two clusters.
From the poorer definition of the cluster along the isochrones in the
right panels of Figure 8, these stars may be distributed over a few
hundred parsecs along the line of sight, centered roughly on the clusters
and probably associated with them. The population of sources roughly
tracking the 13Myr isochrone in J/J-H only slowly disappears by $\sim$ 60'
from the cluster centers (not shown), suggesting a diameter of about 100
pc.  However, while it may be possible that the halo population extends to $\sim$ 60' away from h \& $\chi$ Per,
 carefully quantifying its disappearance, particularly for the upper main sequence at $\sim$ 13 Myr, requires wider angle imaging and deep
spectroscopy and is beyond the scope of this paper.

The existence of a halo population of 
stars at about the same age, distance, and reddening as those within the cores of h \& $\chi$ Persei was considered 
by Schild (1967) based on spectroscopy and photometry of the brightest sources; S02 also noted that, to V $\sim$ 16, the optical 
colors of bright sources beyond 5' of either cluster center and those within 5' of either cluster center appeared quite similar.  
Our result, probing stars slightly fainter than those studied by S02 and over larger spatial area, is broadly consistent 
with both of these references.
For the rest of the paper, we shall restrict our analysis to sources within 25' of the cluster centers.  We
use the V/V-J and J/J-H diagrams to identify sources within this radius that appear to have the 
age, distance, and reddening of h \& $\chi$ Per.

\subsection{JH$K_{s}$ color-color diagrams from 2MASS and Mimir}
Now we investigate the near-IR 2MASS colors of all sources within 25' of the cluster centers ($\sim$ 11,000) with J $\le$ 15.5.
The $J-H/H-K_{s}$ color-color diagram is shown in Figure 9 with the main sequence locus, giant locus, classical T Tauri
locus, and reddening bands.   
Stars with J-H $\approx$
0.0-0.25 and H-$K_{s}$ $\approx$ 0.0-0.3 will include Be stars with near-IR excess from optically-thin gas 
(Dougherty et al. 1991,1994).  Excesses around later-type stars with J-H $\ge$ 0.3 and H-$K_{s}$ $\ge$ 0.35
 are more likely to have near-IR excess emission from warm dust.  The vast majority of sources
have photospheric colors. Sources off the locus are distributed evenly on both sides, suggesting that photometric errors 
may be responsible for any $H$-$K_{s}$ 'excess'.
To characterize the population more accurately and (later) search for near-IR excess sources at $H-K_{s}$
we rely on Mimir data with deeper completeness limits at $JHK_{s}$.

We analyze Mimir sources with
$\sigma$ $\le$ 0.2 and J $\le$ 16.  
Figure 10 shows the $J-H/H-K_{s}$ color-color diagram for all sources in the Mimir survey.
The vast majority of sources fall along the main sequence locus.  
Figure 11 shows data separately for h Persei and $\chi$ Persei; h Persei has more
sources with a slight $H-K_{s}$ excess for small J-H colors ($\le$ 0.3).
The horizontal spread in color is noticeably smaller for sources with $\sigma$ $\le$ 0.1.
Many sources have colors consistent with disk excess emission (H-$K_{s}$ $\sim$ 0.4 or redder).
However, restricting ourselves to sources 
that may be cluster members based on the J/J-H diagram, the IR excess population from H-$K_{s}$ through J=15.5-16 
is \textit{extremely} small: less than $\sim$ $1\%$.
\section{IRAC Analysis: 2MASS -[IRAC] colors \& IRAC-only colors}
To learn whether excesses are more common at longer wavelengths 
we consider the IRAC data. Restricting our analysis to
sources within 25' of the cluster centers, we have
$\sim$ 7000 sources with 5$\sigma$ detections at
[3.6] and [4.5]; $\sim$ 5000 of these have 5$\sigma$
detections at [5.8] and [8.0]. To minimize contamination
due to large errors, we restrict this sample to the
sources with $K_{s}$,[IRAC] $\le$ 14.5.

The H-$K_{s}$/$K_{s}$-[3.6] and H-$K_{s}$/$K_{s}$-[4.5] diagrams
 are shown in Figure 12. 
In each plot, nearly all sources have photospheric
colors, with  $K_{s}$-[3.6], [4.5] $\approx$ 0.0-0.3 or 0.4. 
Sources with $K_{s}$-[3.6], [4.5] much larger than 0.3-0.4 are much redder than normal
stellar photospheres.
The number of 'red' sources is larger at [4.5] than at [3.6] 
and suggests the existence of an IR excess population in h \& $\chi$ Per.

Figure 13 shows the [3.6]-[4.5]/[4.5]-[5.8] and $K_{s}$-[3.6]/$K_{s}$-[5.8] color-color diagrams.  
In previous studies of IRAC colors of pre-main sequence stars (e.g. Hartmann et al. 2005),
a division between class II and class III T Tauri stars occurs roughly at [4.5] -[5.8] $\sim$ 0.2-0.25.  
The h \& $\chi$ Persei color distribution has a red limit of about 
 0.3 resulting from the intrinsic dispersion of colors.
Many sources have colors redder than this limit ($K_{s}$,[4.5]-[5.8] $\ge$ 0.3).
 There are few sources with [4.5]-[5.8] $\le$ -0.3. 
 The common-baseline $K_{s}$-[3.6]/$K_{s}$-[5.8] diagram shows the very red population even more clearly.
Photospheric sources have $K_{s}$-[5.8] $\lesssim$ 0.3-0.4.  The plot also suggests that there may be 
some sources with red colors at $K_{s}$-[5.8] but not $K_{s}$-[4.5], judging from the relatively larger number of sources. 

The distribution at [5.8]-[8] (Figure 14, left panel) continues to show a very gradual, as 
opposed to abrupt, transition in colors from photospheric ([5.8]-[8] $\le$ 0.2) to very red ([5.8]-[8] 
$\ge$ 0.4, see Allen et al. 2004).  In Hartmann et al. (2005) there are very few sources with [5.8]-[8]=0.25-0.4.  
The lack of any gap in h \& $\chi$ Per is likely due to 
larger photometric errors at [8].  The same plot for 10$\sigma$ sources (Figure 14, right panel) shows a much smaller dispersion in 
[5.8]-[8] colors containing sources with [5.8]-[8]=0.25-0.4.  There is also a substantial number of 10$\sigma$ detections 
with [5.8]-[8] $\ge 0.4$.

Constructing colors from the 2MASS $K_{s}$ band and IRAC bands at [5.8] and [8] provides the clearest evidence for an IR excess
population in h \& $\chi$ Persei.  Figure 15 shows the $K_{s}$-[3.6]/$K_{s}$-[8] color-color diagram, which is our 
closest analogue to the K-L/K-N diagram used by Kenyon \& Hartmann (1995) to distinguish Class II and III T Tauri stars 
in Taurus.  
Photospheric sources appear to have $K_{s}$-[8]$\le 0.4$; with $K_{s}$-[8] $\gtrsim$ 0.4 there is a clear population of 
IR excess sources.  The $K_{s}$-[5.8]/$K_{s}$-[8] diagram also shows an excess population for $K_{s}$-[5.8,8] $\ge$ 0.4.  

The population of very red, IR excess sources in $K_{s}$-[IRAC] is statistically significant.  The mean and dispersion in $K_{s}$-[4.5] 
are $\sim$ -0.02 $\pm$ 0.06; 0.07 $\pm$ 0.1; and 0.07 $\pm$ 0.1 at J=12, 14.5, and 15.  For $K_{s}$-[8] color these values are 
0.06 $\pm$ 0.1, 0.15 $\pm$ 0.18, and 0.24 $\pm$ 0.19; the values for $K_{s}$-[3.6] are similar to $K_{s}$-[4.5] and those for 
$K_{s}$-[5.8] are in between $K_{s}$-[4.5] and [8].  The median colors are nearly identical in all cases.  
In all $K_{s}$-[IRAC] colors there exists a substantial population of sources that are more than 
2-5$\sigma$ redder than the mean color while a corresponding blue population does not exist.
It is clear from the IRAC colors that h \& $\chi$ Persei harbors a significant IR excess population.
\section{Analysis of the IR Excess Population}
Now we
quantify the fraction of sources with excesses from 2MASS and IRAC photometry and thus the population of sources 
with circumstellar disks.
With no evidence for 
 IR excess at $K_{s}$, we estimate the fraction of sources with IR excess at [4.5], [5.8], 
and [8], using $K_{s}$ as a common short wavelength baseline and then relate the IR excess population to intrinsic stellar 
properties via the 2MASS J band.  The J filter should have emission dominated by the 
stellar photosphere, especially for sources older than 10 Myr that typically are not actively accreting.  Converting from 
J magnitudes to stellar properties is also more straightforward than with optical filters because J is less affected 
by reddening.  We analyze the IR excess population as a function of J in 0.5 magnitude bins.

In Section 5.1, we describe our two methods for selecting sources as h \& $\chi$ Persei members.  Next, we describe 
our criteria for identifying a source as an IR excess source in section 5.2.  We estimate 
the size of the IR excess population using a highly restrictive model for membership and IR excess identification 
and then using a less restrictive model.  We describe both models in section 5.3.

The estimates from both the restrictive and less restrictive models show excellent agreement.   
The relative size of the IR excess population to the total population increases with J magnitude.  Most of the IR excess sources
have J=14-15; very few excess sources are brighter than J=13.5-14.  This result implies that the 
frequency of disks around stars with ages 10-15 Myr is likely higher for lower stellar masses.  The IR excess population 
is consistently larger at longer IRAC wavelengths than at shorter wavelengths: this behavior is expected if circumstellar disks clear 
from the inside out.
At least $\sim 4-8 \%$ of sources with 
J=14-15 ($\sim 2.2-1.4 M_{\odot}$) have IR excess indicative of a circumstellar disk. 
We also identify sources that have IR excess at both [5.8] and [8].  About $\sim 2-3\%$ of sources have strong IR excess at both 
long wavelength IRAC channels for sources as faint as J=15.0.  
\subsection{Sample Selection}
Our first task is to remove probable non members of h \& $\chi$ Persei.  After removing sources identified 
as AGN/aromatic-emission galaxies we used  
the V/V-J and J/J-H color-magnitudes to remove sources inconsistent with
 the 13 Myr isochrone (Section 2.3).  The most restrictive, cautious approach is to use only the sources with optical photometry
 from S02 because the long-baseline V-J color is better at identifying non members than the shorter-baseline 
J-H color.  For this approach we require sources lying within 0.3 magnitudes ($\sim 5\sigma$ errors added in quadrature) 
of the isochrone in V
\begin{equation}
|V(source)-V(isochrone)|\le 0.3.
\end{equation}
Thus our sources consistent with h \& $\chi$ Per membership should fall within a band 0.3 magnitudes brighter and fainter than 
the nominal reddened isochrone.  However, the V magnitude data from S02 is complete only to 
$\sim 14.5$ in J.  
Thus, restricting our analysis to optical sources eliminates stars with J $>$ 14.5 or
 $M\lesssim$ 1.6 $M_{\odot}$.

A less restrictive approach uses the 2MASS J-H color to identify $\sim 13$ Myr, 2.34 kpc distant stars. 
Sources in unresolved binary systems should be $\le 0.75$ magnitudes too bright for their color.  The faint limit, however, should 
be unchanged.  Our criteria are then:
\begin{equation}
J(source)-J(isochrone) \ge -0.75,\le 0.3.
\end{equation}
This 2MASS sample is complete to J=15.5, which includes stars with $M\gtrsim 1.3 M_{\odot}$.

\subsection{IR Excess Criteria}
A color threshold is set for a source to be considered an IR excess candidate using  
 the 2MASS - [IRAC] color-color diagrams and Kenyon \& Hartmann (1995) photometry and color table as a guide. 
For the [4.5] channel we use 
the $K_{s}$-[3.6]/$K_{s}$-[4.5] diagram, for the [5.8] channel we use $K_{s}$-[3.6]/$K_{s}$-[5.8], and 
for the [8] channel we use $K_{s}$-[3.6]/$K_{s}$-[8] (Figures 11, 12, and 14).   The main distribution 
of sources in these diagrams has $K_{s}$-[IRAC] $\le 0.4$.  For comparison, Kenyon \& Hartmann found
that sources with K-L $\le 0.3-0.4$ were typically class III T Tauri stars with weak $H_{\alpha}$ emission and 
little evidence of circumstellar dust at L band (see also McCabe et al. 2006, figures 1 and 2).  
According to the Siess et al. (2000) isochrones, 13 Myr sources
 more massive than $\sim$ 1$M_{\odot}$ should also not have
K-L$\ge 0.2-0.25$, assuming the extinction derived by BK05.  Sources with $K_{s}$-[IRAC] $\ge 0.4$ are then likely 
IR excess source candidates.  

To avoid defining sources with large photometric errors as excess sources, 
we make the additional 
requirement that a source's color must be redder than the absolute threshold 
plus the source's photometric error, $\sigma$, in the IRAC bands, or 
\begin{equation} 
K_{s}-[IRAC] \ge 0.4 + \sigma(IRAC).
\end{equation}
Fainter IRAC sources ($\sim$ 14-14.5) then 
typically must have $K_{s}$-[IRAC] $\gtrsim$ 0.5-0.6 to be classified as excess sources. 
As a sanity check, we also require that the next longest wavelength IRAC band show at least a marginal excess, $K_{s}$-[IRAC] 
$\ge 0.3$, if a 5$\sigma$ detection was made at such a channel.  

Finally, we remove potential completeness-related bias and calculate errors on the size estimates of the IR excess population.
  First, because the completeness limits from J through [8] vary, we apply
a uniform cutoff across 2MASS and IRAC bands and 
 restrict ourselves to sources with $K_{s}$,[IRAC]$\le 14.5$, as the count rate for 5$\sigma$ detections at [8]  
falls to half-peak values by 14.5.  Because sources with $K_{s}$ $\sim$ 14.5 have J-$K_{s}$ $\sim$ 0.4-0.7, 
we add a further restriction that sources must have J$\le 15.0$.  Estimates for the size of the IR 
excess population in each IRAC band are then drawn from a single population.  
Nominally, the errors for the IR excess fraction in each 0.5 J magnitude bin are calculated from Poisson statistics, 
$\sigma$$\sim \sqrt{\#}$/N, where $\#$ is the number of sources with IR excess and N is the number of sources in the bin with detections 
in 2MASS and at a given IRAC band.  We retain this error estimate as an upper bound on the IR excess fraction.  
However, there may be sources in our coverage that are detected in 2MASS but not in IRAC and are brighter than our J magnitude cutoff.
Thus, these sources would not be included in our calculation for N because of their non-detection in IRAC.  
This could bias our results in favor of detecting more IR excess sources and fewer photospheric 
sources near magnitude 14.5 and thus is another source of error.  

To put a limit on this error, we calculate the number of sources 
in each bin detected at both J and $K_{s}$. If the IRAC sample is 100\% complete, this number M should be equal to N, otherwise it will 
be larger.  We then assume that any source in M not detected in N is not an IR excess source and divide the total number of IR excess 
sources by M.  This procedure yields an absolute lower limit on the fraction of sources with IR excess in each 0.5 magnitude bin.  We subtract this value 
from the nominal fraction to produce an error estimate due to sample completeness.  The larger of the two 
errors, the Poisson error and the completeness error, is chosen as the lower bound on the IR excess fraction.  

\subsection{Two Models for Analyzing the IR Excess Population}
Equipped with our two approaches for sample selection, our method for identifying an IR excess source, 
and our brightness cutoffs in 2MASS \& IRAC, we now describe our two models for analyzing the IR excess population.  
The models are summarized 
in Table 3.
\\
\\
\textbf{Model 1} (Restrictive) - We require that sources have optical counterparts from S02 and that they are within 0.3 
magnitudes of the 13 Myr V/V-J isochrone (Equation 1).  We require that IR excess candidates fulfill the condition set 
in Equation 3.  We investigate the IR excess population at [4.5], [5.8], 
and [8] for sources with $J\le 14.5$.   
Errors for the size of the IR excess population in all models assume Poisson statistics.
\\
\\
\textbf{Model 2} (Less Restrictive) - We require that sources have near-IR colors consistent with the 13 Myr J/J-H isochrone as described 
by Equation 2. We require that the IR excess candidates fulfill the condition set in Equation 3.  
The IR excess population is analyzed through J=15.  

As additional checks of the reliability of the data, we examined
the individual frames for the sources with apparent excesses and considered as
highest weight the excesses detected at {\it both} [5.8] and [8].  We quantify the excess population at both [5.8] and [8] as well.
\subsection{IR Excess Population: Restrictive Model} 
We show the size of the IR excess population relative to the total population as a function of J magnitude for our 
restricted model (Model 1) for $K_{s}$-[4.5], $K_{s}$-[5.8], and $K_{s}$-[8] in Figure 16.  The most striking 
feature in the diagram is the complete lack of IR excess sources at any band for J $\le 13.5$: 0/250 sources with J=11-13.5 have 
IR excess.  We only begin to detect excess sources at [8] by J=13.9.  The IR excess population at [4.5] and [5.8] 
only starts to appear by $J\sim 14.05$.  While our approach may underestimate the fraction of 
IR excess sources near our faint limit of J=14.5 ($M\sim 1.6 M_{\odot}$), sources with J=11-13.5 typically have 
small 
photometric errors ($\le 0.05$ mag) at all 2MASS and IRAC bands.  Thus, we have not missed  
IR excess sources brighter than J=13.5 ($M\sim 2.7 M_{\odot}$) from our sample.

We note two other features in the IR excess population.  The [8] IR excess fraction is consistently larger than at [4.5] and 
[5.8].  This result is consistent with an inside-out clearing of circumstellar disks.  
Stars with J=14-14.5 (2.2-1.6 $M_{\odot}$ also tend to have a larger fraction of IR excess sources than stars with J=13.5-14.
For J=14-14.5, 14/409 (3.4 +0.9,-1\%) sources 
have IR excess at [8]; for J=13.5-14 only 2/169 ($1.2 \pm 0.8\%$) of sources have [8] excess.  This result coupled with the lack 
of excess sources at J$\le 13.5$ suggests that the frequency of IR excess sources is related to stellar properties.
\subsection{IR Excess Population: Less Restrictive Model}
Now we investigate the IR excess population as a function of J magnitude for our less restrictive model (Model 2; Figure 17). 
While this approach may introduce more non members, the larger number of cluster members in this sample (to J$\le$ 15.5)
yields a larger statistical significance for our results.
There are no IR excess sources for J$\le$ 13.5: 0/399 with J=11-13.5 have IR excess at [8] or at shorter wavelengths. 
Thus the lack of IR excess sources at IRAC bands to J=13.5 is model independent.  

The IR excess population from this sample appears remarkably similar to the restrictive sample where the two overlap (to J=14.5).  
In both models the [8] excess population is consistently larger than that at [4.5] and [5.8].  The IR excess population at [8]
increases from $1.6 \pm 0.6 \%$ (8/496) of the total population from J=13.5-14 to 3.5 +0.6,-0.9 \% (34/983) at J=14-14.5.  This result 
 agrees with the estimates (complete to J=14.5) derived in section 5.4.  
Thus to J=14.5, our estimates of the IR excess population's size is not 
 sensitive to whether we use the long-baseline V/V-J or the short-baseline J/J-H diagram to constrain cluster membership. 
For J=14.5-15 the IR excess population at [8] is 
$\sim$ 8.1 +1.1,-4.5\% (50/618).  The population of sources with J=14.5-15,  $K_{s}$$\le 14.5$, and [8] detected with errors $\le 0.2$ is 
 smaller than the total population of sources with J=14.5-15 and $K_{s}$ $\le 14.5$: this accounts for the relatively large lower bound error.  
We note that if the sample were restricted to $K_{s}$,[8] $\le 14.25$, where completeness is better, the percentage of excess sources is 
roughly halfway between our measured value of 8.1\% and the lower limit of $\sim$ 4\% at 
$\sim 6.3$\%.  Thus, the IR excess population at [8] near the brightness limit of our analysis comprises about 4-8\% of the total population.  

The excess population at $K_{s}$-[4.5] is consistently the smallest, ranging from zero through J=13.5 to 0.2 $\pm 0.2 \%$, 
1.4$\pm 0.4\%$ (16/1117), to 1.0 +0.5,-0.6\% (11/1138) of the total population 
from J=13.5-14, 14-14.5, and 14.5-15 respectively.  
Over the same range in J magnitude the [5.8] excess population represents respectively zero, 0.6$\pm 0.3\%$, 1.6$\pm 0.4 
\%$ (19/1214), 
and 3.3 +0.5,-0.6\% (38/1150) of the total population.  

\subsection{Sources with Clear IR Excess at both [5.8] and [8]}
If the IR excess population is dominated by sources with optically thick circumstellar disks extending to the magnetospheric 
truncation radius (Kenyon et al. 1996), then the IR excess population at [5.8] and [8] should be nearly identical.  However, in both models 
for analyzing the IR excess population,
the fraction of sources with IR excess at [8] is consistently larger.
 It is plausible that sources may not have IR excess at short-wavelength 
IRAC bands but have excess beyond $\sim 5-10\mu m$, especially at 10-20 Myr.  For example, TW Hya, a 10 Myr T Tauri star, has photospheric 
colors through [4.5], 
and a clear IR excess only at [8] (Hartmann et al. 2005).  Other evolved T Tauri stars, such as the 'transitional' T Tauri stars 
in Taurus (Kenyon \& Hartmann 1995), have SEDs consist with a star+circumstellar disk where the disk has an inner hole.   Sources
with strong IR excess longward of [5.8] may also be explained by debris produced by planet formation 
(Kenyon \& Bromley 2004; Currie et al. 2006).
However, unlike for TW Hya and similar systems, we do not have constraints beyond 8$\mu m$ on IR excess candidates in h \& 
$\chi$ Per since the MIPS completeness limit is typically too bright to detect these candidates (Currie et al. 2006; Balog \& Currie et al., in prep.).  

Therefore sources with both [5.8] 
and [8] excess set an absolute lower limit on the size of the IR excess population.  We set the excess criteria to 
$K_{s}$-[5.8,8] $\ge 0.4$ and use our sample from Model 2.  The IR excess population at both [5.8] and [8] begins 
at J=13.8 and represents $\sim 1\%$ of the population for sources with J=14-14.5.  This percentage increases to $\sim 2.2\%$ 
for J=14.5-15, slightly less than but still consistent with the excess fraction at [5.8] alone from Model 2.  We also 
recover the same IR excess population vs. J magnitude relation in models 1 and 2, albeit at a lower statistical significance.  

Thus, for sources with $M\sim 1.4-1.6 M_{\odot}$ at least $\sim 2-3\%$ have IR excess at both [5.8] and [8].  If we include 
sources with photospheric colors at all wavelengths short of 8 $\mu m$ then about $\sim 4-8\%$ of sources 
with J=14.5-15 have IR excess.
\subsection{Distribution of IR excess sources on V/V-J and J/J-H Color-Magnitude Diagrams}
Here we consider estimates of the IR excess population without cluster membership criteria. 
 If IR excess sources are distributed 
randomly in V/V-J or J/J-H space our IR excess population estimates might be invalid.

Figure 18 shows our results.  In the top panel of Figure 18 IR excess sources with V band photometry from S02 preferentially lie along the 
13 Myr isochrone. \textit{They are far more clustered around the isochrone than the distribution of all sources.}  
 This trend also prevails for sources without V band photometry as shown in the bottom panel of Figure 18.  Here, 
 the IR excess population is preferentially clustered along the 13 Myr J/J-H isochrone.  Just over half of the 
sources within 25' of h \& $\chi$ Persei are foreground/background stars, so the sizes of the h \& $\chi$ Per population and 
foreground/background populations are comparable.  However, the IR excess population for h \& $\chi$ Per cluster and halo population members is 
clearly larger than the population of non members.  This result is expected if our membership and IR excess criteria are valid.  

\subsection{Statistical Trends in the IR Excess Population of h \& $\chi$ Persei}
To conclude, analyzing the excess populations at $K_{s}$ -[4.5], [5.8], and [8] according to our two models 
leads to the following trends.  First, both models show a complete lack of h \& $\chi$ Persei sources with 
J$\le 13.5$ \textit{and} IR excess at any band.  This result suggests that \textit{almost all inner disks ($\le$ about 1 AU from the star) 
around massive stars ($\ge 2.5-3 M_{\odot}$) disappear 
prior to $\sim$ 10 Myr}. Second, both models show a general 
increase in the IR excess population with J magnitude.  This suggests that \textit{a small number of disks around lower mass stars 
(to $\sim 1.4-1.6 M_{\odot}$) 
have lifetimes longer than $\sim$ 10 Myr}.  Third, to J=14.5 ($\sim 1.6 M_{\odot}$) about 3-4 \% of sources have IR excess in the IRAC bands.  

The deeper completeness and stronger statistical constraints from Model 2 separately yield the following additional results.  The IR excess 
population is larger at [8] than at [5.8] and, especially, [4.5].  This result is consistent with an inside-out clearing of 
disks.  About 4-8\% of sources slightly more massive than the Sun (J=14.5-15, $\sim 1.4-1.6 M_{\odot}$) have IR excess.   
Tighter constraints on the cluster membership from optical spectroscopy and deeper IRAC and MIPS observations will be required to 
more precisely determine the IR excess population.  

\section{Summary, Future Work, \& Discussion}
We have conducted the first deep IR survey of the
double cluster, h \& $\chi$ Per. The
survey combines 2MASS JH$K_{s}$ data with deeper JH$K_{s}$
data from the Lowell Mimir camera and Spitzer IRAC 3.6-8$\mu m$
data covering a region with an area of roughly 0.75 $\deg^{2}$.
To derive the fraction of stars with IR excesses,
we considered the observed stellar surface density
distribution, optical and infrared color-magnitude
diagrams, and infrared color-color diagrams. This
analysis yields the following robust results:
\begin{itemize}
\item For stars with J $\le$ 15.5, the stellar surface density
peaks at 15 $arcmin^{-2}$. In both
clusters, the density falls to background levels of
$\sim 2.7$ $arcmin^{-2}$ $\sim$ 15' away from the center peak.
About $\sim 47\%$ of the stars within 15' are 
cluster members.
Comparisons of color-magnitude diagrams
show that $\sim$ $41\%$  of the stars at 15-25' from the
cluster centers are of approximately the same age, distance, and reddening.  
In both cases, the
Siess et al (2000) evolutionary tracks suggest ages of
$\sim$ 13 Myr for cluster stars.
\item For stars with J $<$ 13.5, the two clusters have essentially
no stars with IR excess at $\lambda \lesssim $ 8 $\mu$m.
This result demonstrates
that disk emission with R $\lesssim$ 1 AU disappears from
stars with M $\gtrsim$ 2.5-3 $M_{\odot}$ on timescales of $\lesssim$ 10-15 Myr.
\item Many stars with J =13.5-15 have IR excesses: $\gtrsim$ 
2-3\% of the stars in this magnitude range have excesses at both [5.8] and
[8.0]. 
Combined with the lack of
detectable excesses for brighter stars, these results suggest
that the disk evolution time depends on the stellar mass,
with lower mass stars having longer disk evolution timescales.
\end{itemize}
 In addition to these firm conclusions, our analysis indicates
the following more tentative results:
\begin{itemize}
\item If the requirement that sources have excesses at both [5.8] and [8] is relaxed, 
the fraction of sources with IR excesses increases from
3-4\% at J = 14-14.5 to 4-8\% at J=14.5-15.  
\item At a fixed brightness,
a larger fraction of sources appear to have excesses at
longer wavelengths, 5.8-8 $\mu$m, than at shorter wavelengths,
4.5 $\mu$m.  This result is consistent with an inside-out clearing of 
protoplanetary disks as proposed in standard theories of planet 
formation.
\end{itemize}

These results demonstrate that h $\&$ $\chi$ Per are an excellent
laboratory for studying the evolution of circumstellar disks
at 10-15 Myr. With $\ge$ 5000 sources more massive than the
Sun and a potentially significant halo population outside the main cluster
boundaries, h \& $\chi$ Per are at least as populous as the Orion Nebula Cluster
(ONC; Hillenbrand 1997) and are the most populous nearby clusters in this age range
within 2-3 kpc. Detailed comparisons of these clusters
with the ONC and other young clusters provide robust tests of the
universality of the IMF (S02) and the evolution of small
scale structures within the clusters (Hillenbrand \& Hartmann 1998,
BK05).

Further progress on understanding the nature of the IR excess
population requires deeper IR imaging surveys and comprehensive
spectroscopic surveys. Current near-IR imagers on 4-10 m class
telescopes can reach 0.5 $M_{\odot}$ stars (J $\sim$ 17.5)
in a few nights. When combined with a deeper IRAC survey ([3.6, 4.5] $\sim 16-17$, [5.8, 8] $\sim 16$),
these data would yield measures of the IR excess for low mass
stars where the predicted timescales and outcomes for disk evolution and planet
formation are much different than at 1-3 $M_{\odot}$ (Plavchan et al. 2005; Laughlin et al. 2004). 
Modern 
multi-object optical spectrgraphs routinely acquire high S/N
spectra of V = 19-20 stars in 30-45 min. With current spectroscopic
samples
complete only to V $\sim$ 15-16 (S02, BK05), a deep
spectroscopic survey enables more reliable measures of cluster
membership and deep searches for H$\alpha$ emission for
stars with masses of $\sim$ 0.5-2 $M_{\odot}$.

\acknowledgements
We thank Lori Allen, Peter Plavchan, Rudy Schild, Rob Gutermuth and Nancy Evans for useful comments, Lionel Siess for
use of the stellar evolution tracks, and 
Amanda Bosh for assistance with data acquired during the Mimir observing run.  We also
 thank Charles Lada for initially suggesting Spitzer observations of h \& $\chi$ Persei.
Finally, we thank the anonymous referee for a thorough review of the manuscript. 
T. Currie is supported by an SAO Predoctoral Fellowship.  Z. Balog received support
from Hungarian OTKA Grants TS049872, T042509 and
T049082.  We acknowledge additional support from the NASA Astrophysics Theory Program grant NAG5-13278
and the Spitzer GO Program (Proposal 20132).
 This work was partially supported by contract
1255094, issued by
JPL/Caltech to the University of Arizona.
This publication makes use of data products from the Two Micron
All Sky Survey, which is a joint project of the University of
Massachusetts and the Infrared Processing and Analysis Center/California
Institute of Technology, funded by the National Aeronautics and
Space Administration and the National Science Foundation.
This research has made use of the NASA/IPAC Infrared Science Archive,
which is operated by the Jet Propulsion Laboratory, California
Institute of Technology, under contract with the National
Aeronautics and Space Administration. 

\clearpage
\input{tab1.tex}

\input{tab2.tex}
\input{tab3.tex}
\begin{figure}
    \centering
   {\centering \resizebox*{0.9\textwidth}{!}{{\includegraphics{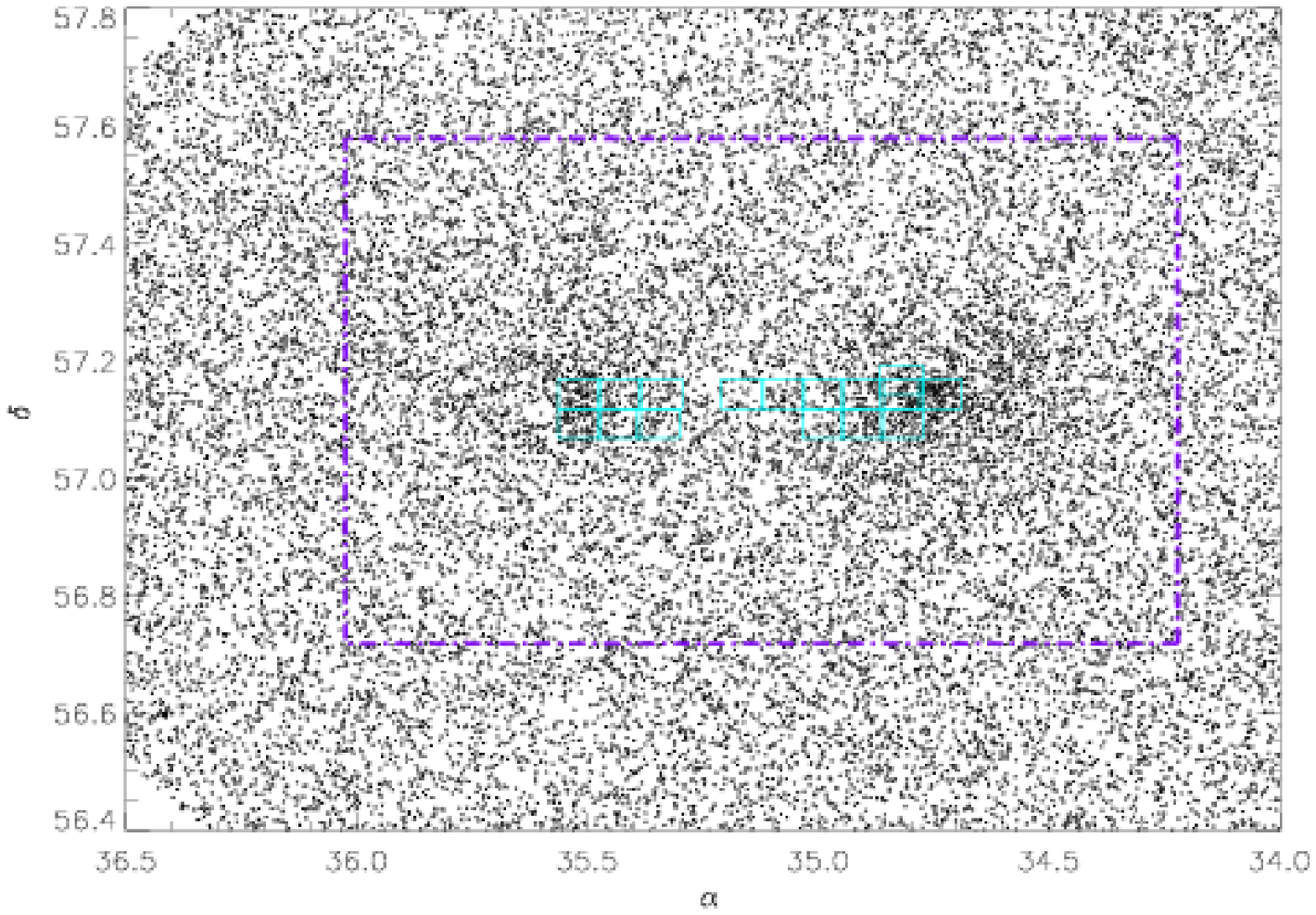}}}\par}
   \caption{Coverage map for Mimir (cyan boxes) and IRAC (purple dash-dotted line) observations plotted over 2MASS sources in h \& $\chi$ Per.
The center of h Persei ($2^{h}$$22^{m}$$4.3^{s}$ $57^{o}$$8'$$25"$) is in the extreme right cyan box and the $\chi$ Persei center 
($2^{h}$$22^{m}$$4.3^{s}$, $57^{o}$$8'$$35"$) is in the two leftmost boxes.  The areas covered by 
Mimir and IRAC are $\sim 144$ arc-min$^{2}$ and $\sim 0.75$ deg$^{2}$ respectively.
}\label{figure 1}
\end{figure}
\clearpage

\begin{figure}
    \centering
   \epsscale{.85}
   \plottwo{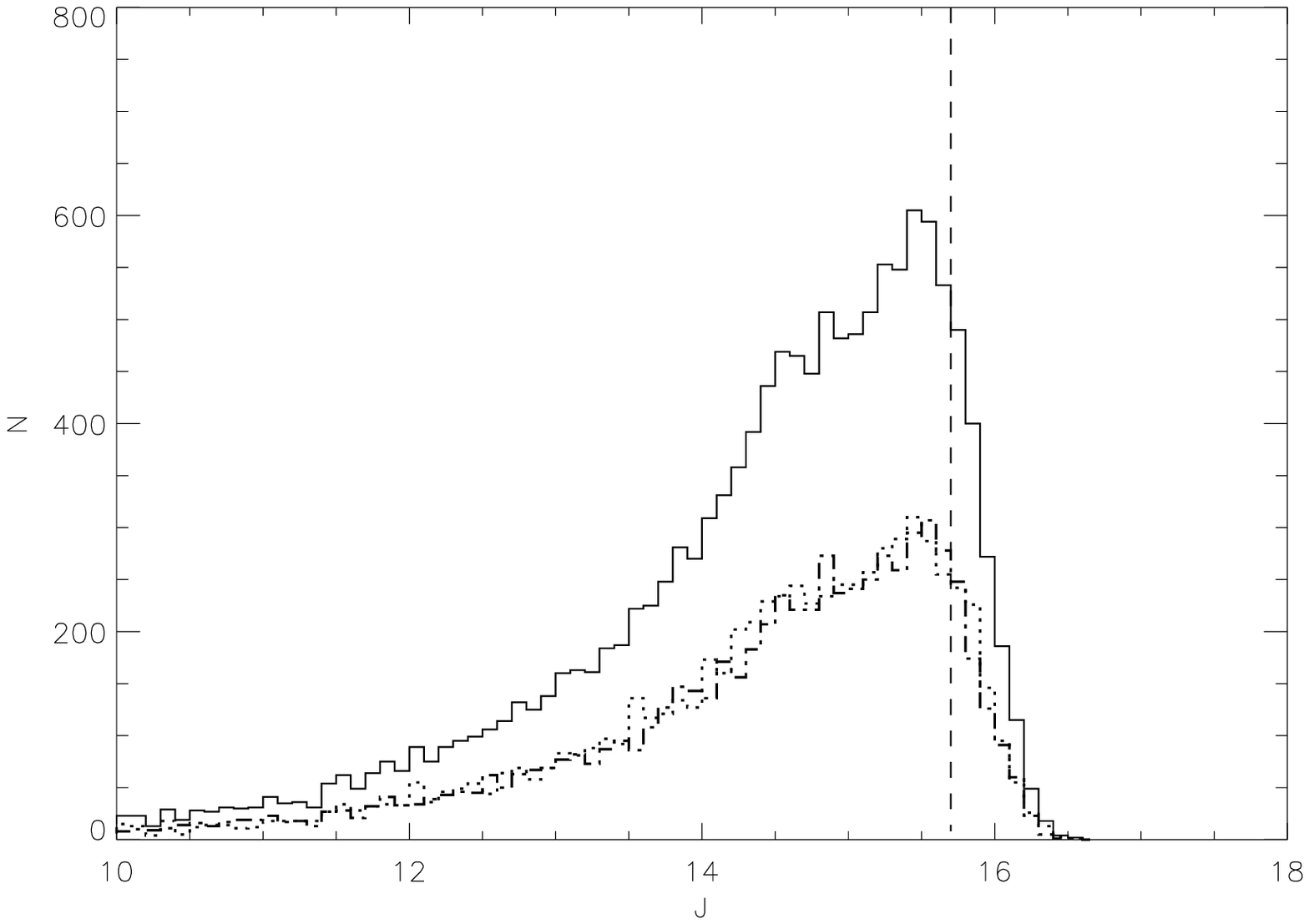}{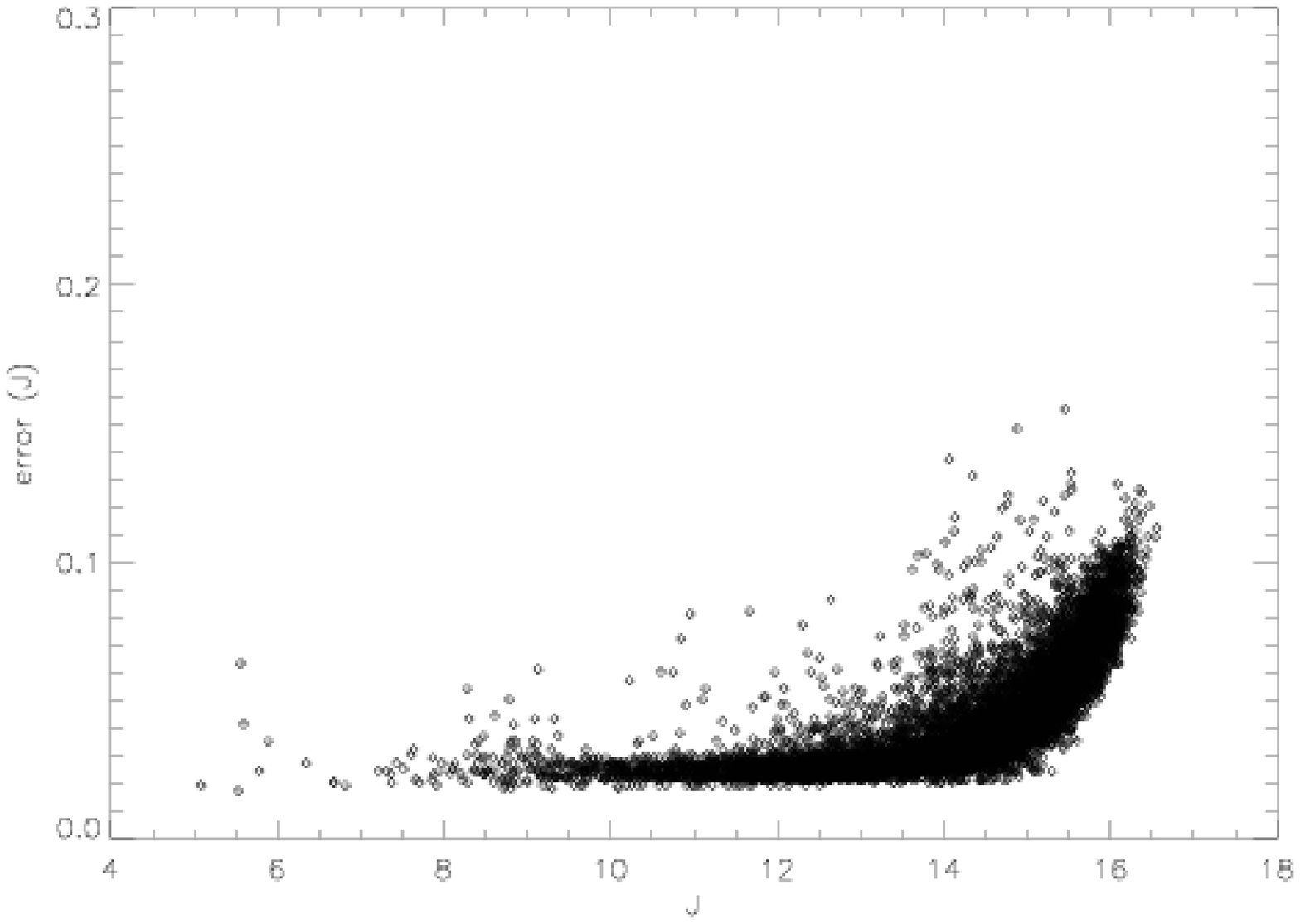}
   \plottwo{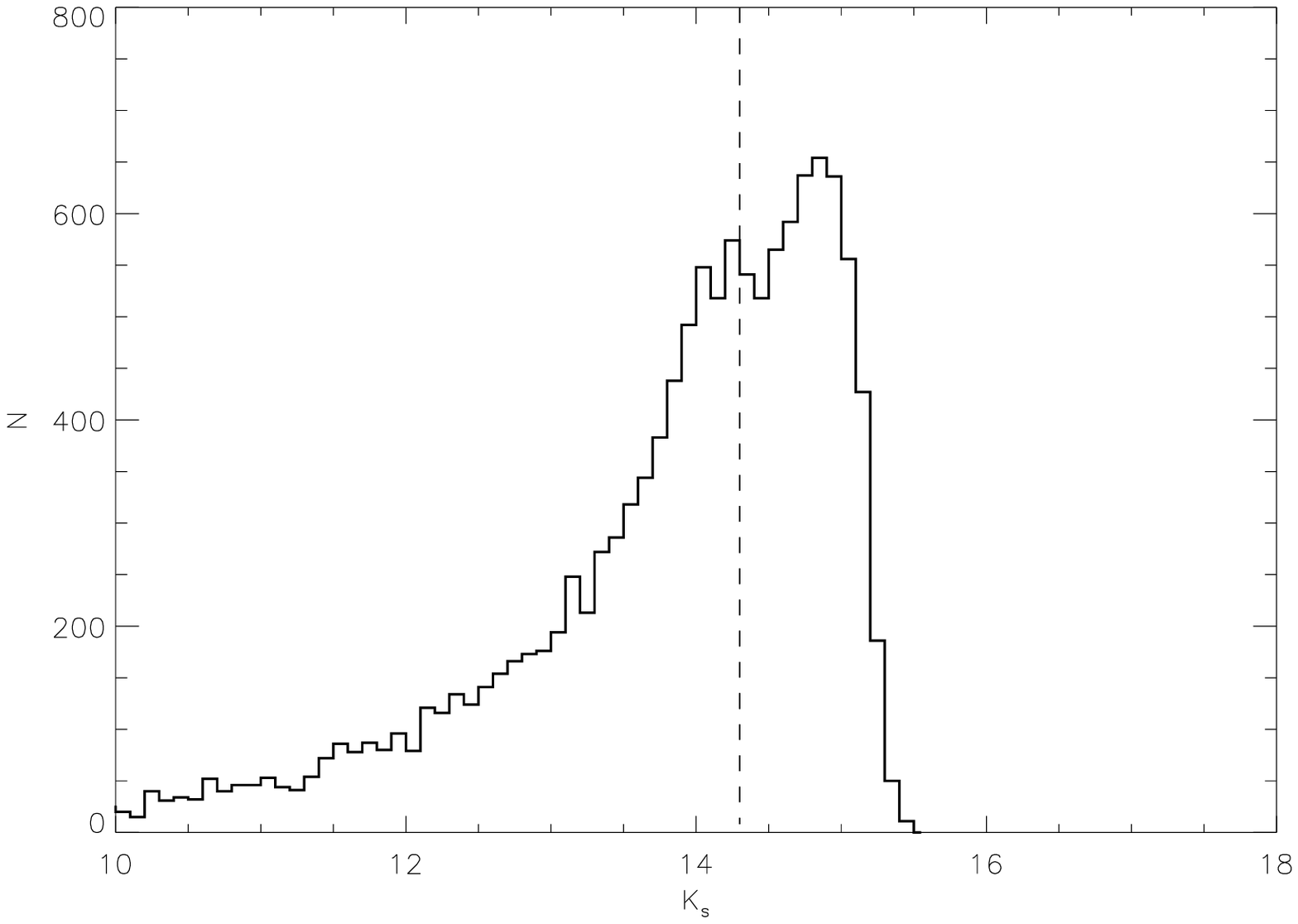}{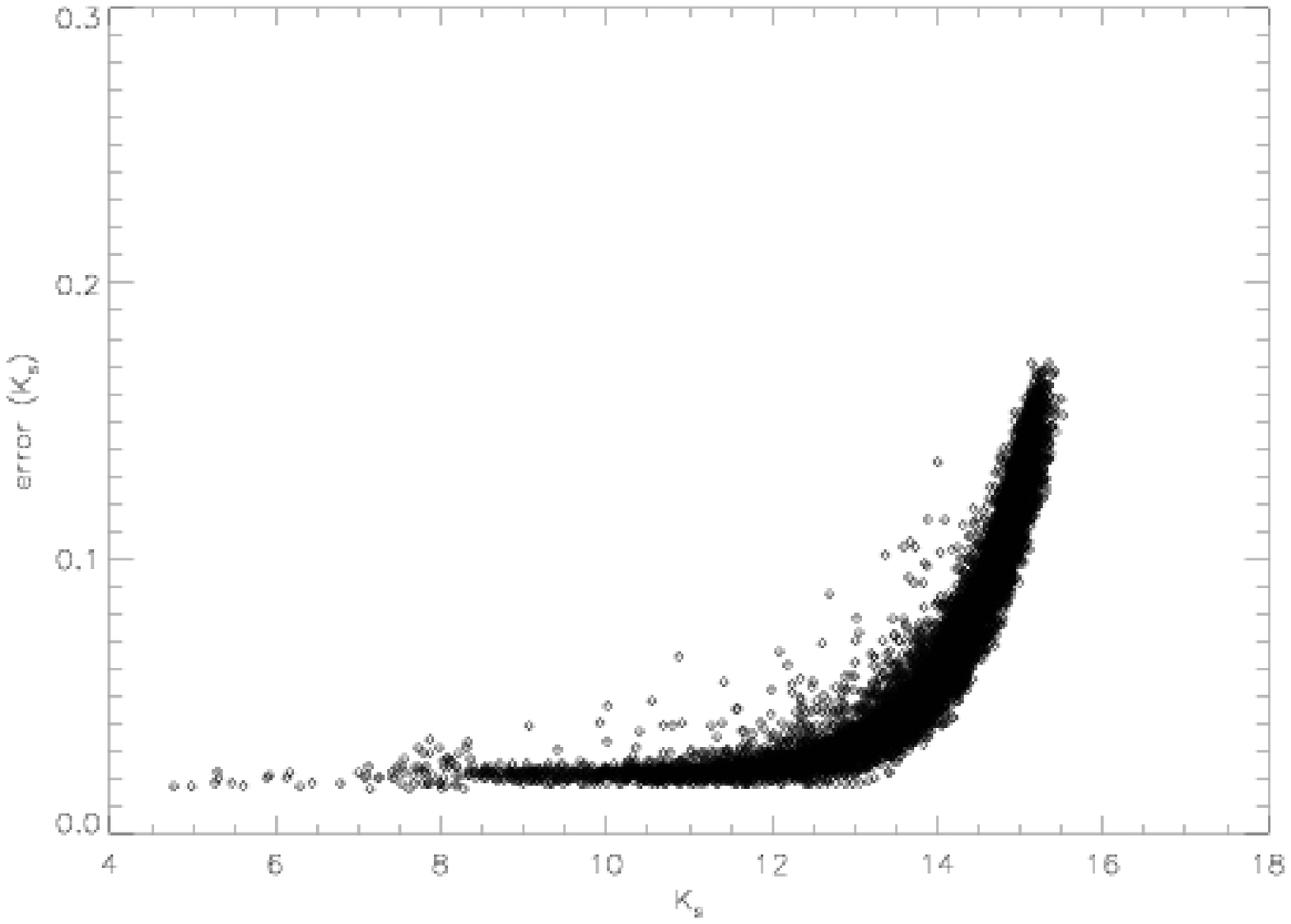}
   \caption{J (top) and $K_{s}$ (bottom) magnitude and error distributions for 2MASS sources within 25' of the h \& $\chi$ Per centers.  
The dotted lines are for 
sources closer to h Persei and the dash-dotted lines are sources closer to $\chi$ Persei. The vertical dashed lines are 
the published 10$\sigma$ sensitivity limits for 2MASS: J=15.8, $K_{s}$=14.3.
  The small turnover near $J\sim 14.5$, $K_{s}\sim 14.25$ results from the
large number of sources near the centers of h \& $\chi$ Per that follow the $\sim 13$ Myr isochrone.  
The total sample is complete to J=15.5 and 
$K_{s}$$\sim$ 15.0 ($\sim 
1.3 M_{\odot}$).  
falls to half the peak value by $J\sim 15.75-16$ and $K_{s}$$\sim15.25$.  Errors at J=15.5 and $K_{s}$
=15.0 are less than $\sim 0.1$ and $\sim 0.12$ respectively.  The distribution in $K_{s}$ 
band shows reasonable agreement with the published 10$\sigma$ limits while those in J fall slightly short 
but are reasonable through J$\sim$ 15.5.
  We restrict the data analysis
 to sources with $J\le 15.5$.  
}\label{figure 2_01}
\end{figure}

\begin{figure}
\epsscale{0.8}\plottwo{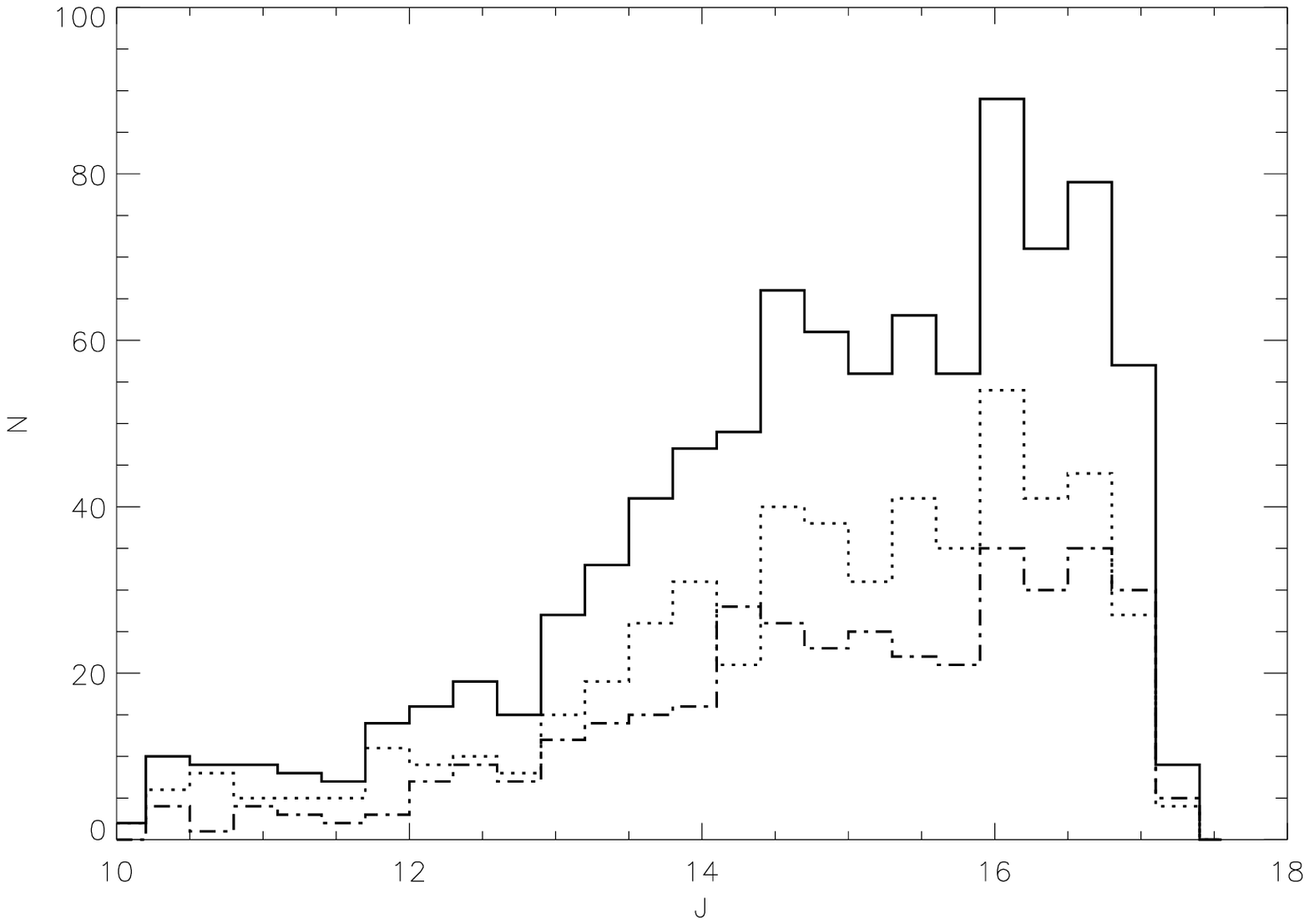}{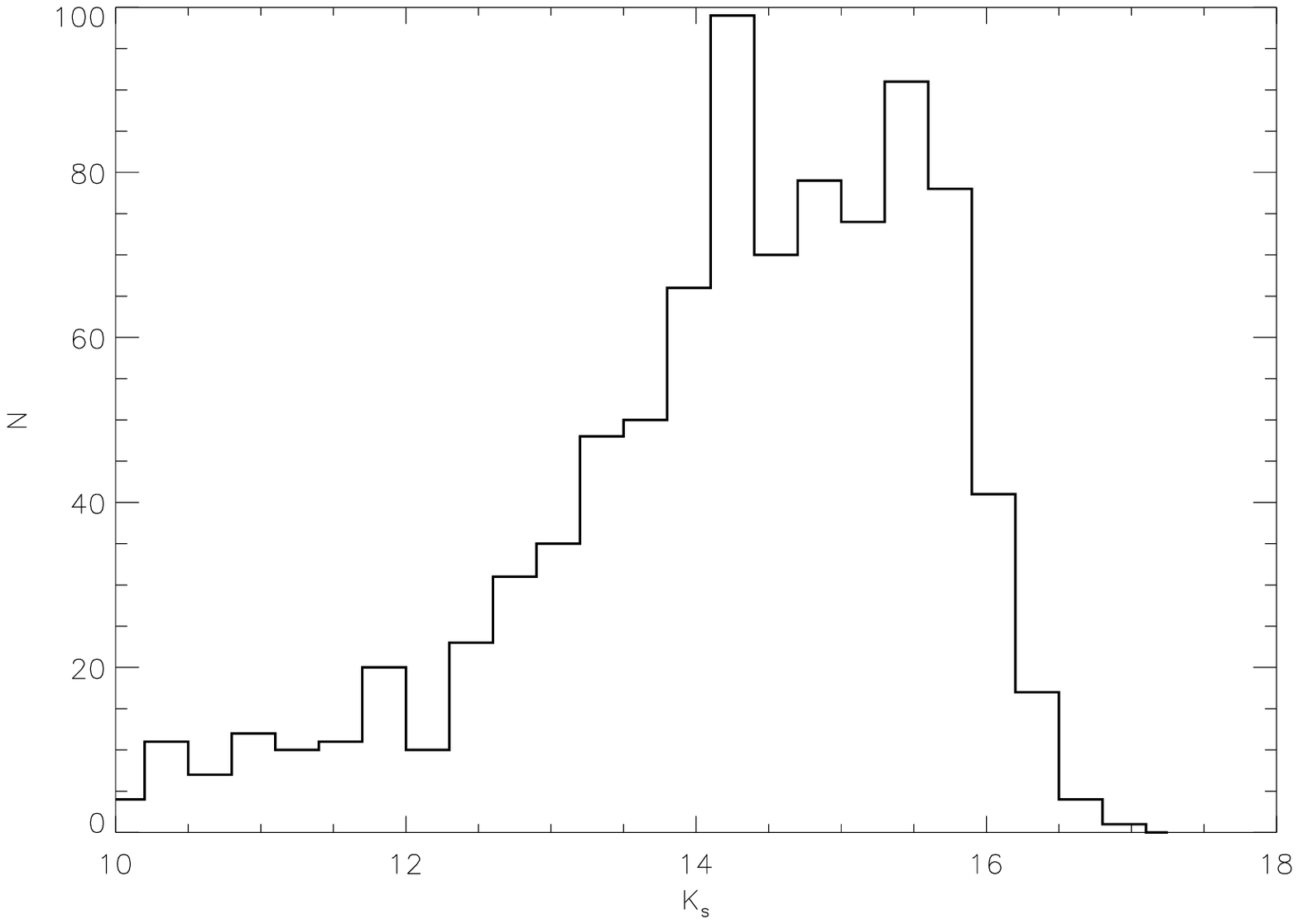}
\epsscale{0.8}\plottwo{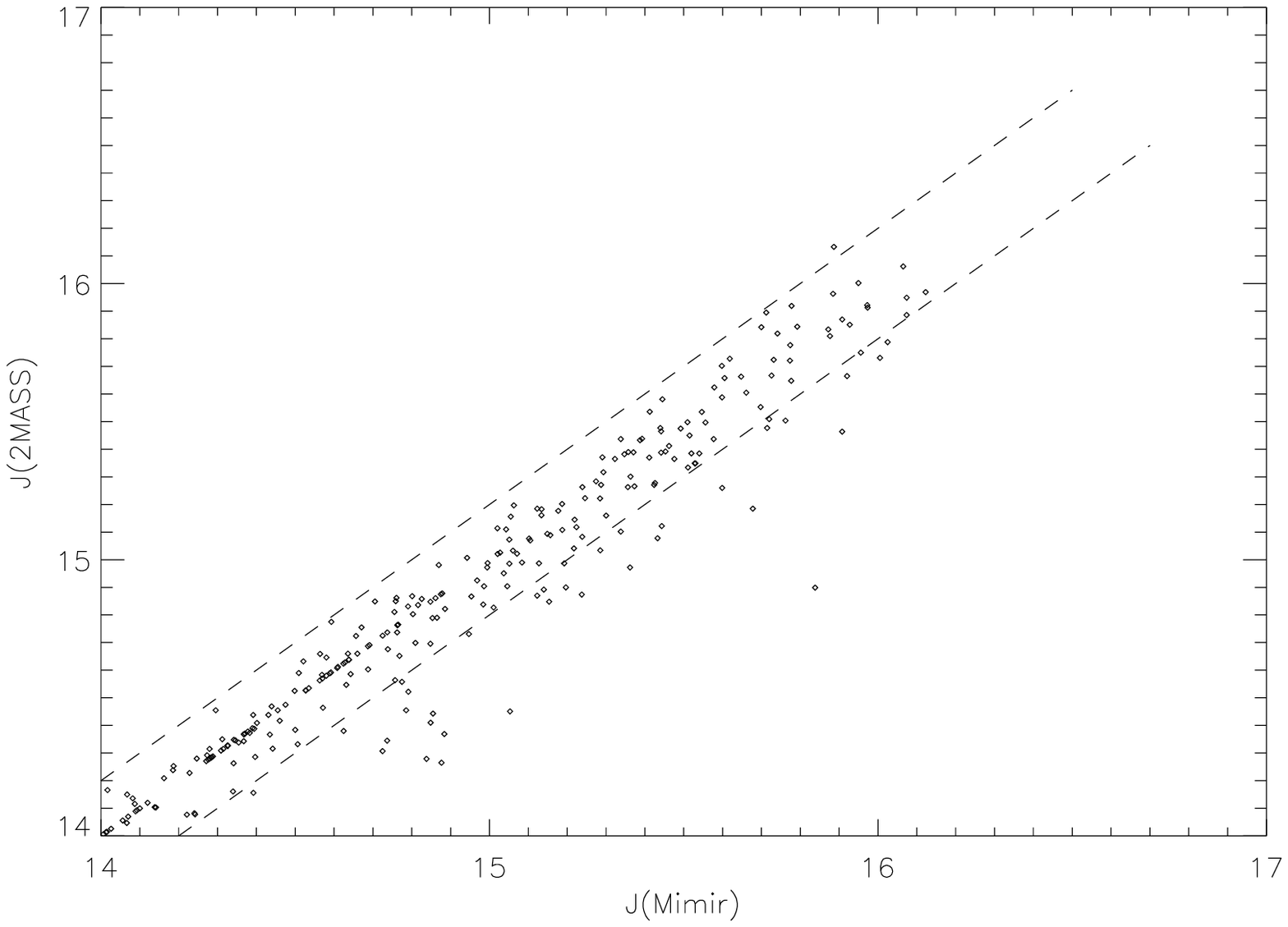}{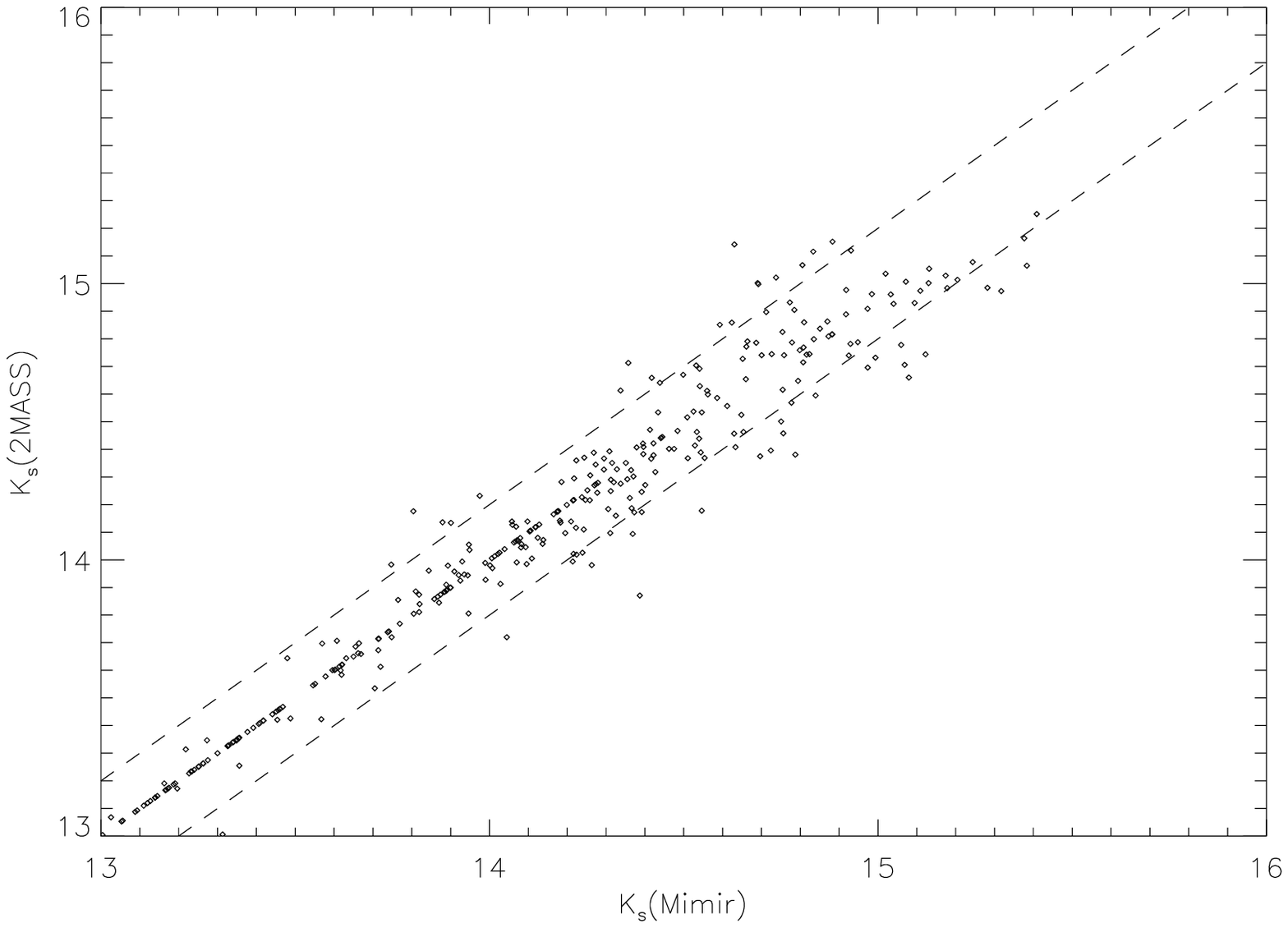}
   \caption{(top-left) J distribution for h Persei sources (dotted line), $\chi$ Persei sources (dot-dashed), 
and all h \& $\chi$ Per sources (solid line) from Mimir 
.  There is a small enhancement of sources near $J\sim 14.0-14.5$ which are cluster sources on a 13 Myr isochrone. 
The number counts in J peak at $\sim 16.0-16.25$ which corresponds to stars with $1.1-1.2 M_{\odot}$ 
at the adopted distance of h \& $\chi$ Per ($\sim 2.34$ kpc).  
However, the distribution does not fall to half its peak value until $J\sim 16.75$.
We see a secondary peak at $J\sim 14.5$.  (top-right) $K_{s}$ distribution for h \& $\chi$ Persei sources.  
This distribution has peaks at 14.25 and 15.5.  
(bottom) The 2MASS J and $K_{s}$ magnitudes vs. Mimir J and $K_{s}$ magnitudes for faint 2MASS sources.     
Dotted lines show $\sim 0.2$ magnitude deviations.  The vast majority of sources fall within the dotted lines; 
the majority of sources falling outside the lines have fainter Mimir magnitudes.  This is consistent with 
resolving binary stars, a likelihood given Mimir's higher spatial resolution.
We restrict our analysis of the Mimir data to sources with $J\le 16$. 
}\label{figure 3}
\end{figure}
\begin{figure}
    \centering
\epsscale{0.8}
\plotone{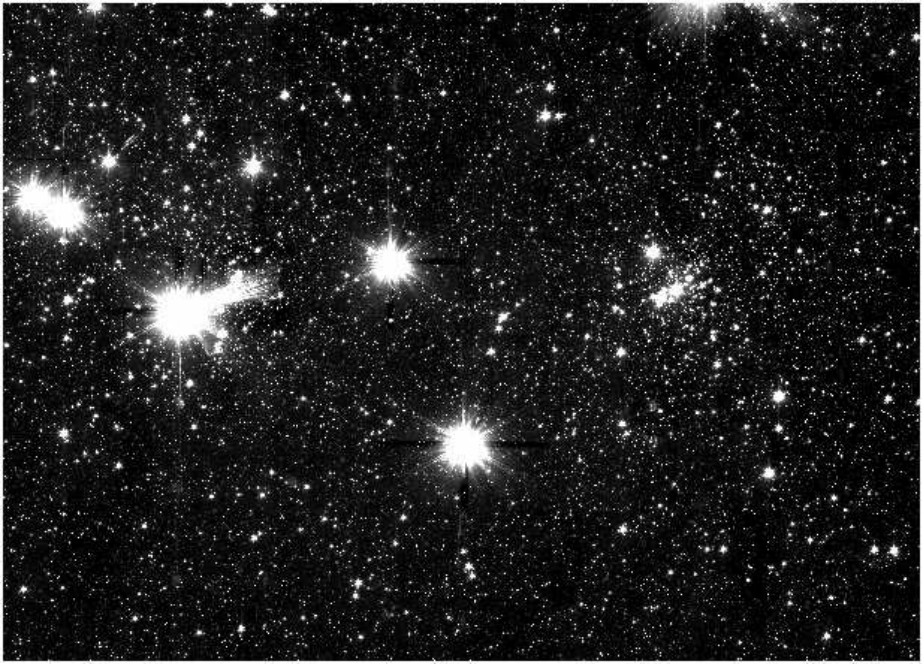}
\caption{IRAC Mosaic image of the h \& $\chi$ Persei region at 3.6 $\mu m$.  The centers of h Persei and $\chi$ Persei
are in the center right and center-left
portions of the image respectively.  The contrast is set to 
98\%.  Diffraction spikes are visible on several of the brightest sources.  
The Spitzer/IRAC coverage is $\sim$ 1 square degree on the sky.}
\end{figure}

\begin{figure}
    \centering
   \plottwo{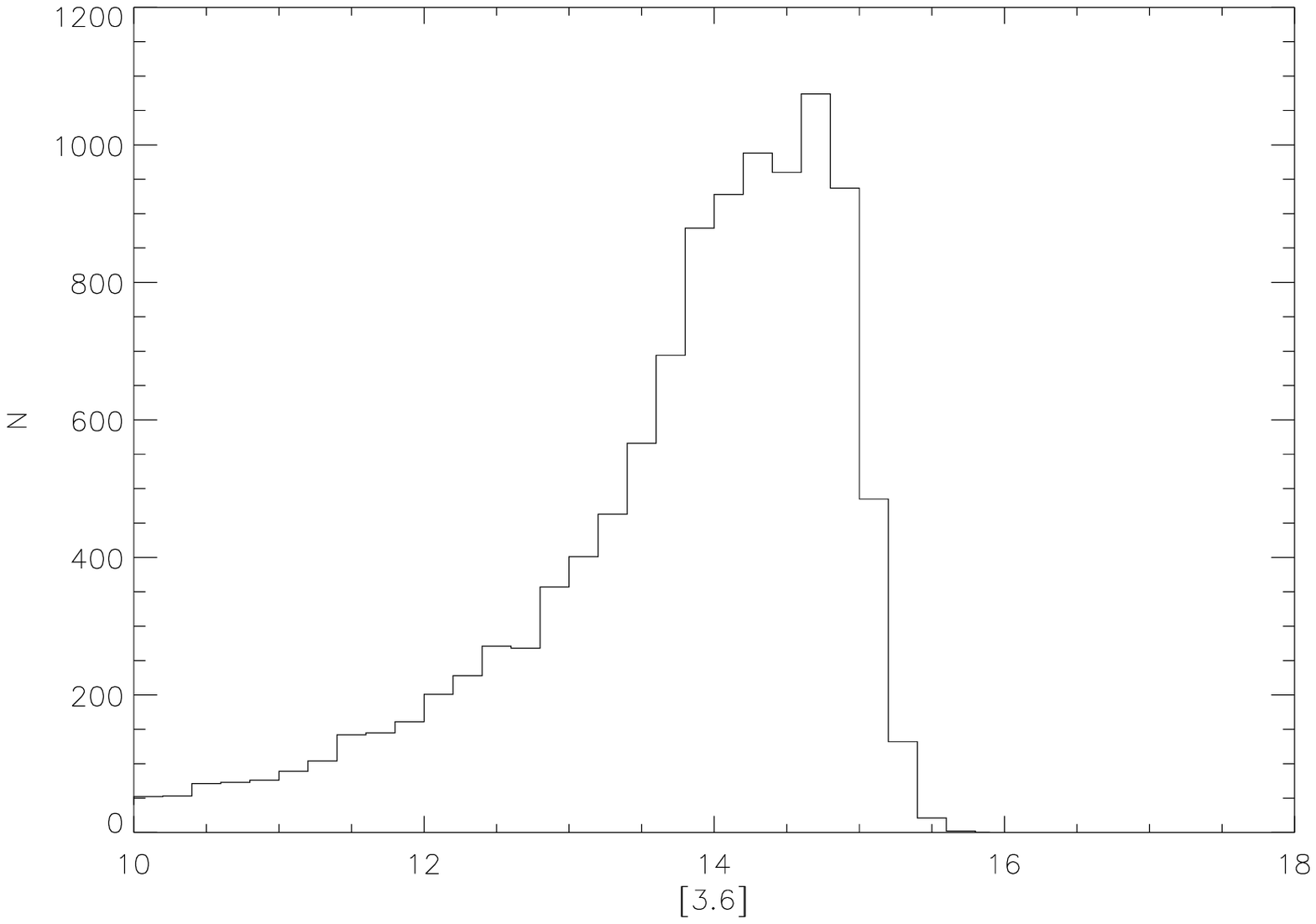}{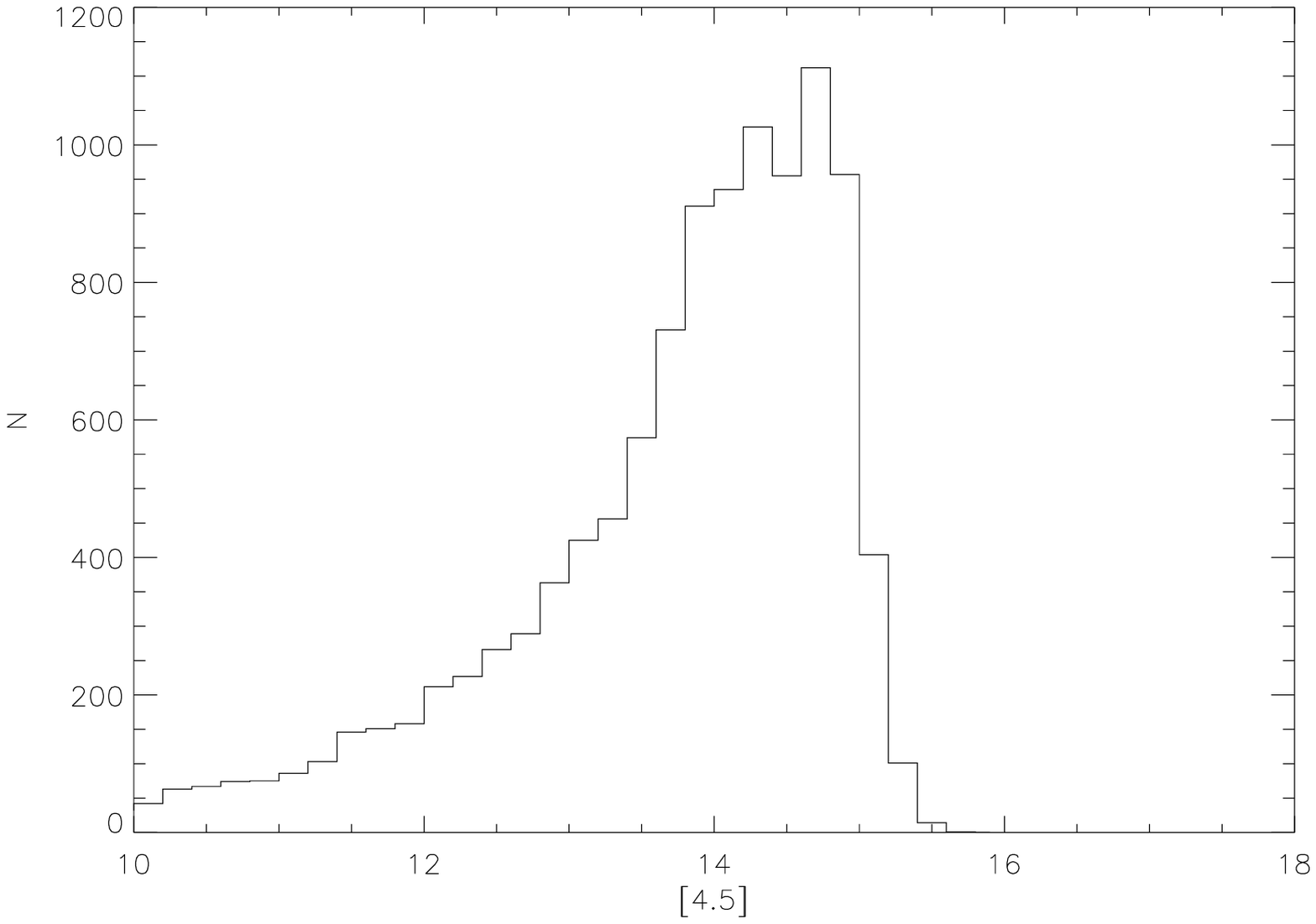}
   \plottwo{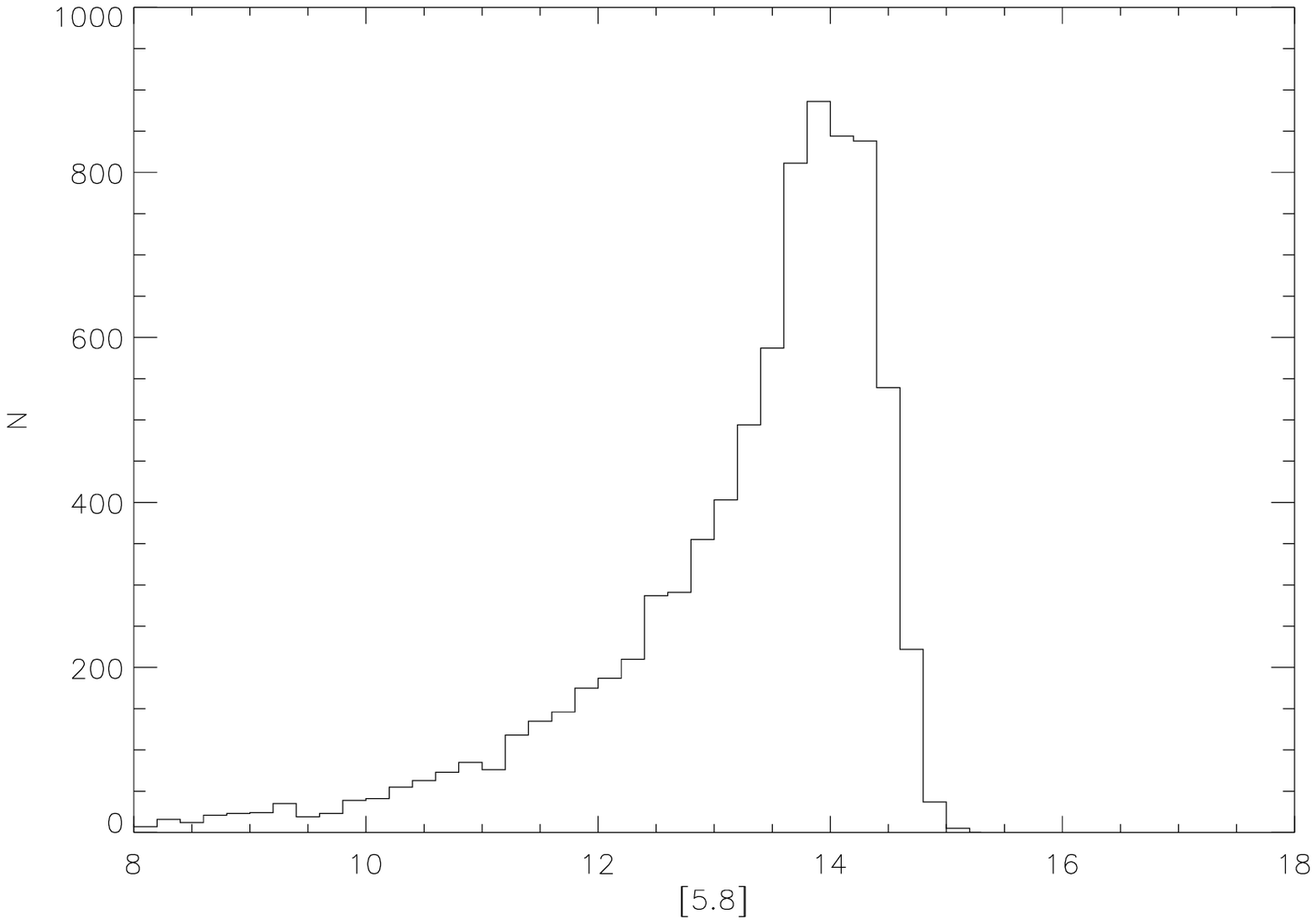}{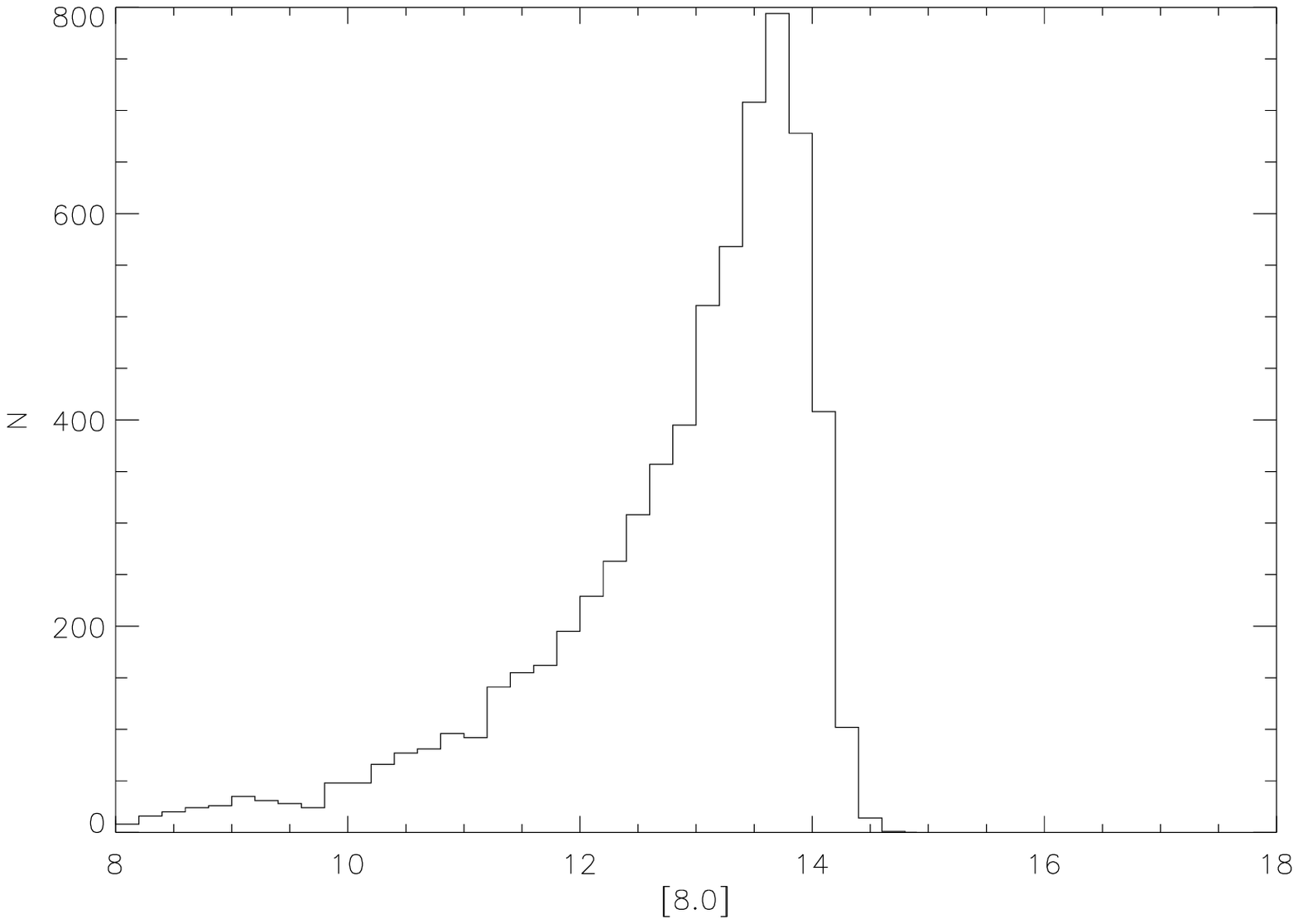}
   \plottwo{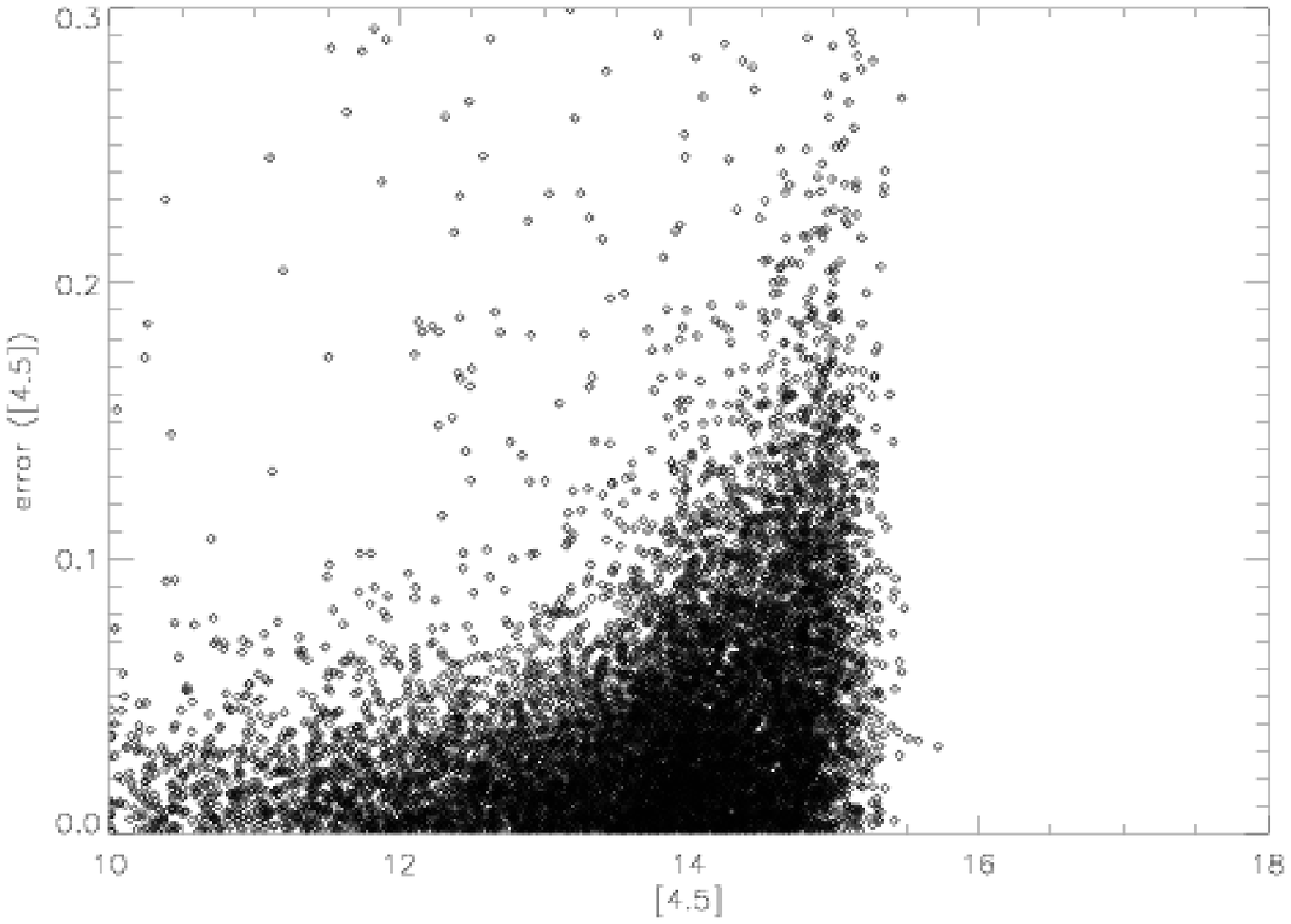}{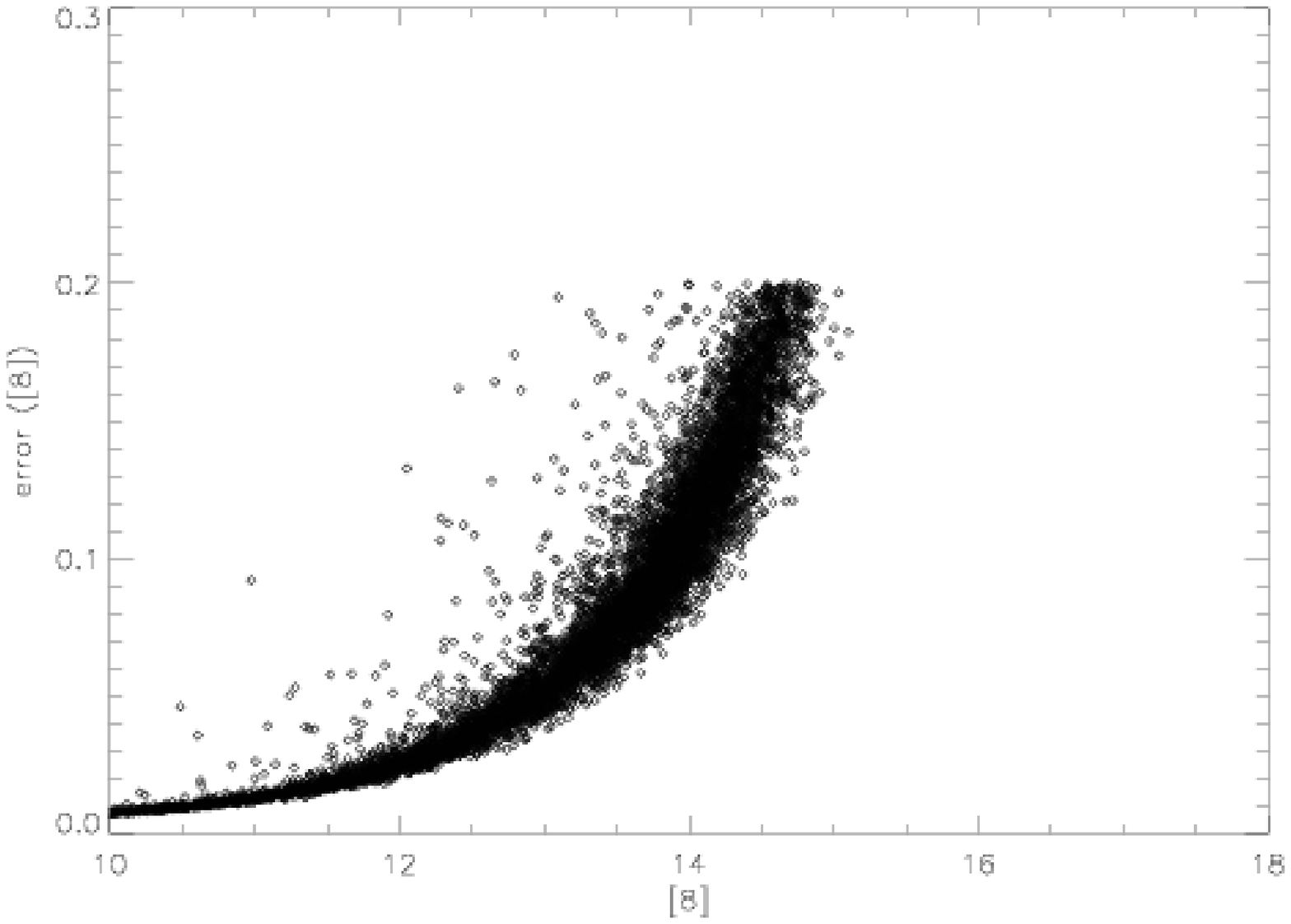}
   \caption{Distributions of 5$\sigma$ detections at [3.6], [4.5], [5.8], and [8]; error distributions at 
[4.5] and [8].
The [3.6] and [4.5] data are complete to $\sim 14.5$ but have a substantial population to m(3.6,4.5)$\sim 15$.
The [5.8] data are complete to $\sim 14.25$ with a substantial population to $\sim 14.5$.  The [8] data are complete to 
$\sim 13.75$ and falls below half the peak value by $\sim 14.5$.  The vast majority of sources in [4.5] have 
errors $\le 0.15$ to m(4.5)=15; sources in [8] have errors $\le 0.2$ to m(8)=14.5.}
\label{magerror}
\end{figure}
\begin{figure}
\plotone{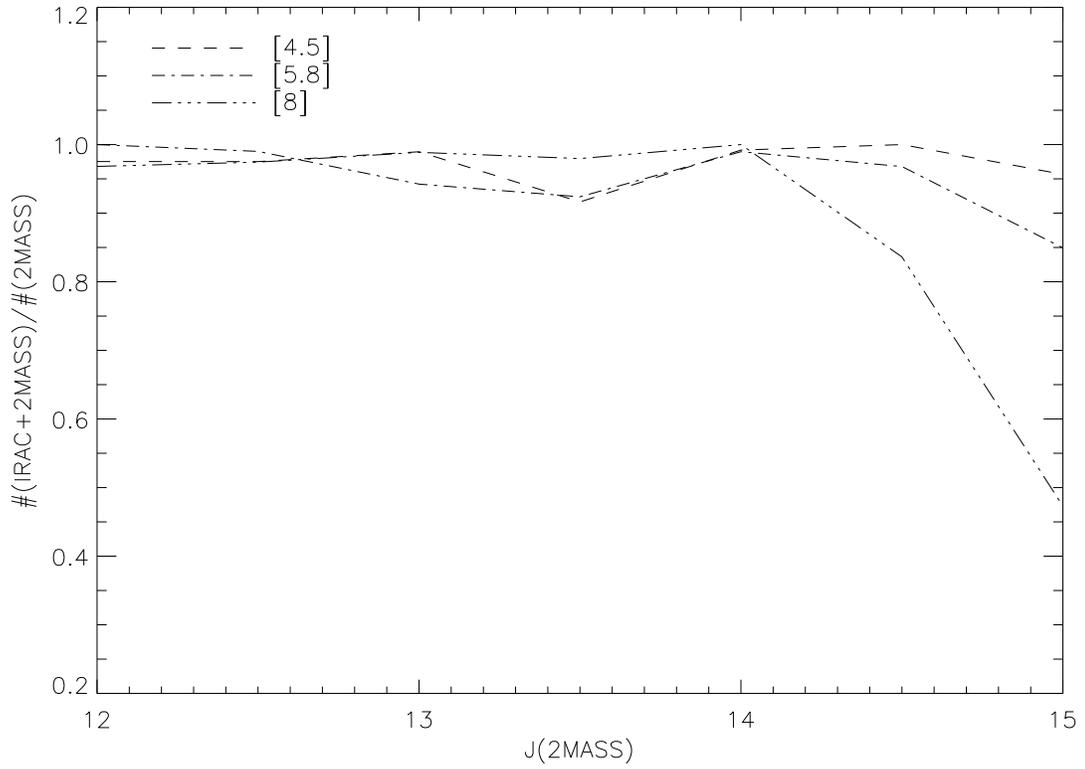}
\caption{The completeness in each IRAC band as a function of 2MASS J magnitude (to J=15.0).  Due 
to the small matching radius used to combine the 2MASS and IRAC data sets, the completeness level for IRAC
oscillates between 90 and 100\% for J$\le 14$.  Except for the [8] band for sources with J=14.5-15, all IRAC bands 
are better than 85\% complete.  In the IRAC analysis section (Sect. 5) we account for 
completeness errors in estimating the IR excess population.}
\label{complete}
\end{figure}
\begin{figure}
\plotone{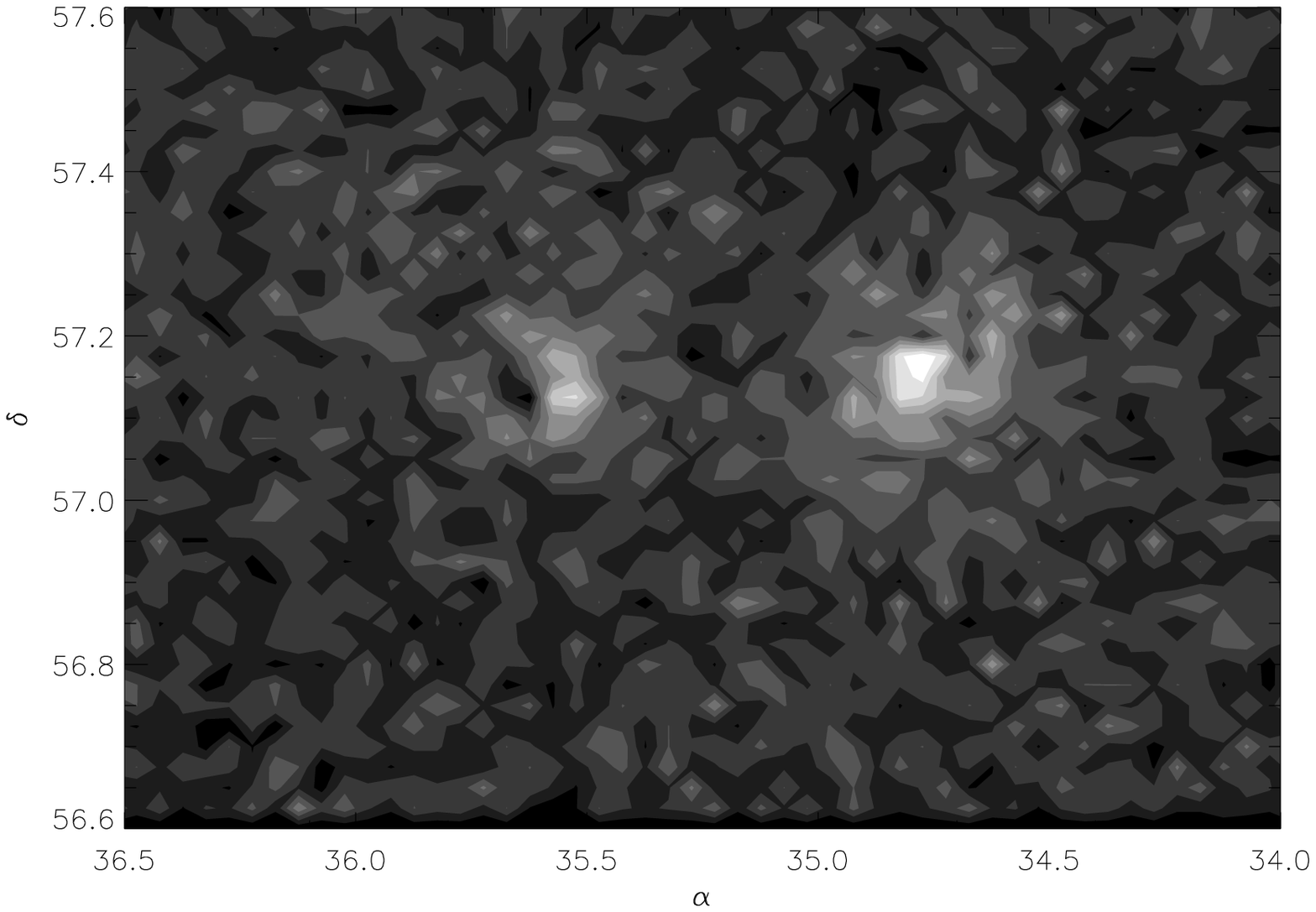}
\plotone{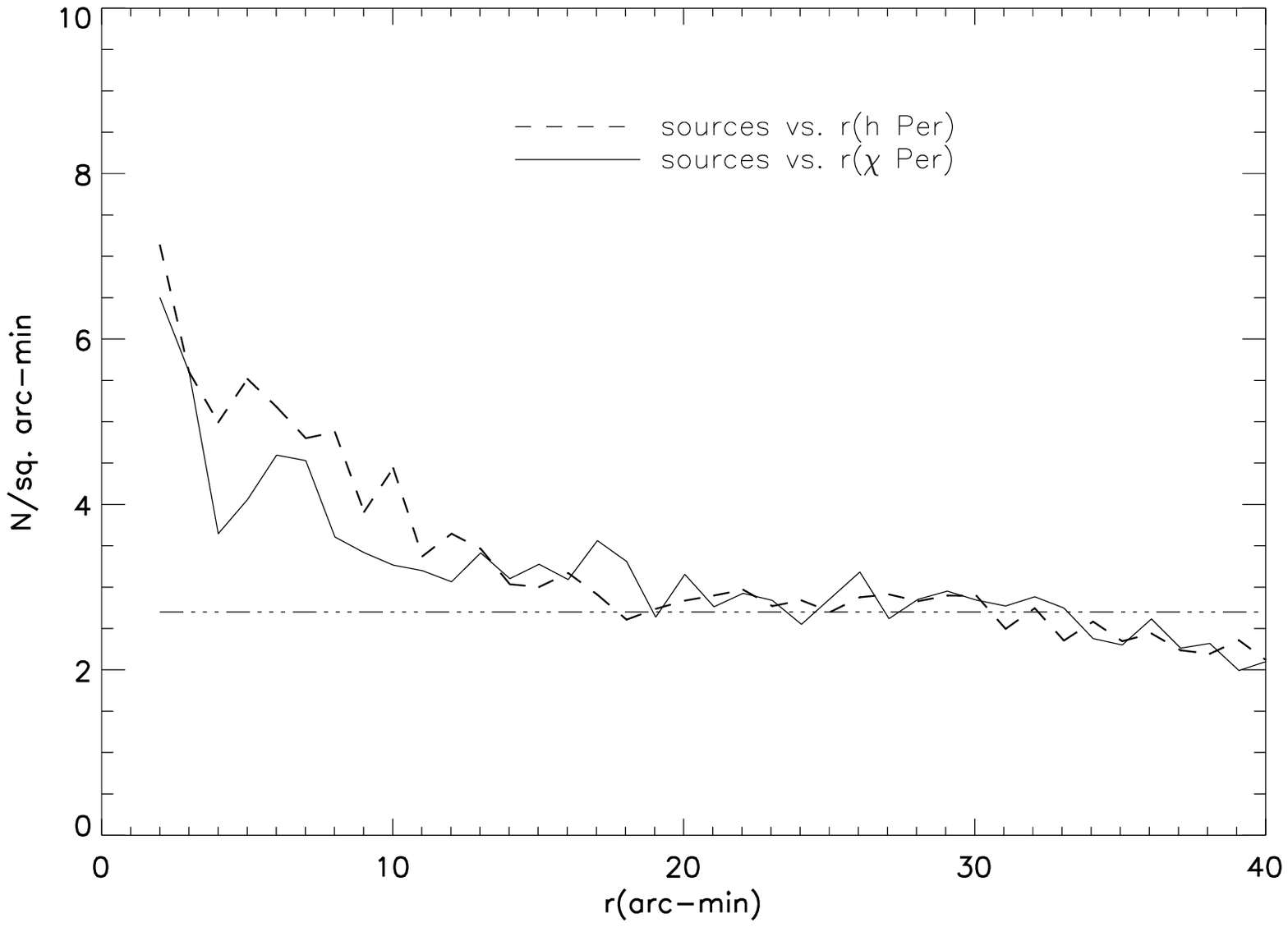}
\caption{(Top) Spatial plot of the star surface density for sources with $J\le 15.5$ from 0 to 90\% in increments of 10\%.   
The centers of both h \& $\chi$ Persei 
(right and left peaks, respectively) are clearly visible and are separated by about $\sim$ 26'.  
The clusters are $\sim$ 20'-30' across with substantial asymmetries, most 
notably close to the $\chi$ Persei core at slightly larger declinations. 
(Bottom) Radial density plot of sources with $J\le 15.5$.  The counts reach the median background density ($\sim 2.7$ $arcmin^{-2}$; 
dash-three dots) 
by $\sim 15-25'$ away from both h Per and $\chi$ Per.}   
\end{figure}
\begin{figure}
   \epsscale{0.98}\plottwo{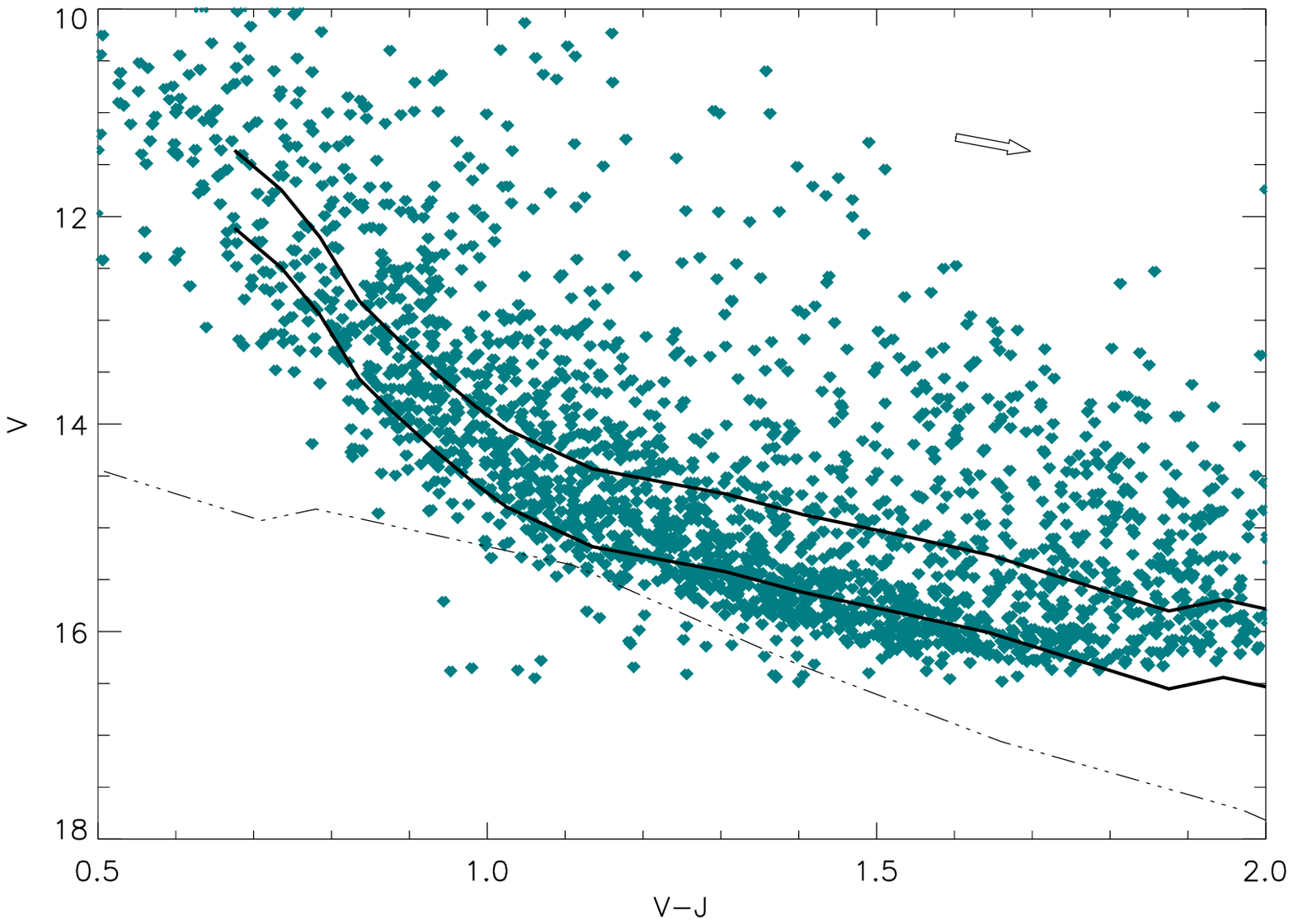}{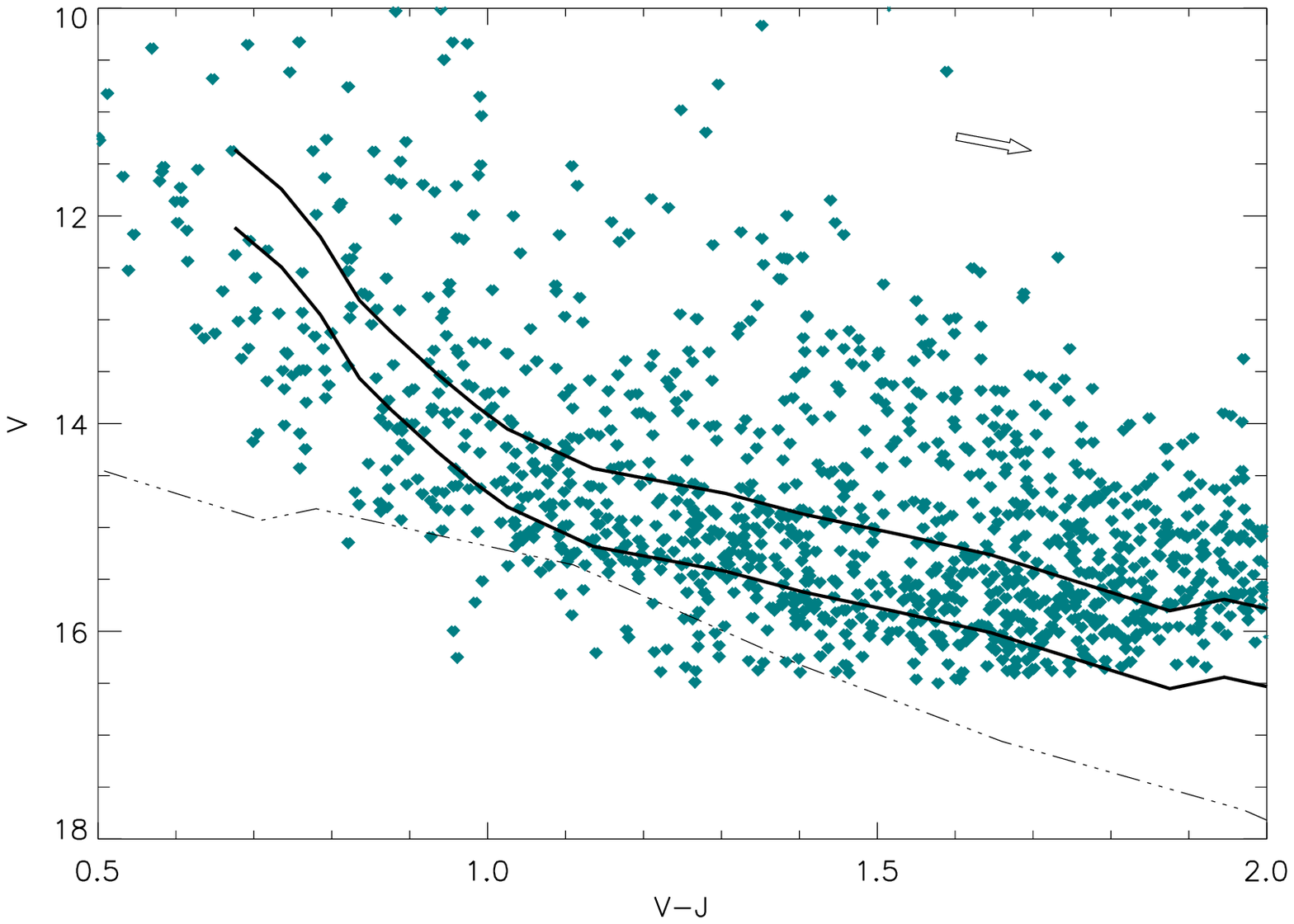}
   \epsscale{0.98}\plottwo{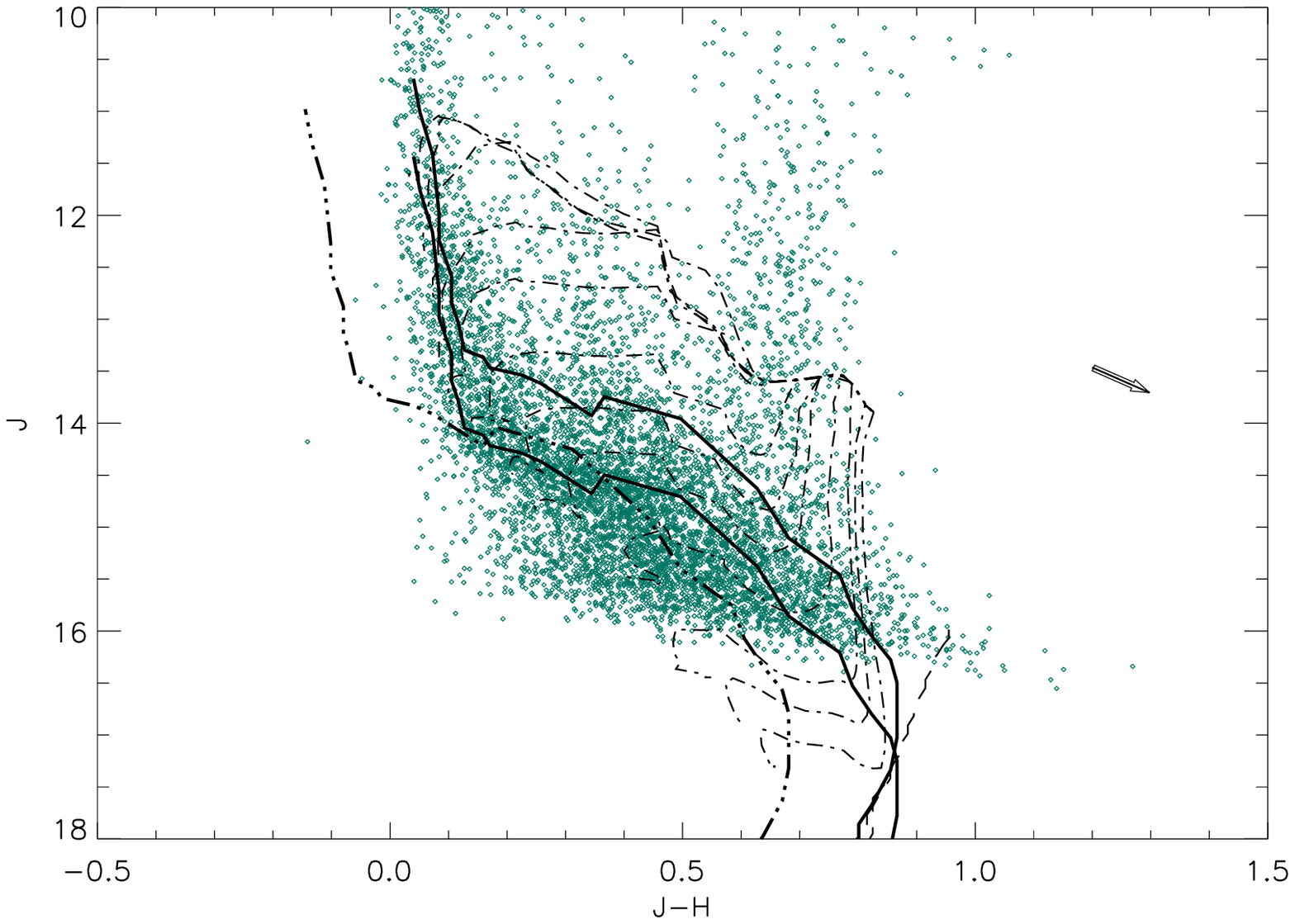}{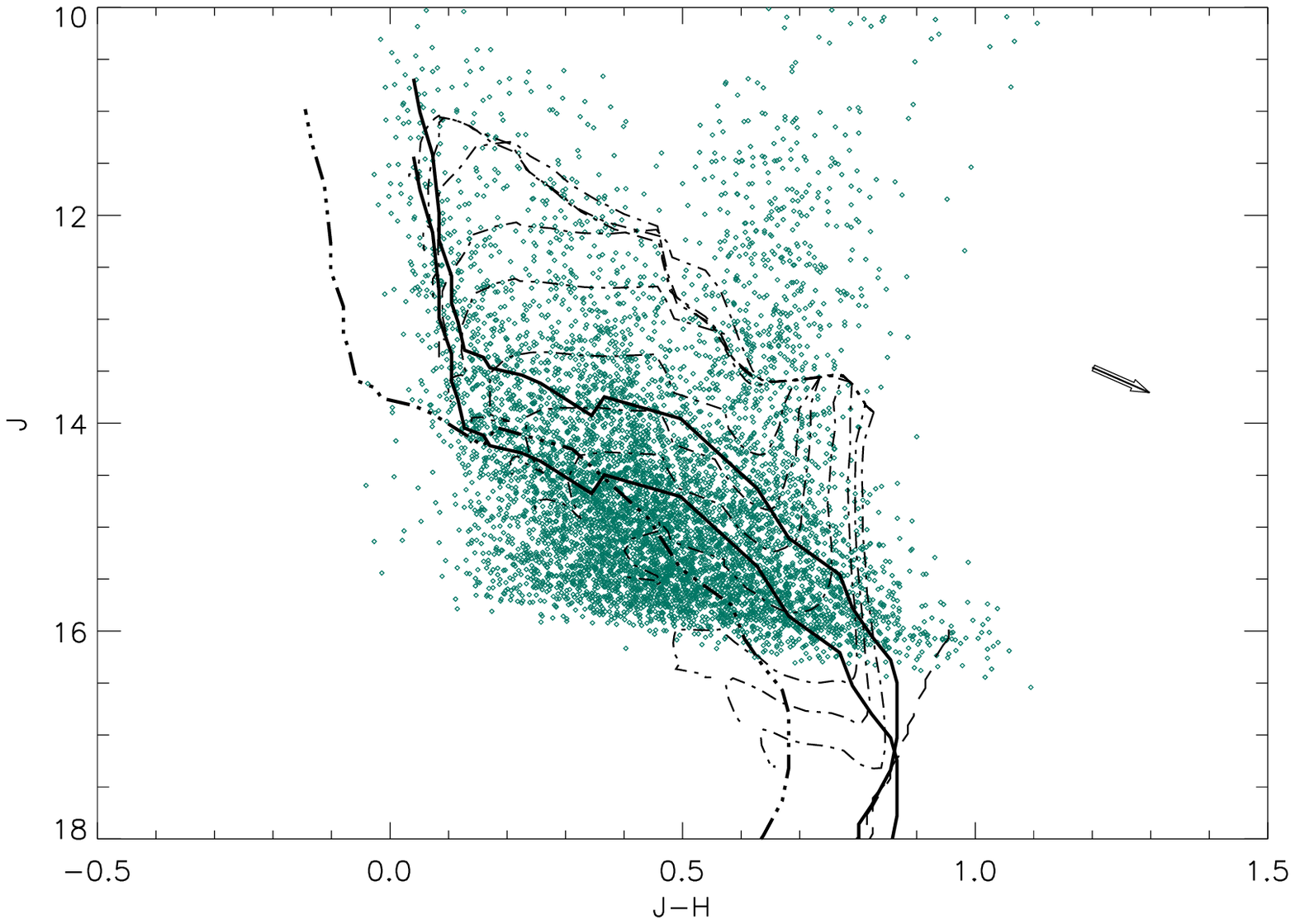}
   \caption{
V/V-J (top) and J/J-H (bottom) color-magnitude diagrams of the 2MASS 
and S02 sources with $\sigma \le 0.2$ for sources within 
15' (left) and those between 15' and 25' (right) from 
the h \& $\chi$ Per cluster centers.  Overplotted are the Siess et al. (2001)
isochrones for a 13 Myr cluster (dash-three dots) with a 'reddened' isochrone (E(B-V)$\sim$ 
0.52; solid line) and appropriate reddening vectors  
(A$_{V}$ $\sim$ 1.4 E(V-J); A$_{J}$ $\sim$ 2.49 E(J-H)).
For the J/J-H diagram, the reddened pre-main sequence tracks for 0.8-7 $M_{\odot}$ 
stars from Bernasconi et al. (1996, dot-dashed) and those for lower
mass stars from Baraffe et al. (1998, long dashes) are also plotted.  Sources between 
15' and 25' away from the cluster centers appear to contain a population tracking the 
isochrone by V$\ge 14.5$ and J$\ge 13.5$.
}\label{VJdist}
\end{figure}
\begin{figure}
    \centering
   {\centering \resizebox*{1.0\textwidth}{!}{{\includegraphics{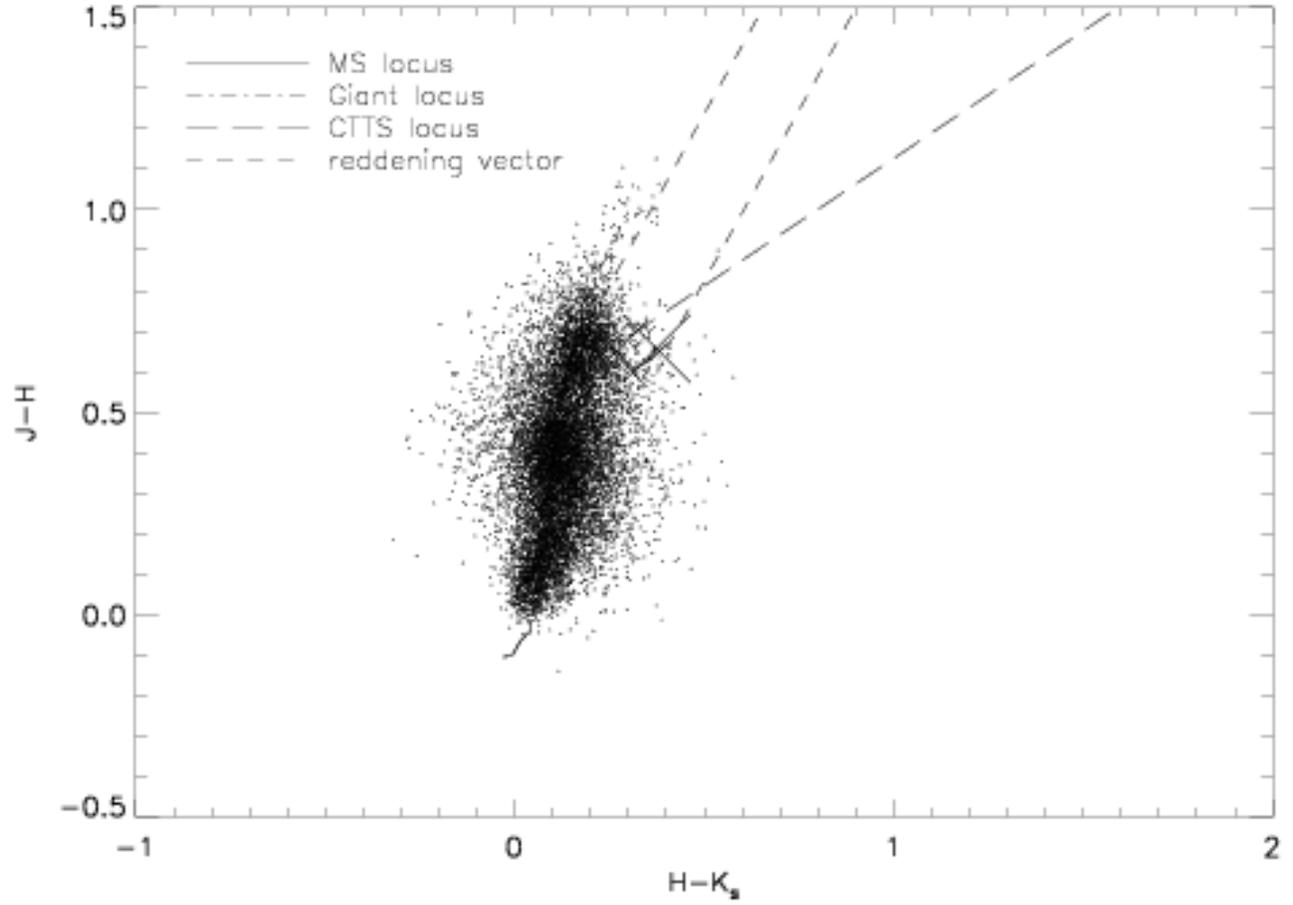}}}\par}
   \caption{J-H vs. H - $K_{s}$ colors from 2MASS for 11048 sources with $J\le 15.5$
.  Overplotted are the classical T Tauri (Meyer et al. 1997) and giant loci (Kenyon \& Hartmann 1995).  
 Crosses represent 
the change in colors that a source with photospheric J-H/H-$K_{s}$ colors of $\sim 0.55/0.1$ would have
if a 1500 K blackbody (from a circumstellar disk) contributed 5\%, 25\%, and 50\% of the total flux at 
3.6 $\mu m$.  
}\label{figure 4}
\end{figure}

\begin{figure}
    \centering
   {\centering \resizebox*{0.8\textwidth}{!}{{\includegraphics{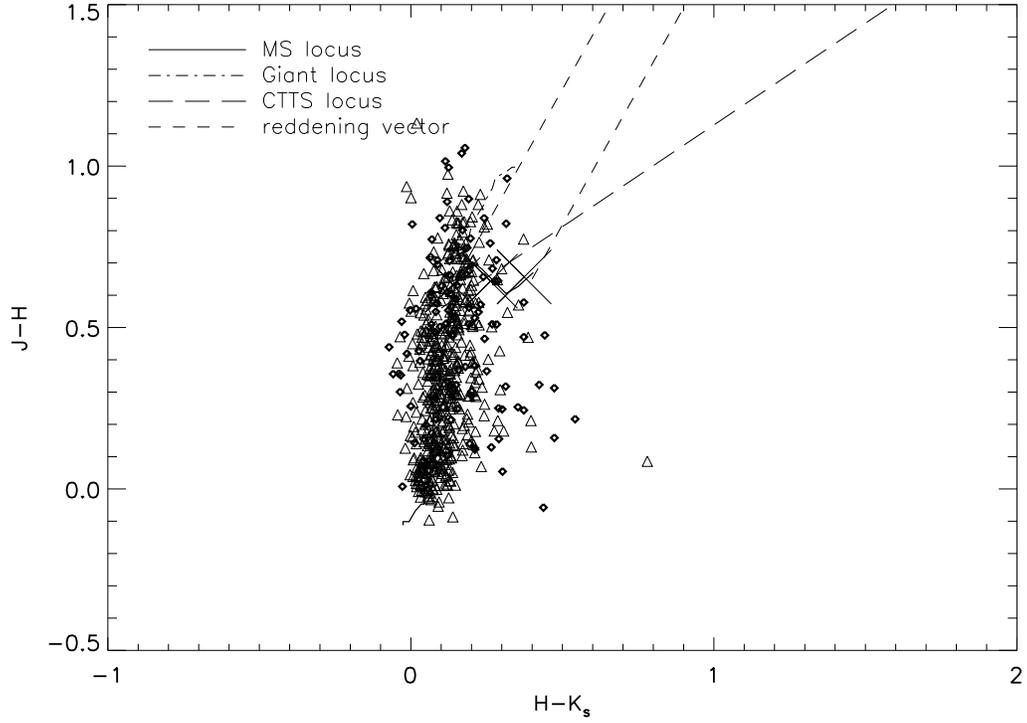}}}\par}
   \caption{J-H vs. H - $K_{s}$ colors from Mimir complete for 657 stars with to J=16 ($\sim 1.1 M_{\odot}$, d$\sim$2.34 kpc) 
.  Filled diamonds represent sources detected at the 10$\sigma$ ($\sigma \le 0.1$) level in all bands, triangles represent 
sources with errors less than $\sigma = 0.2$ in the three bands.
Overplotted are the classical T Tauri (Meyer et al. 1997) and giant (Kenyon \& Hartmann 1995)
loci.  
}\label{jhkmmr}
\end{figure}

\begin{figure}
    \centering
    \epsscale{1.0}\plottwo{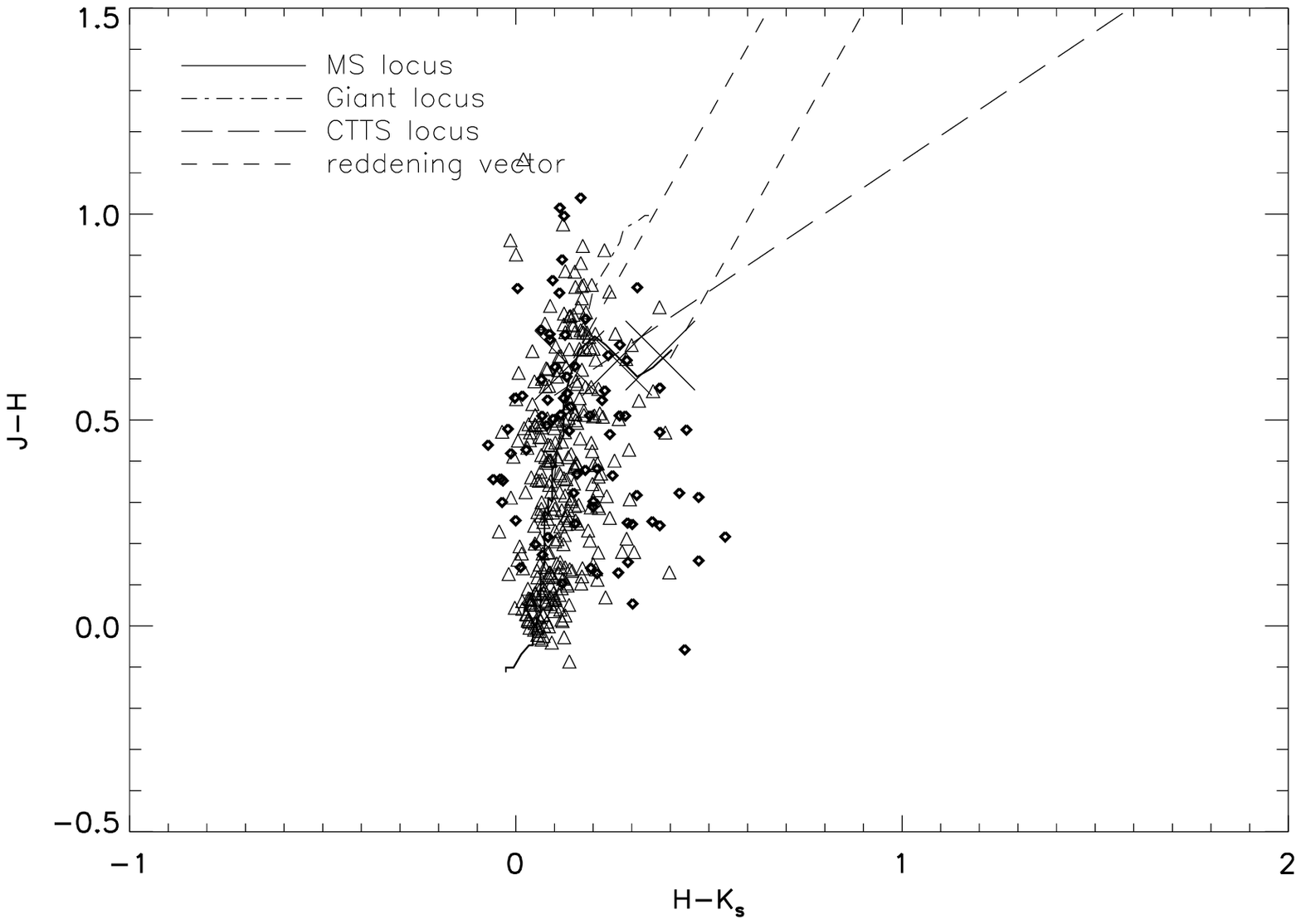}{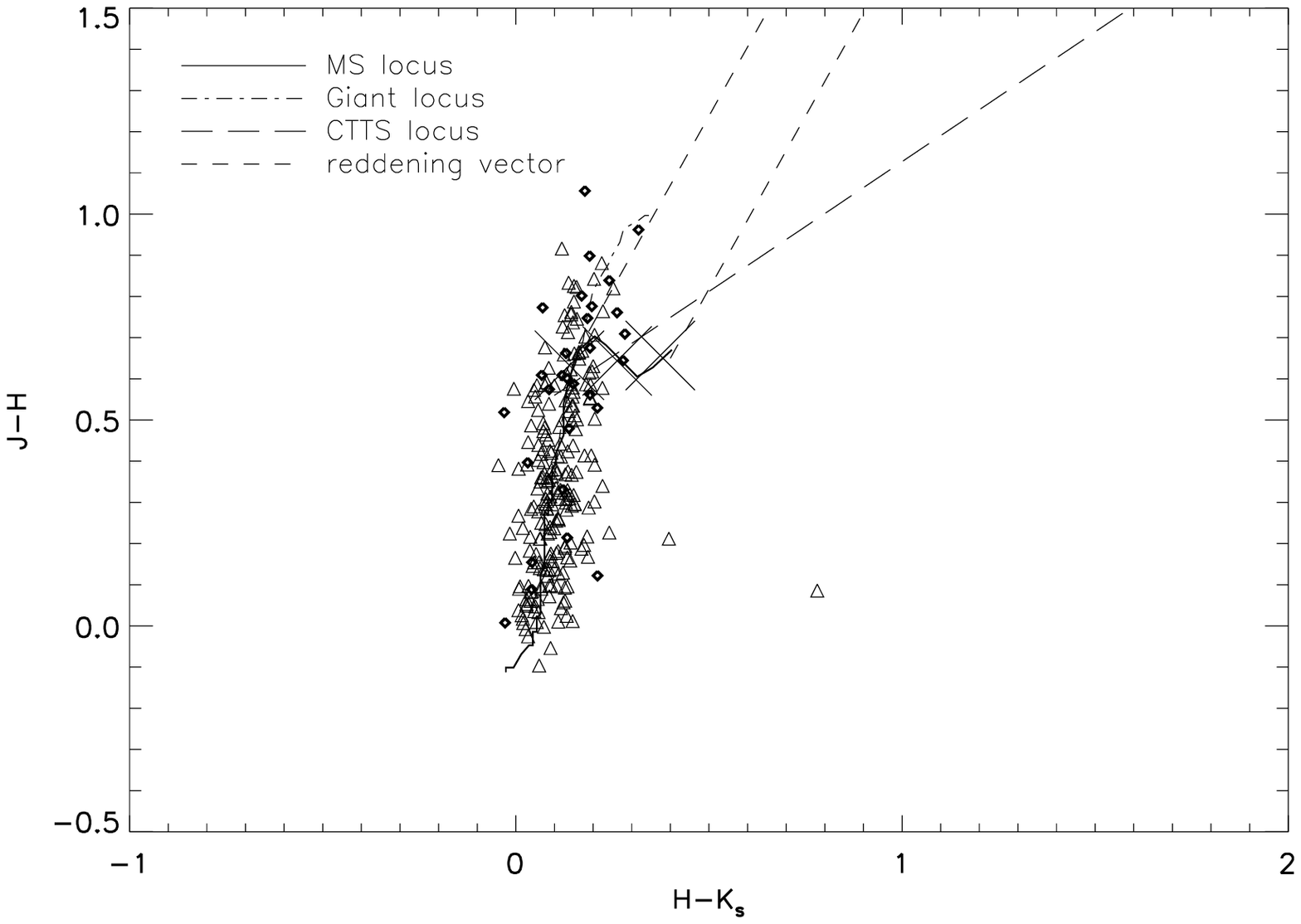}
   \caption{J-H vs. H - $K_{s}$ colors from Mimir 
for h Per (left) and $\chi$ Per (right).
 The symbols are the same as in previous plots.
}\label{jhkmmr2}
\end{figure}

\begin{figure}
    \centering
    \plottwo{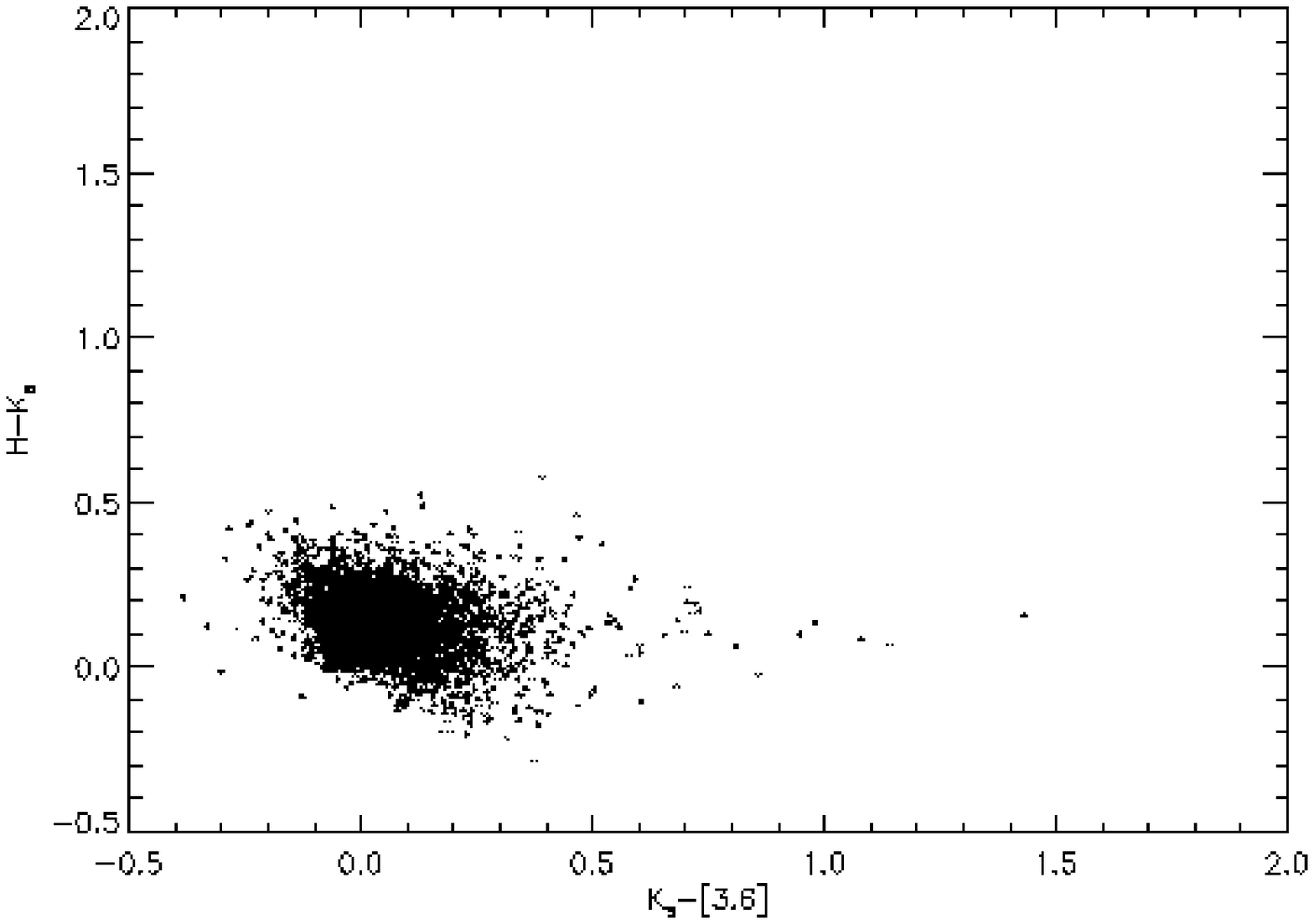}{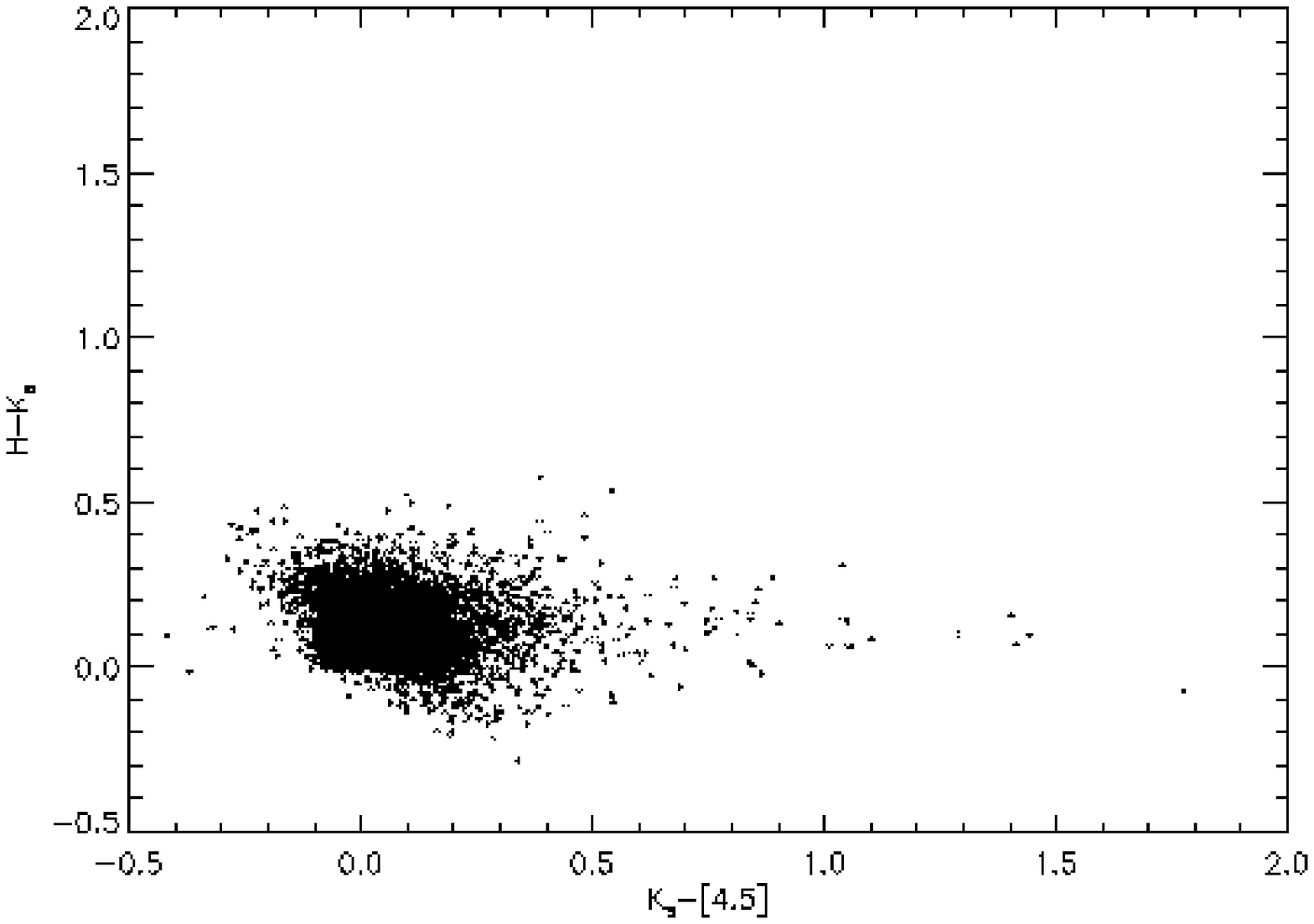}
   \caption{
The H-$K_{s}$/$K_{s}$-[3.6] color-color diagram (left), and
H-$K_{s}$/$K_{s}$-[4.5] diagram (right).  
 The distribution of H-[3.6], H-[4.5], $K_{s}$-[3.6], and $K_{s}$-[4.5] colors show a potential IR excess population 
with $K_{s}$-[IRAC] $\gtrsim$ 0.3-0.4.
}\label{2MIRAC1}
\end{figure}
\begin{figure}
    \centering
\plottwo{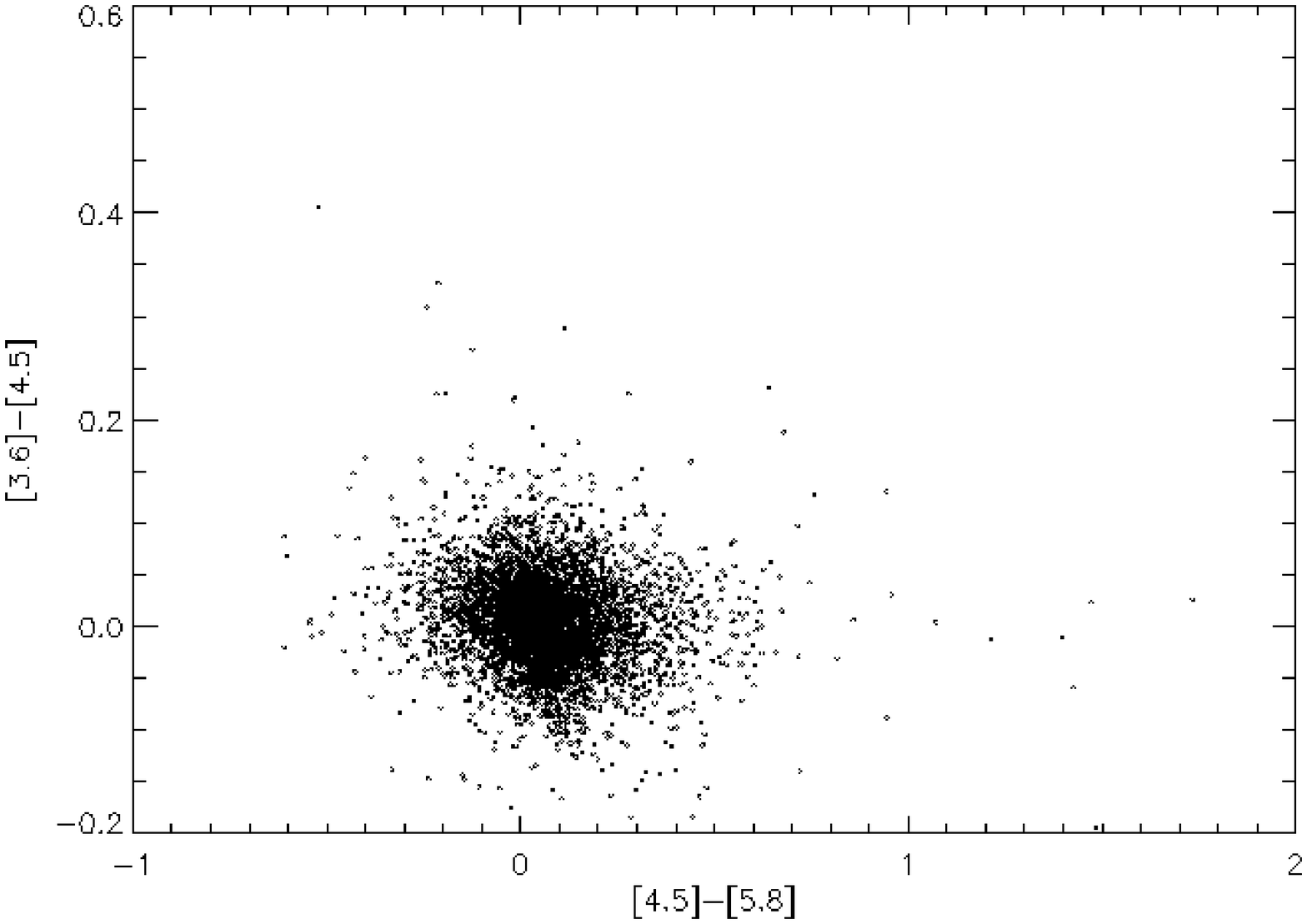}{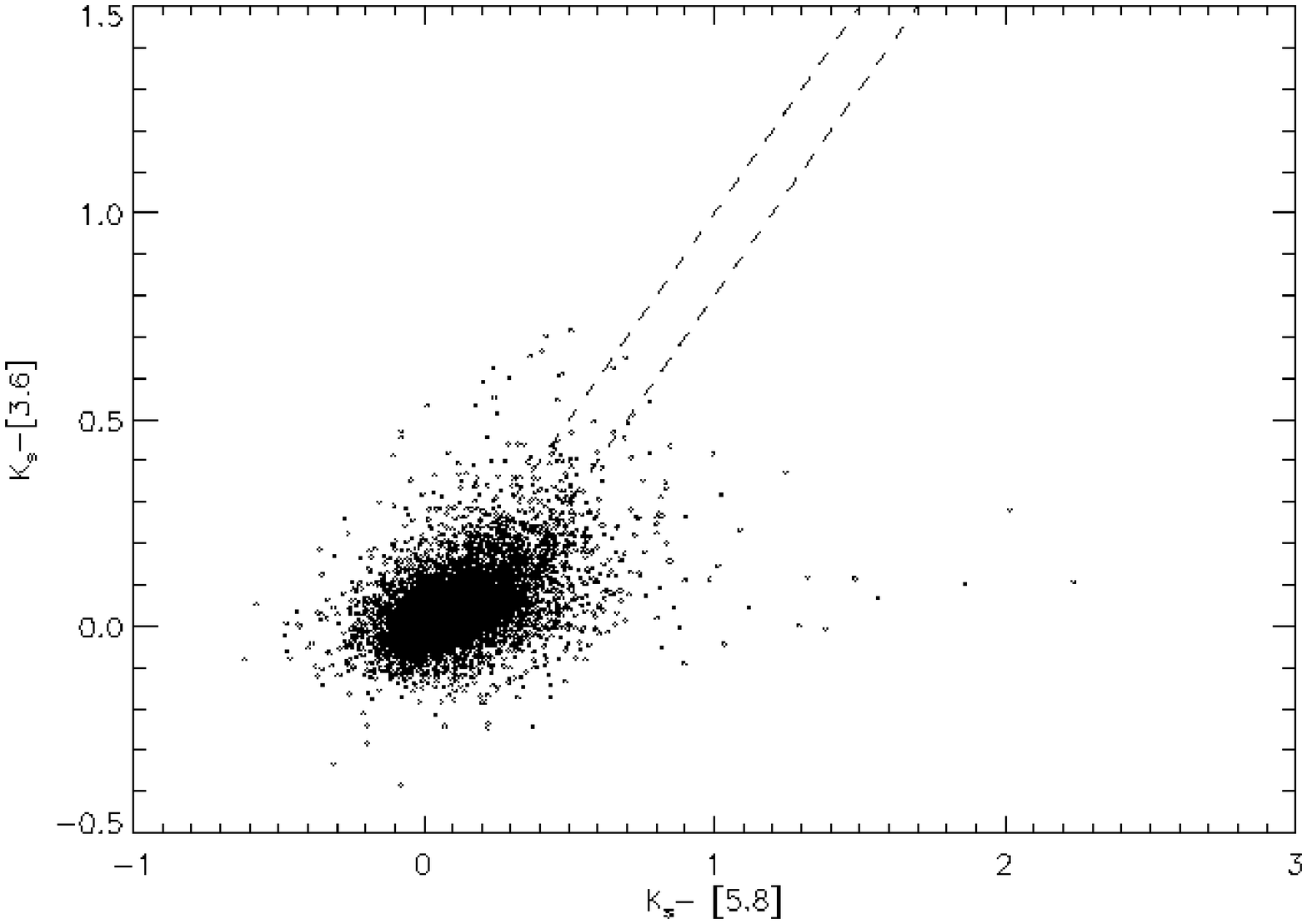}
\caption{[3.6]-[4.5]/[4.5]-[5.8] (left panel) and $K_{s}$-[3.6]/$K_{s}$-[5.8] color-color diagrams (right panel) 
of sources with $J\ge 11, \le 15.5$ and m(5.8)$\le 14.5$.  In the bottom diagram the dotted lines denote sources with [3.6]-[5.8]=0-0.3.
The photospheric population appears to have a red edge [4.5]-[5.8]=0.3 and $K_{s}$-[5.8]=0.4. 
 In Section 5 we count sources with $K_{s}$-[5.8]$\ge 0.5$ as IR excess candidates.
}\label{2MIRAC2}
\end{figure}

\begin{figure}
    \centering
\plottwo{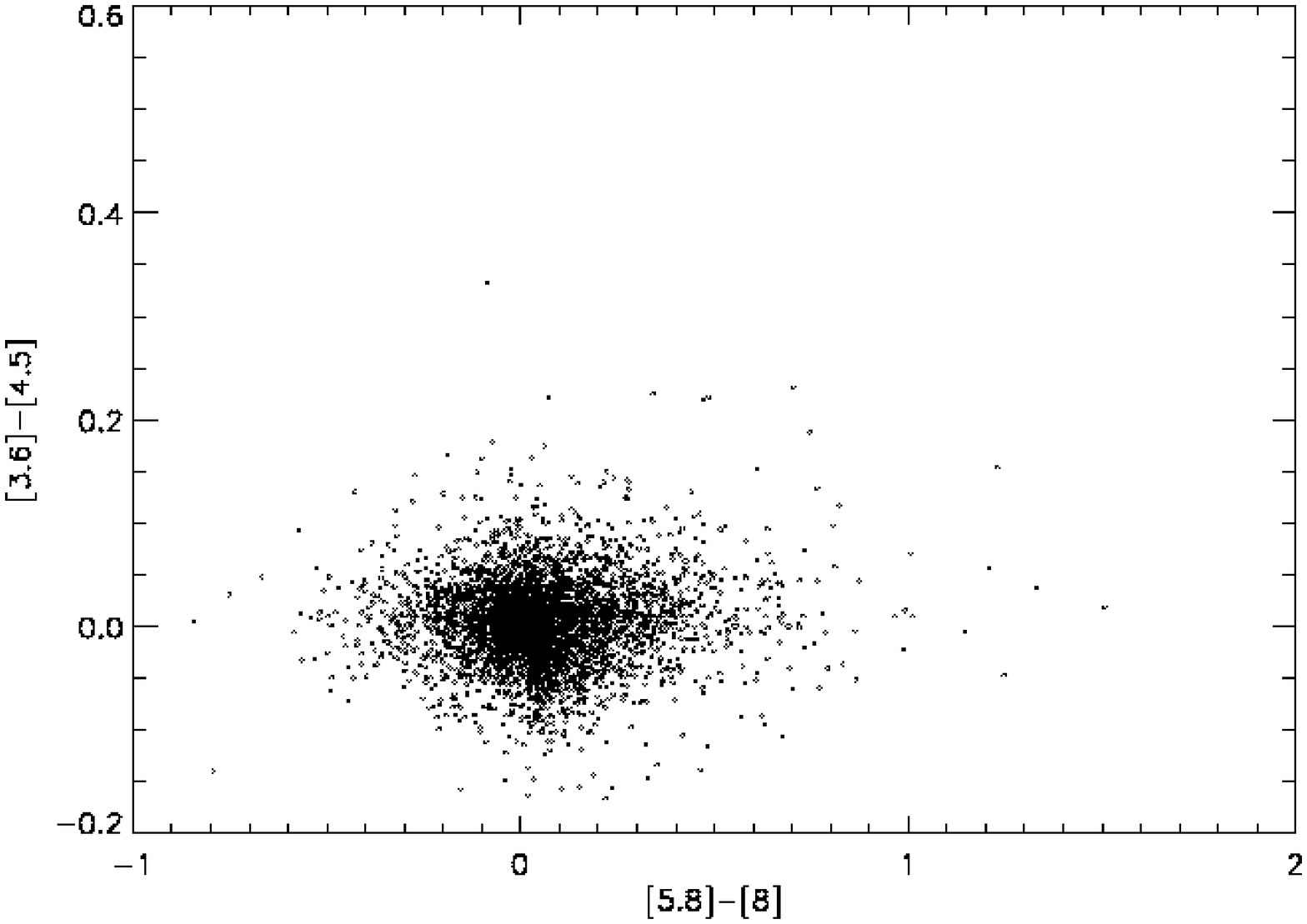}{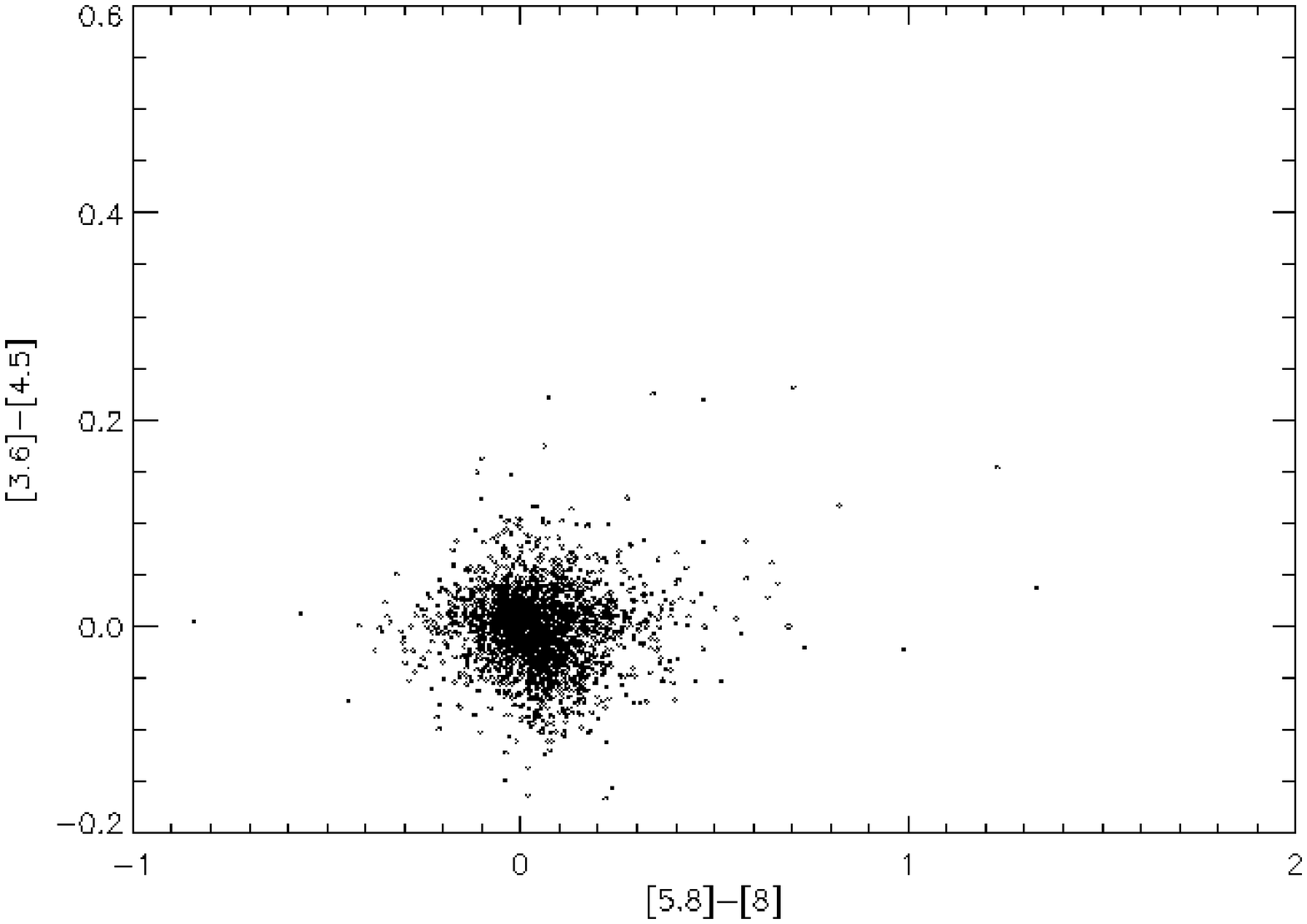}
\caption{(Left panel) Four-channel IRAC colors for sources within 25' of h \& $\chi$ Per centers.  The population of sources in
the clear 'break' between Class II and III T Tauri stars from 0.25-0.4 is large compared to the overall population of 
sources with IR excess (5.8-8 $\ge$ 0.4). (Right panel) The same four-channel IRAC color distribution among sources with
small ($\sigma \le 0.1$) errors in [8].  There are sources in the 'break' between Class II and Class III colors
from [5.8]-[8]=0.25-0.4 and excess sources beyond [5.8]-[8]$\sim 0.4$.  
}\label{2MIRAC3}
\end{figure}
\begin{figure}
    \centering
\plottwo{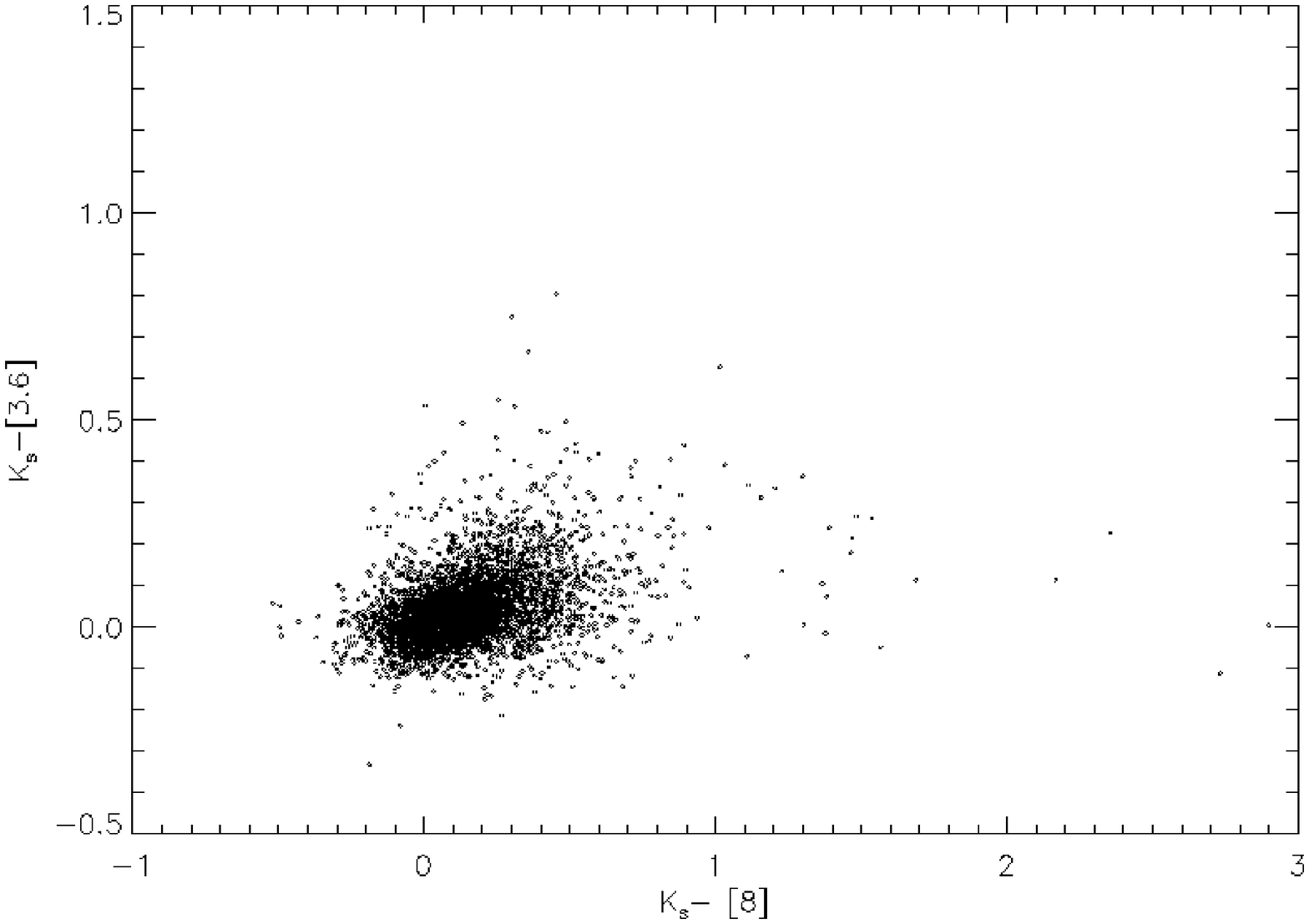}{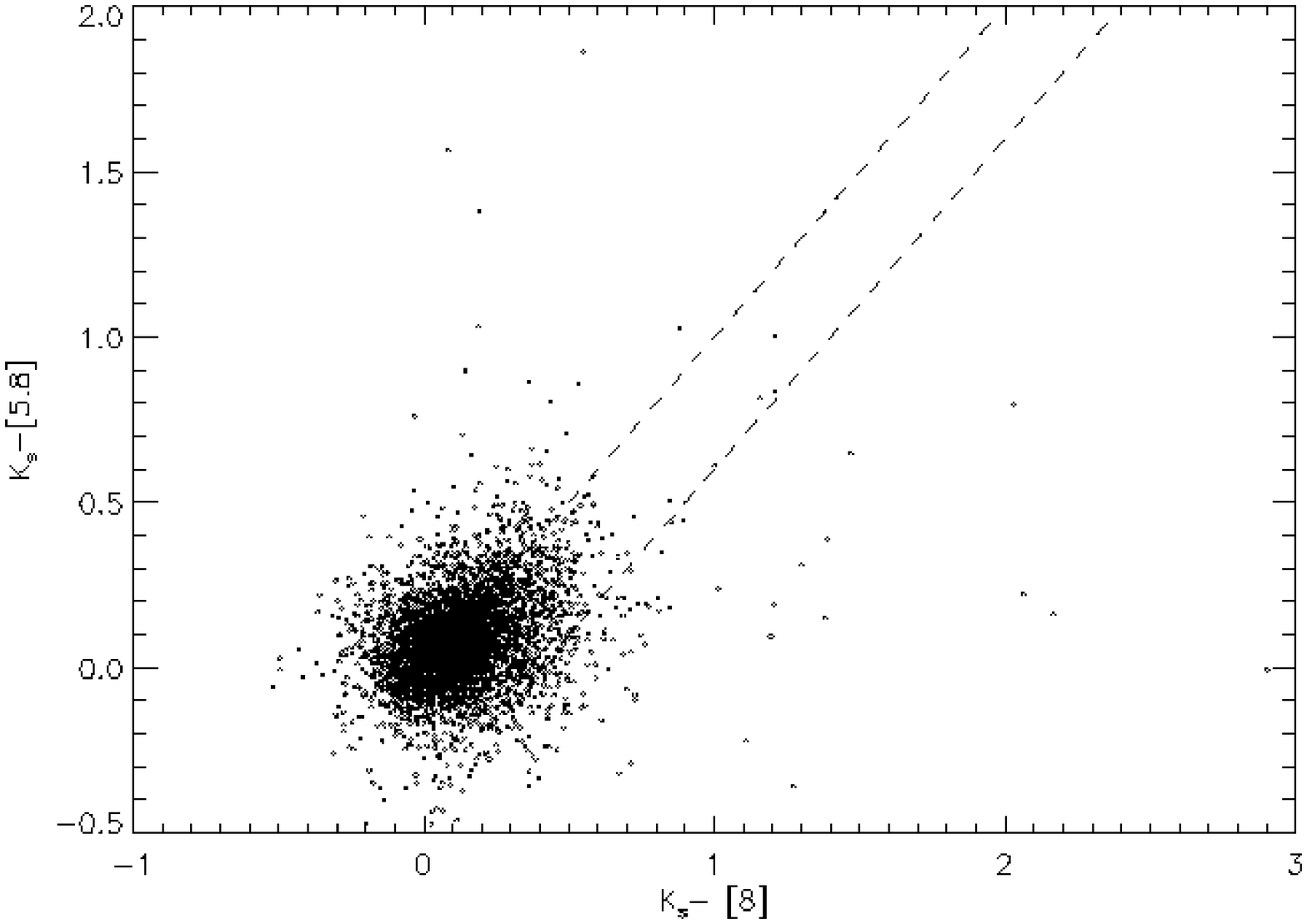}
   \caption{$K_{s}$-[3.6]/$K_{s}$-[8] (left panel) and $K_{s}$-[5.8]/$K_{s}$-[8] (right) color-color diagrams
for sources with J=11-15.5 and [8]$\le$ 14.5.  The dotted lines in 
the bottom panel bound sources with [5.8]-[8]= 0-0.4.
In both plots there is a red edge to the main population at $K_{s}$-[8]$\sim$ 0.4 and many IR excess sources with 
$K_{s}$-[8]$\ge$ 0.4.
}\label{2MIRAC4}
\end{figure}
\clearpage
\begin{figure}
    \centering
   {\centering \resizebox*{1.0\textwidth}{!}{{\includegraphics{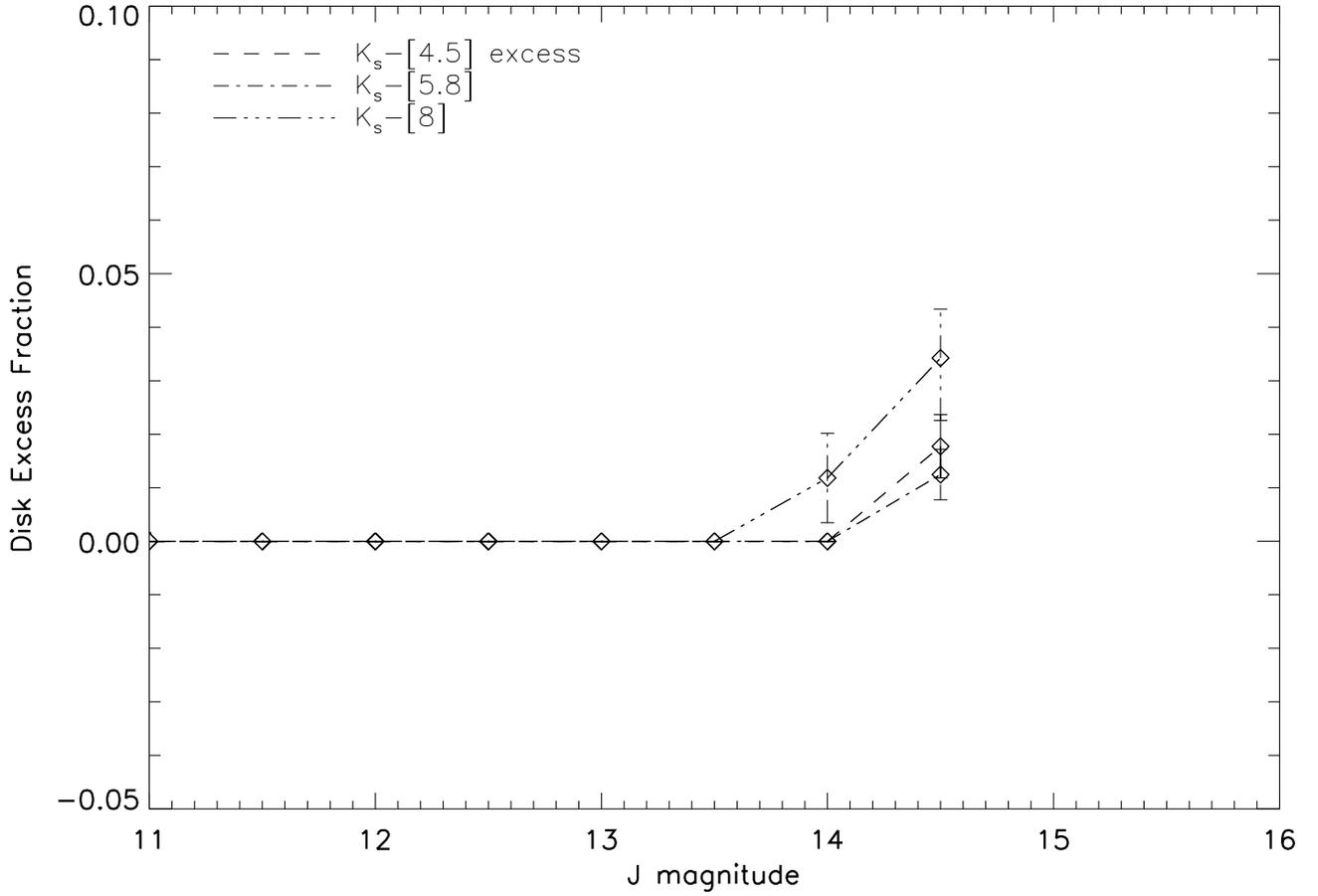}}}\par}
   \caption{The disk fraction at 4.5$\mu m$, 5.8$\mu m$, and 8$\mu$m as a function of J magnitude according to Model 1. 
Fractional excesses are estimated in 0.5 magnitude bins from 10.5 through 14.5, which corresponds 
to a range in mass of 7-1.6 $M_{\odot}$ for d$\sim$ 2.34 kpc.  The IR-excess population is larger for fainter sources.  We detect no 
excess sources with J$\le$ 13.5.  About 3.5\% of sources $\sim 1.6$ $M_{\odot}$ have IR excess and may harbor 
circumstellar disks.}
\label{IRx1}
\end{figure}

\begin{figure}
    \centering
   {\centering \resizebox*{1.0\textwidth}{!}{{\includegraphics{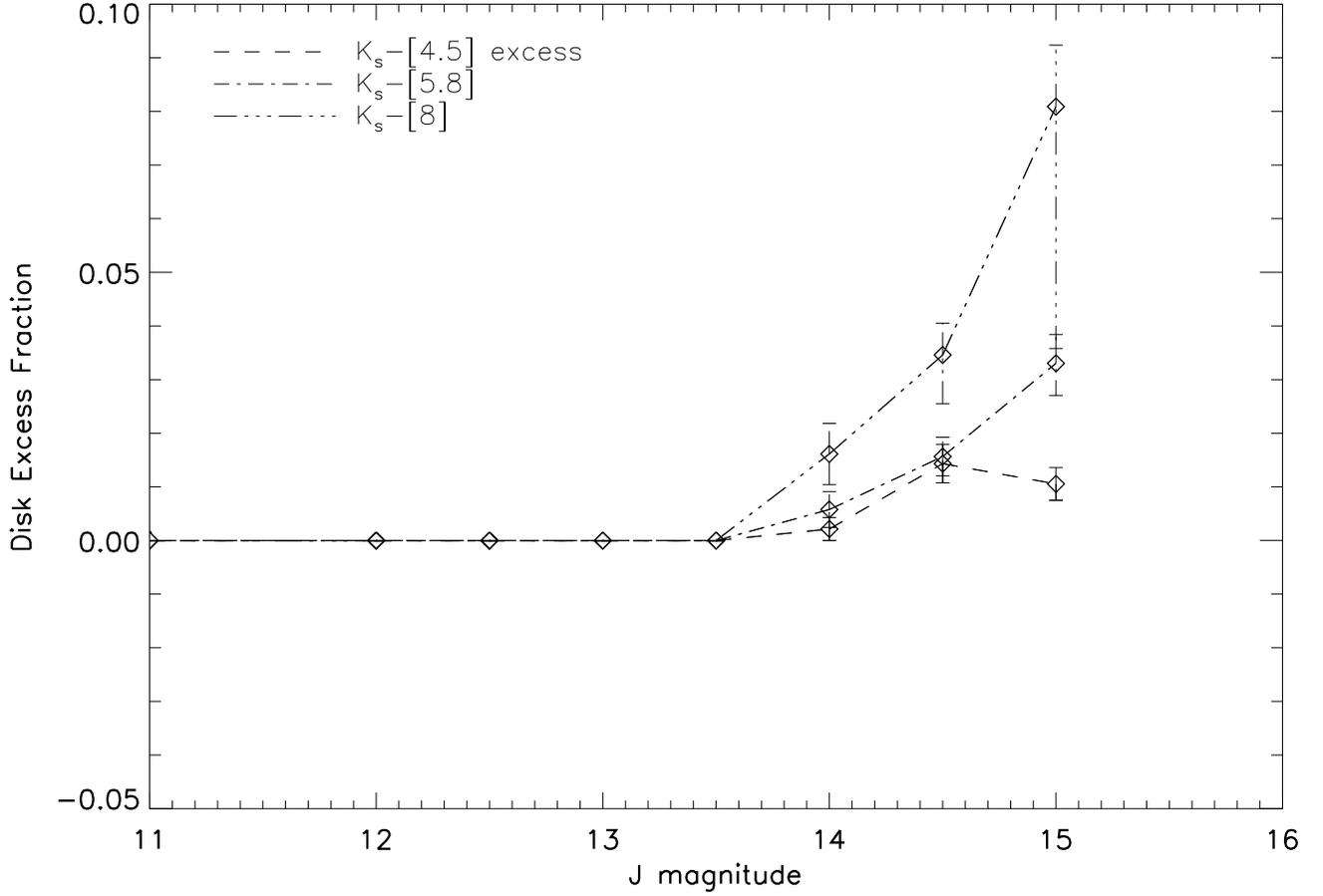}}}\par}
   \caption{The disk fraction at 4.5$\mu m$, 5.8$\mu m$, and 8 $\mu m$ as a function of J magnitude for Model 2. 
Fractional excesses are estimated in 0.5 magnitude bins corresponding to a mass range 
of 7-1.4 $M_{\odot}$.  The results are consistent with those in Figure 16 (Model 1). 
 The IR excess population is clearly larger for fainter sources to J=15.0.  There are no IR excess 
sources for J$\le 13.5$ ($\sim 2.7 M_{\odot}$), which implies that disk lifetimes are larger for lower mass stars 
(to $\sim 1.4 M_{\odot}$).  The IR excess population is also consistently larger at progressively longer wavelengths, as  
 expected if the disk lifetime in the hotter, inner disk is shorter than in the cooler, outer disk.   About 4-8\% of sources 
to J=15 have IR excess and may possess circumstellar disks.
}\label{figure 13_02}
\end{figure}
\clearpage
\begin{figure}
\epsscale{0.7}\plotone{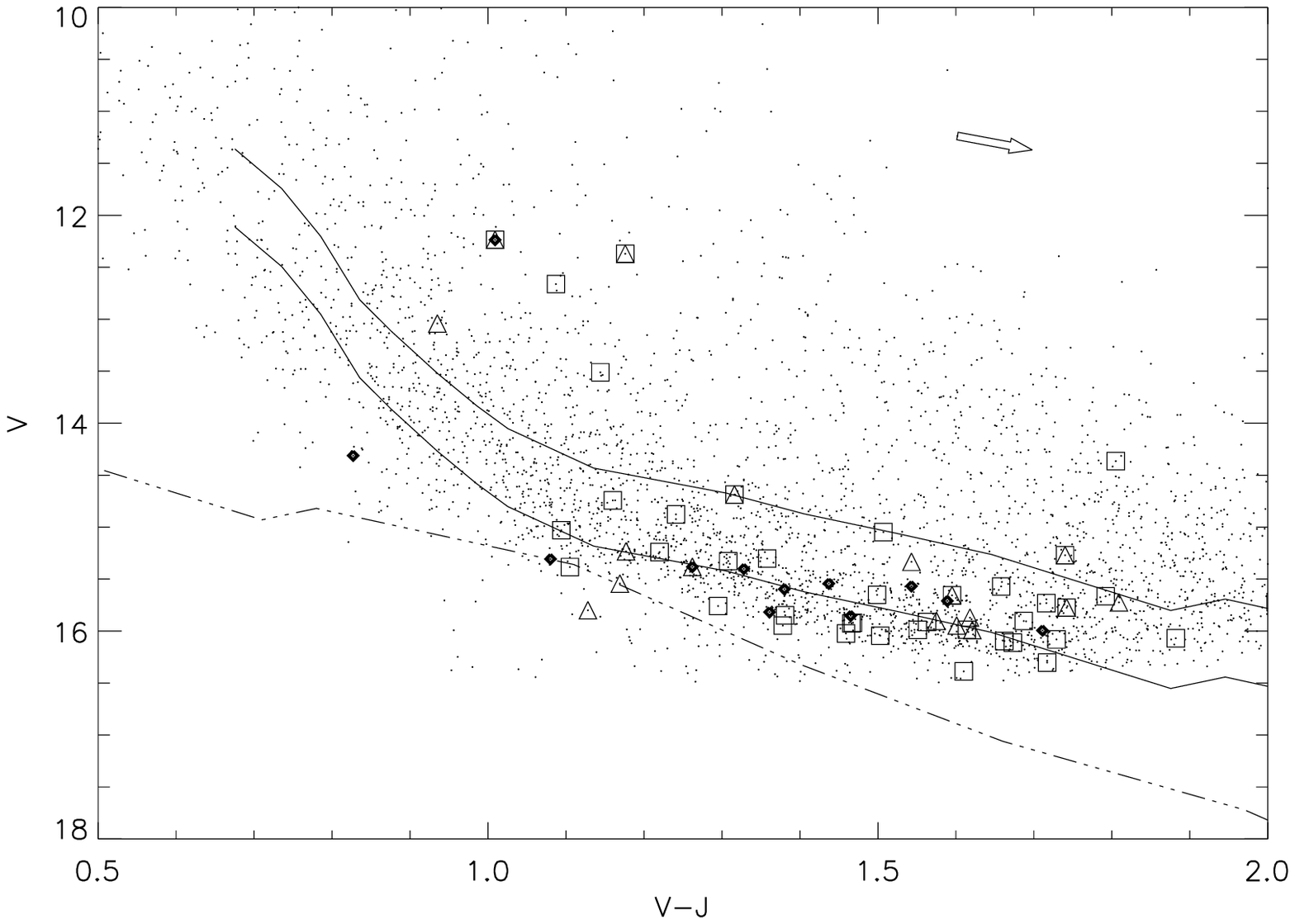}
\epsscale{0.7}\plotone{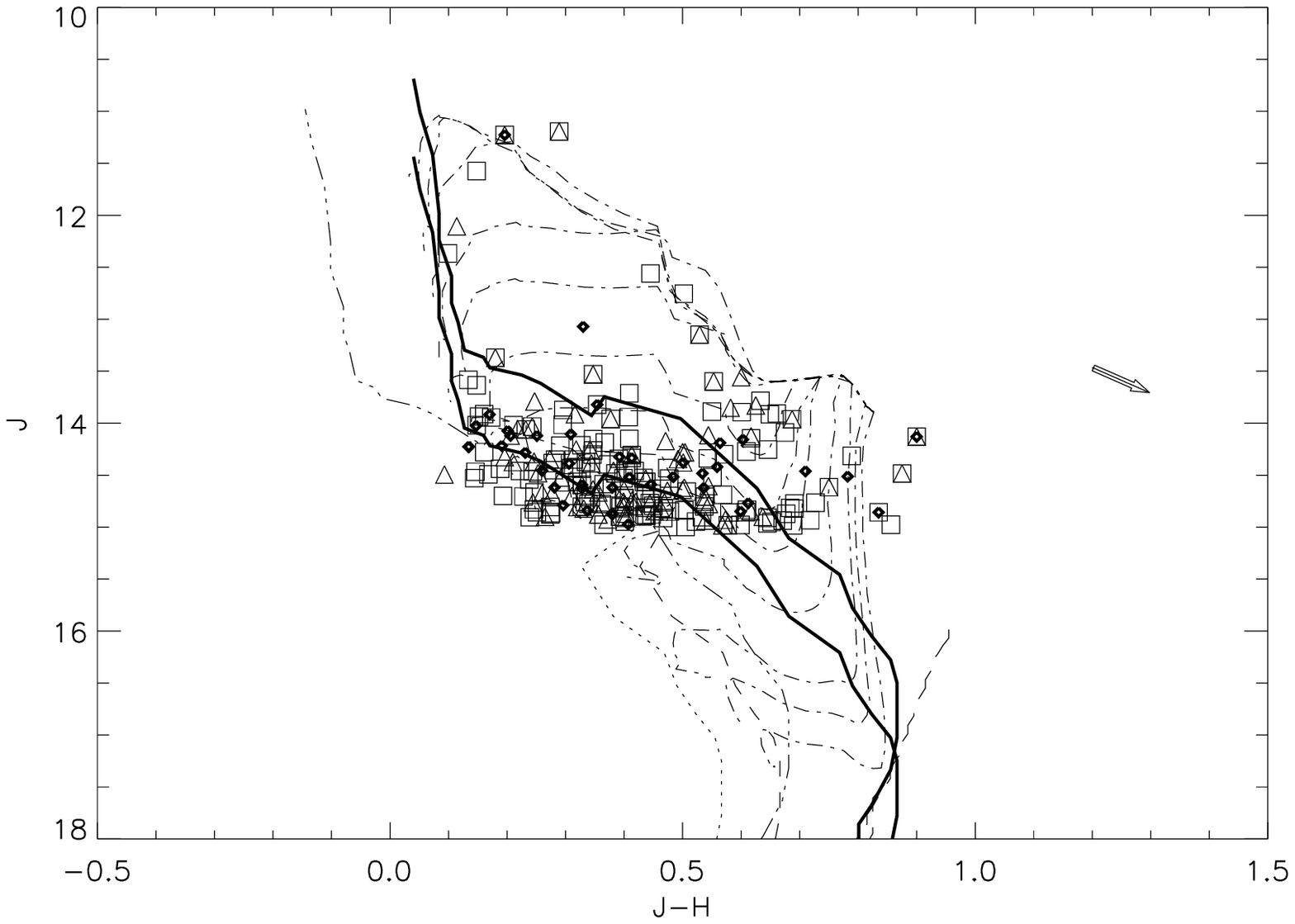}
\caption{The distribution of IR excess sources on color-magnitude diagrams 
of all sources.  Both plots show the 
$K_{s}$-[4.5] (filled diamonds), $K_{s}$-[5.8] (triangles), and $K_{s}$-[8.0] (squares) 
excess and J$\le$15.5.  Arrows correspond to the appropriate reddening vectors as in Figure 8.
In the top panel, the V/V-J distribution for S02 sources with the IR excess sources is overplotted.  The reddened 
13 Myr isochrone (dash-three dots), the isochrone for unresolved binaries (top solid line) and the 
unreddened isochrone (bottom solid line) are shown, assuming $A_{V}$$\sim$3.12 E(B-V)$\sim$ 0.16 and 
E(V-J)$\sim$ 1.16 (Bessel \& Brett 1988).
The vast majority of IR excess sources are consistent with the 13 Myr isochrone. 
The bottom panel shows the 2MASS J/J-H diagram of for IR excess sources with 
.  The reddened 13 Myr isochrone (assuming $A_{J}\sim 0.46$, $E(J-H) \sim 0.185$ extinction)
the 13 Myr isochrone for unresolved binaries are shown as the bottom and top solid black lines, respectively.  
The unreddened 13 Myr isochrone is also shown (dash-three dots).  Again the vast majority 
of IR excess sources are consistent with the 13 Myr isochrone and thus are consistent with 
cluster membership.}
\end{figure}
\end{document}

%% file: tab1.tex
\begin{deluxetable}{rrrr}
\tablecolumns{4}
\tablecaption{Mimir observations coverage}
\tablehead{{Field} & {$\alpha_{o}$} & {$\delta_{o}$} & {Date (2005)}}
\startdata
1 &2h18m56.4s &57$^{o}$8'35" & November 4-5\\
 2  &2h19m16.8s  &57$^{o}$8'8'35"  &December 5 \\
 3 &2h19m37.3s  &57$^{o}$8'35"  &November 4-6 \\
  4  &2h19m57.6s  &57$^{o}$8'35"  &November 6 \\
  5  &2h19m16.8s  &57$^{o}$5'35"  &November 6 \\
  6  &2h19m37.2s  &57$^{o}$5'35"  &November 6  \\
  7  &2h19m57.6s  &57$^{o}$5'35"  &November 6 \\
  15  &2h22m4.3s  &57$^{o}$8'35"  &November 30  \\
  16  &2h21m43.2s  &57$^{o}$8'35"  &December 5  \\
  17  &2h21m22.1s  &57$^{o}$8'35"  &December 1   \\
 19  &2h20m39.7s  &57$^{o}$8'35"  &December 5     \\
 20  &2h20m18.5s  &57$^{o}$8'35"  &December 5     \\
 22  &2h22m4.3s  &57$^{o}$5'35"  &December 1     \\
 23  &2h21m43.9s  &57$^{o}$5'35"  &December 1 \\
 24  &2h21m23.5s  &57$^{o}$5'35"  &December 4  \\
 25  &2h19m17.0s &57$^{o}$10'0" & December 5 \\
\enddata
\end{deluxetable}

%% file: tab2.tex
\begin{deluxetable}{llllllllllllllll}
\tiny
\tabletypesize{\tiny}
\tablecolumns{16}
\tablecaption{h \& $\chi$ Persei data}
\tiny
\tablehead{{$\alpha$}&{$\delta$}&{J}&{H}&{$K_{s}$}&{[3.6]}&{[4.5]}&{[5.8]}&{[8]}&{$\sigma$(J)}&{$\sigma$(H)}&{$\sigma$($K_{s}$)}&{$\sigma$([3.6])}&{$\sigma$([4.5])}&{$\sigma$([5.8])}&{$\sigma$([8])}}
\startdata
34.9281&56.9391&14.394&13.875&13.691&13.613&13.619&13.451&13.583&.034&.038&.049&.003&.428&.07&.118\\
35.0360&56.8215&14.229&13.654&13.656&13.517&13.562&13.883&99.0&.035&.042&.048&.039&.056&.098&99.0\\
35.0483&57.0313&14.516&13.983&13.857&13.752&13.781&13.771&13.563&.034&.037&.057&.006&.017&.104&.124\\
35.0523&56.7621&12.421&12.001&11.922&11.816&11.844&11.920&11.903&.024&.030&.023&.057&.005&.022&.030\\
35.1106&56.8819&8.795&8.049&7.836&8.324&7.9730&7.787&7.731&.018&.024&.018&.014&.168&.002&.003\\
\enddata
\tablecomments{First five entries in our photometry catalogue from 2MASS and IRAC}.
\end{deluxetable}

%% file: tab3.tex
\begin{deluxetable}{llll}
\tablecolumns{4}
\tablecaption{Models for estimating the IR excess population}
\tablehead{{Model} & {Membership Determination} & {Membership Cutoff}& {J cutoff}}
\startdata
1 (more restrictive)& V/V-J colors$^{1}$ & $|$V-V(isochrone)$|$ $\le 0.3$ & 14.5\\\\
2 (less restrictive)& J/J-H colors$^{2}$ & J-J(isochrone) $\ge -0.75$, $\le 0.3$ & 15.0\\\\
\enddata
\tablecomments{Brief description of the two models for identifying IR excess sources.  In the first model we require that
sources have optical data from S02 and use the V/V-J diagram to constrain cluster membership.  We analyze sources to J=14.5.  In the second model we
use the J/J-H diagram from 2MASS to constrain cluster membership.  We analyze sources to J=15.. The membership cutoff refers
to how far away a source can be from the 13 Myr isochrone and still be classified as an h \& $\chi$ Persei member.  The cutoff limits
in the second model include the effect of binarity.}
\tablenotetext{1}{Data from S02 and 2MASS}
\tablenotetext{2}{Data from 2MASS}
\end{deluxetable}

%% file: ms.bbl
\begin{thebibliography}{}
\bibitem[Al (2006)]{Al06} Alexander, R., et al., 2006, MNRAS, 369, 229
\bibitem[Al (2005)]{Al05} Alibert, Y. et al., 2005, ApJL, 626, 57
\bibitem[Al (2004)]{Al04} Allen, L., et al. 2004, ApJS, 154, 363
\bibitem[Ba (1998)]{Ba98} Baraffe, I., et al., 1998, A\&A, 337, 403
\bibitem[Br (1996)]{Br96} Bernasconi, P., 1996, A\&AS, 120, 57
\bibitem[Be (1996)]{Be96} Bertin, E. \& Arnouts, S., 1996, A\&AS, 117, 393
\bibitem[Bi (1987)]{Bi87} Binney, J \& Tremaine, S., 1987, Galactic Dynamics (Princeton: Princeton Univ. Press)
\bibitem[Bo (1964)]{BO64} Borgman, J. \& Blaauw, A., 1964, BAN, 17, 358
\bibitem[Br (2002)]{Br02} Bragg, A. \&  Kenyon, S., 2002, AJ, 124, 3289
\bibitem[Br (2004)]{Br04} Bragg, A., 2004, Ph.D. thesis, Harvard University
\bibitem[BK (2005)]{BK05} Bragg, A. \& Kenyon, S., 2005, AJ, 130, 134
\bibitem[Ca (2002)]{Ca02} Capilla, G. \& Fabregat, J., 2002, A\& A, 394, 479
\bibitem[Ca (2001)]{Ca01} Carpenter, J., 2001, AJ, 121, 2851
\bibitem[Ca (2006)]{Ca06} Carpenter, J., et al., 2006, astro-ph/0609372
\bibitem[Cl (2001)]{Cl01} Clarke, C., et al., 2001, MNRAS, 328, 485
\bibitem[Cr (1970)]{Cr70} Crawford, D., et al.,1970, AJ, 75, 822
\bibitem[Cu (2005)]{Cu05} Currie, T., 2005, ApJ, 629, 549
\bibitem[Cu (2007a)]{Cu07a} Currie, T., et al., 2007, in prep.
\bibitem[Cu (2006b)]{Cu07b} Currie, T., et al., 2007, in prep.
\bibitem[Do (1991)]{Do91} Dougherty, S., et al., 1991, AJ, 102, 1753
\bibitem[Do (1994)]{Do94} Dougherty, S., et al., 1994, A\&A, 290, 609
\bibitem[Du (2005)]{Du05} Dullemond, C. \&  Dominik, C., 2005, A\&A, 434, 971
\bibitem[Fa (2004)]{Fa04} Fazio, G. G., et al., 2004, ApJS, 154, 10
\bibitem[Gu (2004)]{Gu04} Gutermuth, R., et al., 2004, ApJS, 154, 374
\bibitem[Ha (2001)]{Ha01} Haisch, K., et al., 2001, ApJL, 553, 153
\bibitem[Ha (2005)]{Ha05} Hartmann, L., et al., 2005, ApJ, 629, 881
\bibitem[Hi (1997)]{Hi97} Hillenbrand, L., 1997, AJ, 113,1733
\bibitem[Hi (1998)]{Hi98_01} Hillenbrand, L., et al., 1998, ApJ, 116, 1816
\bibitem[Hl (1998)]{Hi98_02} Hillenbrand, L. \&  Hartmann, L., 1998, ApJ, 492, 540
\bibitem[Hi (2005)]{Hi05} Hillenbrand, L., 2005, astro-ph/0511083
\bibitem[Ke (2001)]{Ke01} Keller, S.C., et al., 2001, AJ, 122, 248
\bibitem[Ke (1996)]{Ke96} Kenyon, S., et al., 1996, ApJ, 462, 439
\bibitem[KB (2004)]{Kb04} Kenyon, S. \&  Bromley, B., 2004, ApJL, 602, 133
\bibitem[KH (1995)]{Kh95} Kenyon, S. \&  Hartmann, L., 1995, ApJS, 101, 117
\bibitem[La (2004)]{La04} Laughlin, G., et al., 2004, ApJL, 612, 72
\bibitem[Ma (2004)]{Ma04} Mamajek, E., et al., 2004, ApJ, 612, 496
\bibitem[MB (2001)]{Mb01} Marco, A. \& Benabeau, 2001, A\&A, 372, 477
\bibitem[Massey(2003)]{2003ARA&A..41...15M} Massey, P.\ 2003, \araa, 41, 15
\bibitem[MC (2006)]{Mc06} McCabe, C., et al., 2006, ApJ, 636, 932
\bibitem[Oo (1937)]{Oo37} Oosterhoff, P., 1937, Ann. Sternw. Leiden, 17, 1
\bibitem[Pl (2005)]{Pl05} Plavchan, P., et al., 2005, ApJ, 631, 1161
\bibitem[Qu (2004)]{Qu04} Quijada, M., et al., 2004, SPIE, 5487, 244
\bibitem[Re (2005)]{Re05} Reach, W., et al., 2005, PASP, 117, 978
\bibitem[Sc (1967)]{Sc67} Schild, R., 1967, ApJ, 148, 449
\bibitem[Si (2000)]{Si00} Siess, L., et al., 2000, A\&A, 358, 593
\bibitem[Sk (2006)]{Sk06} Skrutskie, M. F., et al., 2006, AJ, 131, 1163
\bibitem[Sl (2002)]{Sl02} Slesnick, C., et al., 2002, ApJ, 576, 880
\bibitem[St (1989)]{St89} Strom, K., et al., 1989, AJ, 97, 1451
\bibitem[Ta (1984)]{Ta84} Tapia, M. et al., 1984, RMxAA, 9, 65
\bibitem[Vo (1971)]{Vo71} Vogt, N., et al. 1971, A\&A, 11, 359
\bibitem[Wi (1964)]{Wi64} Wildey, R., 1964, ApJS, 8, 439
\bibitem[YC (2004)]{Yc04} Youdin, A. \&  Chiang, E., 2004, ApJ, 601, 1109
\bibitem[YS (2002)]{Ys02} Youdin, A. \&  Shu, F., 2002, ApJ, 580, 494
\bibitem[Yo (2004)]{Yo04} Young, E., et al., 2004, ApJS, 154, 428
\end{thebibliography}
